\documentclass[prb,onecolumn,showpacs]{revtex4}
\usepackage{amssymb}
\usepackage{graphicx}

\input{tcilatex}

\begin{document}

\title{PHASE STRING THEORY FOR DOPED ANTIFERROMAGNETS}
\author{Zheng-Yu Weng}
\affiliation{\textit{Center for Advanced Study, Tsinghua University, Beijing 100084, China%
}}
\date{{\small \today}}

\begin{abstract}
The recent developments of the phase string theory for doped
antiferromagnets will be briefly reviewed. Such theory is built upon a
singular phase string effect induced by the motion of holes in a doped
antiferromagnet, which as a precise property of the $t$-$J$ model dictates
the novel competition between the charge and spin degrees of freedom. A
global phase diagram including the antiferromagnetic, superconducting, lower
and upper pseudogap, and high-temperature \textquotedblleft
normal\textquotedblright\ phases, as well as a series of anomalous physical
properties of these phases will be presented as the self-consistent and
systematic consequences of the phase string theory.
\end{abstract}

\maketitle

\section{INTRODUCTION}

Since the discovery of high-$T_{c}$ superconductivity two decades ago, a
great effort has been put into the search for a unified microscopic
mechanism for both superconductivity as well as anomalous spin and charge
properties in the cuprates.

In this paper, I shall review a systematic endeavor along a particular line
approaching the doped Mott insulator/doped antiferromagnet, which has been
proposed\cite{anderson1} as the unique property of the cuprates due to the
strong Coulomb interaction. It has gradually become a consensus that the
doped Mott insulator physics holds the key to understanding the cuprate
superconductor, and distinguishes the latter from a conventional BCS
superconductor.

The present line, known as the phase string theory, will be
characteristically different from the main-stream approach to a doped Mott
insulator. The latter has been mainly built on the resonating valence bond
(RVB) pairing of \emph{fermionic }spins and the spin-charge separation in
the early proposals,\cite{anderson1,anderson2,anderson3} the latest
developments of which have been summarized and reviewed in Refs. \cite%
{pBCSwf,f-rvb}.

The phase string theory, on the other hand, has been built on a singular
nonlocal effect hidden in a typical doped Mott insulator, \emph{e.g.}, the $%
t $-$J$ model. This effect is largely omitted in other approaches, but is
critical in constructing a self-consistent theory of the doped
antiferromagnet evolving continuously from the half-filling
Mott-antiferromagnetic insulator. I shall present a rich phase diagram with
complex phenomena as the physical consequences of the phase string effect.
One will see that the RVB and spin-charge separation concepts remain
essential, but they will also acquire a distinct mathematical
characterization in this theory.

Although the leading order results of the phase string theory share some
striking similarity with the experimental measurements in the cuprates, in
this review, a direct comparison with experiment will be \emph{minimal},
partly due to the length limitation and partly because I wish to emphasize
that the physical consequences naturally flow from the \emph{theoretical
structure}, not from a phenomenology based on the experiment. I will not be
able to discuss many interesting theoretical efforts along \emph{different }%
lines of thought in this brief review, also due to the space limitation,
which may be found through our original papers in the references.

The rest of the paper will be organized as follows. In Sec. 2, I will
discuss some important general properties of the $t$-$J$ model, including
the Marshall sign rule, bosonic RVB description, phase string effect, and
exact phase string formalism. In Sec. 3, based on the phase string
formalism, I will describe how an effective theory, known as the phase
string model, is constructed. In Sec. 4, the physical consequences of the
phase string model will be given which cover the superconducting phase,
lower and upper pseudogap phases, high-temperature normal state, as well as
low-doping antiferromagnetic state. Finally, a synthesis and perspectives
are presented in Sec. 5.

\section{GENERAL PROPERTIES OF THE $t$-$J$\ MODEL}

The $t$-$J$ model is defined by $H_{t-J}=H_{t}+H_{J}$:%
\begin{equation}
H_{t}=-t\sum_{\left\langle ij\right\rangle }c_{i\sigma }^{\dagger
}c_{j\sigma }+h.c.  \label{ht0}
\end{equation}%
and
\begin{equation}
H_{J}=J\sum_{\left\langle ij\right\rangle }\left( \mathbf{S}_{i}\cdot
\mathbf{S}_{j}-\frac{n_{i}n_{j}}{4}\right)  \label{hj0}
\end{equation}%
with a restricted Hilbert space under the no double occupancy constraint
\begin{equation}
\sum_{\sigma }c_{i\sigma }^{\dagger }c_{i\sigma }\leq 1.  \label{constraint}
\end{equation}

Due to the no double occupancy constraint, the $t$-$J$ model describes a
Mott insulator at half filling with $n_{i}\equiv \sum_{\sigma }c_{i\sigma
}^{\dagger }c_{i\sigma }=1$, where the hopping term $H_{t}=0$ and the
superexchange term $H_{J}$ depicts the Heisenberg antiferromagnetic (AF)
interaction in the unfrozen spin degrees of freedom. Away from the half
filling, charge carriers are introduced with removing or injecting electrons
into the system, which are known as the hole or electron-doped Mott
insulators. Since the half-filling case is an AF spin state, the doped
system can be also properly called a doped antiferromagnet.

For a bipartite lattice with only the nearest-neighboring (nn) hopping and
superexchange couplings [denoted by $\left\langle ij\right\rangle $ in Eqs. (%
\ref{ht0}) and (\ref{hj0})], there exists a particle-hole symmetry and one
may only focus on the hole-doped side without loss of generality. Note that
the next nn hopping may be important in real materials and its effect will
be commented on in Sec. 5.

\subsection{Half-Filling: A Bosonic RVB Description}

At half-filling, the $t$-$J$ model reduces to the Heisenberg model, where
the physical properties are fairly well understood, as a consensus, in
contrast to the doped case. This model predicts a long-range N\'{e}el order
in the ground state as well as low-lying spin-wave excitations, consistent
with experiment.

It is important to point out that in the study of the doped case, one needs
to have a correct description of spin correlations not only at long
distance, but at \emph{all ranges}. This is because the hole hopping in $%
H_{t}$ involves the nn sites, which\ will be generally quite sensitive to
the short-range spin correlations. Thus, as a starting point, a precise
description of both long and short range spin correlations at half-filling
is essential.

\subsubsection{Marshall sign rule}

As proven by Marshall,\cite{marshall} the ground-state wave function of the
Heisenberg model for a bipartite lattice must be real and satisfies a sign
rule. This sign rule dictates that the flip of a pair of antiparallel spins
at two opposite sublattice sites will always be accompanied by a sign change
in the wave function.

This Marshall sign rule may be easily understood as below. Define a spin
basis state with the built-in Marshall signs as%
\begin{equation}
|\phi \rangle =(-1)^{N_{A}^{\downarrow }}|\uparrow {\tiny \cdot \cdot \cdot }%
\downarrow \uparrow {\tiny \cdot \cdot \cdot }\downarrow \rangle
\label{marshall}
\end{equation}%
where $N_{A}^{\downarrow }$ denotes the total number of down spins at the $A$
sublattice such that the aforementioned Marshall sign rule is always
satisfied. Then it is straightforward to verify that the matrix element of $%
H_{J}$ under the complete set $\{|\phi \rangle \}$ is negative definite%
\begin{equation}
\langle \phi ^{\prime }|H_{J}|\phi \rangle \leq 0  \label{matrix}
\end{equation}%
so that for the ground state
\begin{equation}
|\Psi _{0}\rangle =\sum_{\{\phi \}}\chi _{\phi }|\phi \rangle  \label{gs0}
\end{equation}%
the wave function $\chi _{\phi }$ is always real and positive (except for a
trivial global phase).

\subsubsection{Liang-Doucot-Anderson wave function}

While the exact ground state of the Heisenberg model in two dimensions (2D)
is not known, the best variational state proposed by Liang, Doucot and
Anderson is given by\cite{lda}
\begin{equation}
|\Psi _{0}\rangle =\sum_{i\in Aj\in B}W_{i_{1}j_{1}}\ldots
W_{i_{n}j_{n}}(i_{1}j_{1})\ldots (i_{n}j_{n})~,  \label{lda}
\end{equation}%
where $(ij)$ stands for a singlet spin pairing at two opposite sublattice
sites $i$ and $j$, and $W_{ij}$, the positive weight factor associated with
it. Since each singlet bond $(ij)$ automatically satisfies the Marshall sign
rule, the total wave function can be easily shown to obey the sign rule as
well.

The variational wave function in Eq. (\ref{lda}) can produce\cite{lda} an
unrivaled accurate ground-state energy ($-0.3344J$ per bond as compared to
the exact numerical value of $-0.3346J$ per bond for the Heisenberg model).
Since the energy of the Heisenberg model is directly related to the nn
spin-spin correlation, a good variational energy also means a good
description of short-range spin correlations. On the other hand, this wave
function possesses an AF long-range order (AFLRO) in the spin correlation
function with a similarly accurate magnetization as the system size is
extrapolated to infinity.\cite{lda}

\subsection{Doping: Phase String Effect}

Now let us consider doping. In the above we have seen that the ground state
wave function of the Heisenberg model satisfies the Marshall sign. In fact,
such a Marshall sign rule would hold even at arbitrary doping, if holes
remain static on lattice sites. Consider the single hole case for example.
Define the following spin basis incorporating the Marshall signs similar to
Eq. (\ref{marshall})
\begin{equation}
|\phi ;(n)\rangle =(-1)^{N_{A}^{\downarrow }}|\uparrow {\tiny \cdot \cdot
\cdot }\downarrow o\uparrow {\tiny \cdot \cdot \cdot }\downarrow \rangle
\label{marshall1}
\end{equation}%
where $n$ denotes the hole site. Then it is straightforward to check that
\begin{equation}
\langle \phi ^{\prime };(n)|H_{J}|\phi ;(n)\rangle \leq 0  \label{matrix-j}
\end{equation}%
still holds to ensure the Marshall sign rule.

But once the hole starts to move, the Marshall sign rule will be scrambled
by the hopping of the hole. Indeed, based on Eq. (\ref{marshall1}) the
hopping matrix is given by

\begin{equation}
\langle \phi ;(m)|H_{t}|\phi ;(n)\rangle =-t\sigma _{m}  \label{matrix-h}
\end{equation}%
where $\sigma _{m}$ is the site $m$ spin index in the state $|\phi
;(n)\rangle $, and $|\phi ;(m)\rangle $ is different from $|\phi ;(n)\rangle
$ by an exchange of the spin $\sigma _{m}$ with the hole at site $n.$ Since $%
\sigma _{m}=\pm 1$, the hopping matrix element is no longer sign definite.

Based on Eq. (\ref{matrix-h}), a hole hops on the lattice will pick up a
product of sequential $+$ and $-$ signs,\cite{WSCT97}

\begin{eqnarray}
\prod\limits_{c}\sigma _{m} &=&(+1)\times (-1)\times (-1)\times ...
\nonumber \\
&\equiv &(-1)^{N_{c}^{\downarrow }}  \label{pstring}
\end{eqnarray}%
where $N_{c}^{\downarrow }$ is the number of $\downarrow $ spins which are
exchanged with the hole during its hopping on a given path $c$. Because of
Eq. (\ref{matrix-j}), the superexchange interaction cannot \textquotedblleft
repair\textquotedblright\ such a \emph{phase string effect} created by the
nn hole hopping.\cite{WSCT97}

For example, the single-hole propagator $G(j,i;E)=\left\langle \Psi
_{0}|c_{j\sigma }^{\dagger }G(E)c_{i\sigma }|\Psi _{0}\right\rangle $ with $%
G(E)=1/(E-H_{t-J})$ may be formally expressed as

\begin{equation}
G(j,i;E)\propto \sum\limits_{c,N_{c}^{\downarrow }}(-1)^{N_{c}^{\downarrow
}}W(c;N_{c}^{\downarrow };E)  \label{propagator}
\end{equation}%
in which for each path $c$ connecting $i$ and $j,$ there is a phase string
factor $(-1)^{N_{c}^{\downarrow }}$ weighted by $W(c;N_{c}^{\downarrow };E)$
with
\begin{equation}
W(c;N_{c}^{\downarrow };E)\geq 0  \label{W}
\end{equation}%
at $E<E_{G}^{0}$, whose proof is based on Eq. (\ref{matrix-j}).\cite{WSCT97}
Here $E_{G}^{0}$ denotes the ground-state energy when the hole remains \emph{%
static}. Similarly, the total energy $E_{\mathbf{k}}$ of the state at a
momentum $\mathbf{k}$ can be also formally expressed in terms of the
Wigner-Brillouin expansion as\cite{WMST01}
\begin{equation}
E_{\mathbf{k}}=E_{G}^{0}-\frac{t}{N}\sum_{ij}e^{i\mathbf{k}\cdot (\mathbf{r}%
_{i}-\mathbf{r}_{j})}M_{ij}~~  \label{ge}
\end{equation}%
where
\begin{equation}
M_{ij}\equiv \sum_{c,N_{c}^{\downarrow }}(-1)^{N_{c}^{\downarrow
}}M[c;N_{c}^{\downarrow }]~~  \label{m}
\end{equation}%
with a positive semi-definite weight functional
\begin{equation}
M[c;N_{c}^{\downarrow }]\geq 0.  \label{M}
\end{equation}

Physically, the phase string effect represents the \emph{transverse spin
mismatches} created by the motion of the hole on the AF background, in
additional to the mismatches in the $S^{z}$ component.\cite{WSCT97} The
irreparableness of the phase string means that the three components of the
spin defect induced by the motion of the hole cannot be \emph{simultaneously}
repaired via the spin flips in the superexchange term. It has been concluded
that either the quasiparticle weight vanishes\cite{WSCT97} or the doped hole
is self-localized\cite{WMST01} due to the phase string effect in a spin
background with long-range AF correlation (in the half-filling limit).

One can similarly demonstrate the irreparable phase string effect in an
\emph{arbitrary} multi-hole case although the formulation will be slightly
modified with the emergence of an extra sign factor $(-1)^{N_{c}^{\mathrm{ex}%
}}$ in additional to $(-1)^{N_{c}^{\downarrow }}$, where $N_{c}^{\mathrm{ex}%
} $ is the number of the exchanges between the \emph{holes} and $c$ here
denotes the multi-hole paths.\cite{WW07}

The significant role of the phase string factor $(-1)^{N_{c}^{\downarrow }}$
in Eqs. (\ref{propagator}) and (\ref{m}) is to cause strong \emph{phase
interference} between different hole paths and thus greatly influence the
charge dynamics. But since the spin degrees of freedom are involved in $%
N_{c}^{\downarrow }$, the spin correlations will also get \emph{%
simultaneously} reshaped to minimize the total kinetic and superexchange
energy. Therefore, the irreparable phase string effect will play a role to
\emph{mediate} nonlocal mutual influence between the charge and spin degrees
of freedom as a new kind of \emph{interaction} which emerges from the
strongly correlated system of doped Mott insulator/antiferromagnet.

The factor $(-1)^{N_{c}^{\downarrow }}=\pm 1$ is very singular, as a
fluctuation in $N_{c}^{\downarrow }$ by $\pm 1$ can result in a total sign
change. So the quantum fluctuations are normally extremely strong especially
for long paths, and there is no well-controlled quantum many-body method to
directly handle such an effect. Alternatively a unitary transformation $e^{i%
\hat{\Theta}}$ can be explicitly introduced\cite{WSCT97}
%e^{iT}=exp[-i?_{i,l}n_{i}^{h}(Imln(z_{i}-z_{l}))n_{l?}^{b}]
to precisely keep track of the phase string effect. Then, with the Marshall
sign basis $|\phi \rangle $ [\emph{cf}, Eq.\ (\ref{marshall})] being changed
to $|\bar{\phi}\rangle =e^{i\hat{\Theta}}|\phi \rangle $, the new
ground-state wave function $\bar{\chi}_{\phi }$ in\ $|\Psi _{G}\rangle
=\sum_{\{\phi \}}\bar{\chi}_{\phi }|\bar{\phi}\rangle $ should become more
or less \textquotedblleft conventional\textquotedblright\ as the singular
phase string effect is now sorted out into $|\bar{\phi}\rangle $.

The $t$-$J$ Hamiltonian in this new representation, known as the phase
string formalism,\cite{WSCT97} is expected to be perturbatively treatable as
the singular part of the phase string effect is \textquotedblleft gauged
away\textquotedblright\ by the unitary transformation. In the following, we
shall present such an exact reformulation of the $t$-$J$ Hamiltonian at
arbitrary doping.

\subsection{Phase string formalism}

The phase string formalism is equivalent to \textquotedblleft
bosonizing\textquotedblright\ the electron operator as follows\cite{WSCT97}
\begin{equation}
c_{i\sigma }=h_{i}^{\dagger }b_{i\sigma }e^{i\hat{\Theta}_{i\sigma }}
\label{mutual}
\end{equation}%
where \textquotedblleft holon\textquotedblright\ $h_{i\sigma }^{\dagger }$
and \textquotedblleft spinon\textquotedblright\ $b_{i\sigma }$ operators are
both \emph{bosonic} fields, satisfying the constraint
\begin{equation}
h_{i}^{\dagger }h_{i}+\sum_{\sigma }b_{i\sigma }^{\dagger }b_{i\sigma }=1~.
\label{constr}
\end{equation}%
The nonlocal phase string factor $e^{i\hat{\Theta}_{i\sigma }}$ in Eq. (\ref%
{mutual}) is defined by
\begin{equation}
e^{i\hat{\Theta}_{i\sigma }}\equiv e^{i\frac{1}{2}\left[ \Phi _{i}^{s}-\Phi
_{i}^{0}-\sigma \Phi _{i}^{h}\right] }~(\sigma )^{\hat{N}_{h}}(-\sigma )^{i},
\label{phase}
\end{equation}%
where
\begin{eqnarray}
\Phi _{i}^{s} &=&\sum_{l\neq i}\mbox{Im ln
$(z_i-z_l)$}\left( \sum_{\alpha }\alpha n_{l\alpha }^{b}\right) ~,
\label{phis} \\
\Phi _{i}^{0} &=&\sum_{l\neq i}\mbox{Im ln
$(z_i-z_l)$}~,  \label{phi0}
\end{eqnarray}%
and
\begin{equation}
\Phi _{i}^{h}=\sum_{l\neq i}\mbox{Im ln
$(z_i-z_l)$}n_{l}^{h}~,  \label{phih}
\end{equation}%
in which $n_{l\alpha }^{b}$ and $n_{l}^{h}$ are spinon and holon number
operators respectively, at site $l$ with $z_{l}=x_{l}+iy_{l}$ a complex
coordinate on the lattice.

It is easily verified that the fermionic statistics of $c_{i\sigma }$ is
automatically ensured by $e^{i\hat{\Theta}_{i\sigma }}$, in which the factor
$(\sigma )^{\hat{N}_{h}}$ ($\hat{N}_{h}$ is the total holon number operator)
guarantees anticommutation relations between opposite spins, and the factor $%
(-1)^{i}=\pm 1$ (for $i\in $ even/odd) is a staggered factor added for
convenience. Furthermore, the equality (\ref{constr}) replaces the original
no double occupancy constraint (\ref{constraint}) imposed on the electron
operator. Therefore, this is an exact representation of the electron
operator in the Hilbert space constrained by the no double occupancy
condition.

\subsubsection{Nontrivial gauge structure}

Rewriting the $t-J$ model using this new electron decomposition (\ref{mutual}%
), one gets\cite{WSCT97}
\begin{equation}
H_{t}=-t\sum_{\langle ij\rangle \sigma }h_{i}^{\dagger }h_{j}b_{j\sigma
}^{\dagger }b_{i\sigma }e^{i\left( A_{ij}^{s}-\phi _{ij}^{0}-\sigma
A_{ij}^{h}\right) }+h.c.~,  \label{ht}
\end{equation}%
and%
\begin{equation}
H_{J}=-\frac{J}{2}\sum_{\langle ij\rangle }~\left( \hat{\Delta}%
_{ij}^{s}\right) ^{\dagger }\hat{\Delta}_{ij}^{s}~,  \label{hj}
\end{equation}%
with
\begin{equation}
\hat{\Delta}_{ij}^{s}\equiv \sum_{\sigma }e^{-i\sigma A_{ij}^{h}}b_{i\sigma
}b_{j-\sigma }~.  \label{brvb}
\end{equation}

Here the three link fields defined on the nn sites are given by
\begin{equation}
A_{ij}^{s}\equiv \frac{1}{2}\sum_{l\neq i,j}%
\mbox{Im ln
$[\frac{z_{i}-z_{l}}{z_{j}-z_{l}}]$}\left( \sum_{\sigma }\sigma n_{l\sigma
}^{b}\right) ~,  \label{as}
\end{equation}%
\begin{equation}
\phi _{ij}^{0}\equiv \frac{1}{2}\sum_{l\neq i,j}%
\mbox{Im ln
$[\frac{z_{i}-z_{l}}{z_{j}-z_{l}}]$},  \label{sphi0}
\end{equation}%
and
\begin{equation}
A_{ij}^{h}\equiv \frac{1}{2}\sum_{l\neq i,j}%
\mbox{Im ln
$[\frac{z_{i}-z_{l}}{z_{j}-z_{l}}]$}n_{l}^{h}~.  \label{ah}
\end{equation}%
The strengths of these link fields can be obtained as follows
\begin{equation}
\sum_{c}A_{ij}^{s}=\pi \sum_{l\in \Sigma _{c}}\left( n_{l\uparrow
}^{b}-n_{l\downarrow }^{b}\right) ,  \label{cond1}
\end{equation}%
and
\begin{equation}
\sum\nolimits_{c}A_{ij}^{h}=\pi \sum_{l\in \Sigma _{c}}n_{l}^{h},
\label{cond2}
\end{equation}%
for an arbitrary closed loop $c$ such that the fluxes enclosed, $\sum_{c}$ $%
A_{ij}^{s}$ and $\sum\nolimits_{c}A_{ij}^{h}$, are determined by the number
of spinons and holons respectively, in the region $\Sigma _{c}$ enclosed by
the loop $c$. Furthermore, the phase $\phi _{ij}^{0}$ describes a constant
flux with a strength $\pi $ per plaquette:
\begin{equation}
\sum_{{\large \Box }}\phi _{ij}^{0}=\pm \pi .  \label{fphi}
\end{equation}

The unique feature in the above phase string formalism of the $t-J$ model is
the emergence of three link fields: $A_{ij}^{s}$, $A_{ij}^{h},$ and $\phi
_{ij}^{0}$. Without them, there should be \emph{no} nontrivial sign problem
in the Hamiltonian, because $h$ and $b$ are both bosonic fields. Namely the
matrix elements of $H_{t-J}$ would be real and negative-definite in the
occupation number representation of $h$ and $b$. Consequently, the ground
state expanded in terms of these bosonic fields would have real and positive
coefficients, which is the case at half-filling as discussed above and the
one-dimensional (1D) case to be discussed below.

It is easy to see that the Hamiltonian $H_{t-J}$ is invariant under \textrm{%
U(1)}$\times $\textrm{U(1)} gauge transformations:
\begin{equation}
h_{i}\rightarrow h_{i}e^{i\varphi _{i}}~,\text{ \qquad \quad }%
A_{ij}^{s}\rightarrow A_{ij}^{s}+(\varphi _{i}-\varphi _{j})~,  \label{u(1)1}
\end{equation}%
and
\begin{equation}
b_{i\sigma }\rightarrow b_{i\sigma }e^{i\sigma \theta _{i}}~,\text{ \qquad }%
A_{ij}^{h}\rightarrow A_{ij}^{h}+(\theta _{i}-\theta _{j})~.  \label{u(1)2}
\end{equation}%
Thus $A_{ij}^{s}$ and $A_{ij}^{h}$ are gauge fields, seen by holons and
spinons respectively, as the latter carry their gauge charges according to (%
\ref{u(1)1}) and (\ref{u(1)2}).

Here $A_{ij}^{s}$ and $A_{ij}^{h}$ are not independent gauge fields with
their own dynamics. Rather they are directly connected to the matter fields
as a pair of \emph{mutual} topological gauge fields. The term
\textquotedblleft mutual\textquotedblright\ refers to the fact that $%
A_{ij}^{s}$ describes quantized $\pi $ fluxoids attached to the spinons,
coupled to the holons. Conversely, $A_{ij}^{h}$ describes quantized $\pi $
fluxoids bound to the holons, coupled to the spinons.

By the construction, the phase string formalism is defined in a Hilbert
space where the total $S^{z}$ is an eigen operator.\cite{WSCT97} So the
total numbers of $\uparrow $ and $\downarrow $ spinons are conserved
respectively, such that the topological gauge field $A_{ij}^{s}$ behaves
smoothly as defined in (\ref{cond1}). It is also consistent with the gauge
invariance under (\ref{u(1)2}). Different $S^{z}$ states can be connected by
the spin flip operators, defined in the phase-string representation as
\begin{equation}
S_{i}^{+}=\left[ (-1)^{i}e^{i\Phi _{i}^{h}}\right] b_{i\uparrow }^{\dagger
}b_{i\downarrow },  \label{s+}
\end{equation}%
(a factor $(-1)^{\hat{N}_{h}}$ has been dropped for simplicity) and $%
S_{i}^{-}=(S_{i}^{+})^{\dagger }$, and $S_{i}^{z}=\sum_{\sigma }\sigma
b_{i\sigma }^{\dagger }b_{i\sigma }$. These definitions follow from (\ref%
{mutual}). The nonlocal phase $\Phi _{i}^{h}$ in (\ref{s+}) will play a
crucial role in restoring the spin rotational symmetry.

Finally, the superconducting order parameter can be expressed in the phase
string representation as follows

\begin{eqnarray}
\hat{\Delta}_{ij}^{\mathrm{SC}} &\equiv &\sum_{\sigma }\sigma c_{i\sigma
}c_{j-\sigma }\ \   \nonumber \\
&=&e^{i\frac{1}{2}\left( \Phi _{i}^{s}+\Phi _{j}^{s}\right) }\hat{\Delta}%
_{ij}^{0}~,  \label{sco}
\end{eqnarray}%
with the amplitude operator given by
\begin{equation}
\hat{\Delta}_{ij}^{0}\equiv \left[ (-1)^{j}e^{-i\Phi _{j}^{0}-i\phi
_{ij}^{0}}\right] h_{i}^{\dagger }h_{j}^{\dagger }\hat{\Delta}_{ij}^{s}\text{
}~  \label{sca}
\end{equation}%
(again the factor $(-1)^{\hat{N}_{h}}$ is omitted).

\subsubsection{One-dimensional case}

In the 1D case, one may define

\begin{equation}
\mbox{Im ln
$(z_i-z_l)$}=\QATOPD\{ . {{{\pm \pi }\text{ \ \ \ \ }{\text{\ \textrm{if }}%
i<l,}}}{{0\text{ \ \ \ \ \ \ }{\text{\textrm{if }}i>l,}}}  \label{1dtheta}
\end{equation}%
such that
\begin{equation}
A_{ij}^{s}=\phi _{ij}^{0}=A_{ij}^{h}=0.
\end{equation}%
Thus there is no sign problem in the phase string representation of the $t$-$%
J$ model. It implies that the Hamiltonian may be treated within a
\textquotedblleft mean field\textquotedblright\ approximation.\cite{WSCT97}
Namely the holons and spinons defined in the phase string representation of
the $t$-$J$ model may be regarded as the true \textquotedblleft
free\textquotedblright\ elementary excitations.

However, the correlation functions will be highly nontrivial because of the
singular phase string effect, which is now precisely kept in the phase
factor of the decomposition (\ref{mutual}) with\cite{WSCT97}

\begin{equation}
c_{i\sigma }=h_{i}^{\dagger }b_{i\sigma }e^{\pm i\left[ \sigma \Theta
_{i}^{h}+\Theta _{i}^{b}\right] }~(\sigma )^{\hat{N}_{h}},  \label{mutual-1d}
\end{equation}%
in which
\begin{equation}
\Theta _{i}^{h}=\frac{\pi }{2}\sum\limits_{l>i}(1-n_{l}^{h}),  \label{thetah}
\end{equation}%
and%
\begin{equation}
\Theta _{i}^{b}=\frac{\pi }{2}\sum\limits_{l>i,\alpha }\alpha n_{l\alpha
}^{h}.  \label{thetab}
\end{equation}

Thus, to create a hole by $c_{i\sigma },$ according to Eq. (\ref{mutual-1d}%
), means the creation of a pair of holon and spinon excitations together
with a\emph{\ nonlocal phase shift}. Denoting the average hole concentration
$\left\langle n_{l}^{h}\right\rangle =\delta $, the phase string factor in
Eq. (\ref{mutual-1d}) can be rewritten as%
\begin{equation}
e^{\pm i\left[ \sigma \Theta _{i}^{h}+\Theta _{i}^{b}\right] }\propto e^{\pm
i\sigma k_{f}x_{i}}e^{\pm i\Delta \Phi _{i}}
\end{equation}%
where%
\begin{equation}
k_{f}=\frac{\pi }{2a}(1-\delta )
\end{equation}%
is the Fermi momentum ($a$ is the lattice constant) and

\begin{equation}
\Delta \Phi _{i}=-\frac{\pi }{2}\sum\limits_{l>i}\sigma (n_{l}^{h}-\delta )+%
\frac{\pi }{2}\sum\limits_{l>i,\alpha }\alpha n_{l\alpha }^{h}
\end{equation}%
with $\left\langle \Delta \Phi _{i}\right\rangle =0$. While the leading term
of the phase string factor reproduces the \emph{correct} Fermi momentum $%
k_{f}$ for the electron system, the fluctuations in $\Delta \Phi _{i}$ will
be responsible for reproducing\cite{WSCT97} the \emph{correct} Luttinger
liquid behavior known from the large-$U$ Hubbard model.

The important connection between the phase string effect and the Luttinger
liquid in 1D has been first established previously in a path-integral study%
\cite{WSTS91} of the large-$U$ Hubbard model.

\subsubsection{Two-dimensional case}

At half-filling, $H_{t}$ has no contribution due to the no double occupancy
constraint, and under a proper gauge choice one may set $A_{ij}^{h}=0$ in $%
H_{J}$. In this limit, there is no nontrivial sign problem in the 2D
Hamiltonian which is fully \emph{bosonized}. This is the case previously
discussed in Sec. 2.1, where a precise bosonic RVB description of spin
correlations in all ranges of length scale is available, which can serve a
very good starting point for the doped case in 2D.

In contrast to the full bosonization case at half-filling, as well as in the
1D case, the nontrivial phase structure emerges at finite doping in 2D,
which are represented by the link fields, $A_{ij}^{s}$, $\phi _{ij}^{0}$,
and $A_{ij}^{h}.$ These link phases can no longer be \textquotedblleft
gauged away\textquotedblright\ here and they \emph{completely} capture the
essential sign problem (\emph{i.e.}, the phase string effect) of the doped $%
t $-$J$ model. These gauge fields are generally well controlled in the
regimes of our interest: $\phi _{ij}^{0}$ is a non-dynamic phase describing
a constant $\pi $ flux per plaquette; $A_{ij}^{s}$ is cancelled when spinons
are RVB paired at low-temperature phases; $A_{ij}^{h}$ remains weak at small
doping or well behaves if the holons are coherent. Therefore, these gauge
fields will be well tractable at least in low doping and low temperature
regimes.

It is noted that the \emph{bosonization} decomposition (\ref{mutual}) was
actually first obtained\cite{WST95} based on optimizing a slave-boson
mean-field state using a \textquotedblleft flux binding\textquotedblright\
scheme. Similar procedure has been also employed recently to get essentially
the same bosonization decomposition in Ref. \cite{wang}. This bosonization
decomposition may be also regarded as the \emph{mutual-semion} scheme as
described in Ref. \cite{WSCT97} without explicitly breaking the
time-reversal symmetry. It is thus clearly distinguished\cite{WST95} from an
earlier flux-binding construction leading to a slave-semion type of
formulation,\cite{WST94} or a variant of it in a more complicated semionic
representation proposed\cite{marchetti} in literature.

\subsection{Wave function structure}

In the above, we have discussed how the \emph{intrinsic} \emph{phase
structure }of the $t$-$J$ model can be revealed in the exact phase string
formalism. In the following we further examine the corresponding wave
function structure.

A wave function $\psi _{e}$ in the electron $c$-operator representation can
be related to $\psi _{b}$ in the full bosonic $h$ and $b$ representation of
the phase string formalism by\cite{WZM05}
\begin{equation}
\psi _{e}(i_{1},\cdot \cdot \cdot ,i_{M};j_{1},\cdot \cdot \cdot
,j_{N_{e}-M})=\mathcal{K}\text{ }\psi _{b}(i_{1},\cdot \cdot \cdot
,i_{M};j_{1},\cdot \cdot \cdot ,j_{N_{e}-M};l_{1},\cdot \cdot \cdot
,l_{N_{h}})~  \label{wavefunction}
\end{equation}%
where the $\uparrow $ spin electron sites, $\{i_{u}\}=i_{1},\cdot \cdot
\cdot ,i_{M},$ and the $\downarrow $ spin sites, $\{j_{d}\}=j_{1},\cdot
\cdot \cdot ,j_{N_{e}-M},$ and $\{l_{h}\}=l_{1},\cdot \cdot \cdot ,l_{N^{h}}$
denote the empty sites that are \emph{not} independent from $\{i_{u}\}$ and $%
\{j_{d}\}$ under the no double occupancy constraint. Here and below, we use $%
i$ to specify an $\uparrow $ spin, $j$ a $\downarrow $ spin, and $l$, a
holon, where the subscripts $u$, $d$, and $h$ label the sequences of the $%
\uparrow $ spins, $\downarrow $ spins, and holons, respectively.

According to Eq. (\ref{mutual}), the $\mathcal{K}$ factor is given by\cite%
{WZM05}

\begin{equation}
\mathcal{K}=\mathcal{JG}~,  \label{k2}
\end{equation}%
where
\begin{equation}
\mathcal{J}\equiv \prod_{u<u^{^{\prime }}}(z_{i_{u}}^{\ast }-z_{i_{u^{\prime
}}}^{\ast })\prod_{d<d^{^{\prime }}}(z_{j_{d}}^{\ast }-z_{j_{d^{\prime
}}}^{\ast })\prod_{ud}(z_{i_{u}}^{\ast }-z_{j_{d}}^{\ast
})\prod_{h<h^{\prime }}\left\vert z_{l_{h}}-z_{l_{h^{\prime }}}\right\vert
\prod_{uh}|z_{i_{u}}-z_{l_{h}}|\prod_{dh}|z_{j_{d}}-z_{l_{h}}|~  \label{jh}
\end{equation}%
and
\begin{equation}
\mathcal{G}\equiv \mathcal{C}^{-1}(-1)^{N_{A}^{\uparrow }}\prod_{uh}\frac{%
z_{i_{u}}^{\ast }-z_{l_{h}}^{\ast }}{|z_{i_{u}}-z_{l_{h}}|}~,  \label{g0}
\end{equation}%
in which the coefficient $\mathcal{C}$ is given by
\begin{equation}
\mathcal{C}=\left\vert \mathcal{J}\right\vert =\prod_{k<m}|z_{k}-z_{m}|~,
\label{C}
\end{equation}%
with $k$ and $m$ running through all lattice sites such that $\mathcal{C}$
is a \emph{constant}.

It is easily seen that the Jastrow-like factors in $\mathcal{J}$
automatically enforce the single occupancy constraint: $\mathcal{J}$
vanishes if two spinons (or holons) occupy the same site, or if a holon and
a spinon occupy the same site. The factor $\mathcal{J}$ further explicitly
captures the fermionic statistics of the electrons. Therefore, the no double
occupancy constraint, which has been considered as one of the most important
effects but difficult to tackle with in the $t$-$J$ model, is no longer
important in the phase string representation $\psi _{b}$, since $\mathcal{J}$
in (\ref{k2}) naturally plays the role of a \emph{projection} operator. This
may be understood in the following way. In the phase string representation,
the effect of $\mathcal{K}$ in the original $\psi _{e}$, is transformed into
the topological gauge fields$,$ $A_{ij}^{s}$ and $A_{ij}^{h},$ in the
Hamiltonians, (\ref{ht}) and (\ref{hj}), which describe spinons and holons
as mutual vortices, as perceived by each other. This clearly implies a
mutual \emph{repulsion} between two species, since a spinon cannot stay at
the center of its vortex (where a holon is located), and \textit{vice versa}%
. Thus the constraint that a holon and a spinon cannot occupy the same site
is now reflected in the \emph{interactions} present in the new Hamiltonian,
and the condition (\ref{constr}) is not needed as an extra condition to
enforce. Note that the constraint (\ref{constr}) also requires the hard core
conditions among the holons or spinons themselves. But since both holon and
spinon fields are bosonic fields, local hard core exclusions usually do not
involve the sign change of the wave function. Hence, in the phase string
representation, the local constraint (\ref{constr}) is neither crucial nor
singular, as far as low energy physics is concerned.

Finally, the singular phase string effect is captured by the factor $%
\mathcal{G}$ in $\mathcal{K}$. Firstly the sign factor $(-1)^{N_{A}^{%
\uparrow }}$ can be identified with the Marshall sign, and $N_{A}^{\uparrow
} $ denotes the total number of $\uparrow $ spins in sublattice $A$ (note
that it is equivalent to the previous definition using $(-1)^{N_{A}^{%
\downarrow }},$ by a trivial global sign factor). Then the phase factor $%
\prod_{uh}\frac{z_{i_{u}}^{\ast }-z_{l_{h}}^{\ast }}{|z_{i_{u}}-z_{l_{h}}|}$
will describe the phase string effect -- disordered Marshall sign. Note that
it is asymmetric with regard to $\uparrow $ and $\downarrow $ spins: it only
involves an $\uparrow $ spin complex coordinate $z_{i_{u}}^{\ast }$ and a
holon coordinate $z_{l_{h}}^{\ast }$, and then will acquire the following
additional phase as a hole moves through a closed path $c$, $\mathcal{%
G\rightarrow G}\times (-1)^{N_{c}^{\uparrow }}$, with the displaced spins
being restored to the original configuration by the exchange term $H_{J}$.%
\cite{WZM05}

\section{Phase String Model: Effective Theory}

The exact phase-string formalism of the $t$-$J$ Hamiltonian provides a new
starting point to construct an effective theory which can be smoothly
connected to the better-understood half-filling limit. The \emph{gauge
structure }in the phase-string formalism is a very useful guide for such a
construction as it generally cannot be spontaneously broken according to the
Elitzur's theorem.

\subsection{Phase string model}

Based on the $t$-$J$ Hamiltonian in the phase string formalism, a \emph{%
minimal} effective model may be written down as follows\cite{WST98,WZM05}%
\begin{equation}
H_{\mathrm{string}}=H_{h}+H_{s}  \label{hstring}
\end{equation}%
with
\begin{eqnarray}
H_{h} &=&-t_{h}\sum_{\langle ij\rangle }\left(
e^{iA_{ij}^{s}+ieA_{ij}^{e}}\right) h_{i}^{\dagger }h_{j}+h.c.  \label{hh} \\
H_{s} &=&-J_{s}\sum_{\langle ij\rangle \sigma }\left( e^{i\sigma
A_{ij}^{h}}\right) b_{i\sigma }^{\dagger }b_{j-\sigma }^{\dagger }+h.c.
\label{hs}
\end{eqnarray}%
This model remains invariant under the gauge transformations, (\ref{u(1)1})
and (\ref{u(1)2}), and is thus a gauge model, known as the phase string
model.

The \textrm{U(1)}$\times $\textrm{U(1) }gauge invariance here corresponds to
the charge and spin $S^{z}$ conservations of the holons and spinons,
respectively, which ensures the correct quantum numbers in such a
spin-charge separation description. This is in contrast to the slave-boson
\textrm{U(1)} gauge theory\cite{lee3} where both holon and spinon carry
partial charges. In Eq. (\ref{hh}), an external electromagnetic gauge
potential $A_{ij}^{e}$ is explicitly introduced which couples to the holon
field carrying an electric charge $+e.$ By contrast, the spinon field does
not carry the electric charge and thus describes a charge neutral and spin-$%
1/2$ object, which can directly couple to the external magnetic field $B$
only by a Zeeman term
\begin{equation}
H_{s}^{\mathrm{ZM}}\equiv -\mu _{\mathrm{B}}B\sum_{i\sigma }\sigma
n_{i\sigma }^{b}.  \label{zeeman}
\end{equation}%
Note that, without loss of generality, the magnetic field will be always
added along the spin quantization $S^{z}$ axis due to the requirement of the
$S^{z}$ conservation in the phase string formulation.

The global conditions of
\begin{eqnarray}
\sum_{i}n_{i}^{h} &=&N\delta ,  \label{global-1} \\
\sum_{i\sigma }n_{i\sigma }^{b} &=&N(1-\delta ),  \label{global-2}
\end{eqnarray}%
can be added to $H_{h}$ and $H_{s}$ by the Lagrangian multipliers, $\lambda
_{h}$ and $\lambda ,$ respectively. Due to the relaxation of the local no
double occupancy constraint, to avoid the short-distance uncertainty at each
center of a $\pi $-flux tube, on the right-hand sides of Eqs. (\ref{cond1})
and (\ref{cond2}), the distribution of a holon or spinon at site $l$ should
be understood as being slightly smeared within a \emph{small} area centered
at $l$.

Based on the spin operators defined in the phase string representation like
Eq. (\ref{s+}), It is straightforward to verify the spin rotational
invariance of the phase string model
\begin{equation}
\left[ H_{\mathrm{string}},\mathbf{S}\right] =0
\end{equation}%
where $\mathbf{S=}\sum_{i}\mathbf{S}_{i},$ by noting that $\Phi
_{i}^{h}-\Phi _{j}^{h}=2A_{ij}^{h}$ (using the fact that the core of each
flux-tube being slightly smeared within a small area as mentioned above).
The time-reversal symmetry at $A_{ij}^{e}=0$ can be also easily verified by
noting that $b_{i\sigma }^{\dagger }\rightarrow \sigma b_{i-\sigma
}^{\dagger }$, $h_{i}^{\dagger }\rightarrow h_{i}^{\dagger }$, $%
A_{ij}^{h}\rightarrow A_{ij}^{h}$, and $A_{ij}^{s}\rightarrow -A_{ij}^{s},$
according to their definitions, under the time-reversal transformation.

\subsection{Topological gauge structure and mutual Chern-Simons description}

The phase string model is uniquely featured by the two topological gauge
fields, $A_{ij}^{s}$ and $A_{ij}^{h}$. According to Eqs. (\ref{cond1}) and (%
\ref{cond2}), the holons in (\ref{hh}) feel the presence of the spinons as
quantized $\pi $ fluxoids through $A_{ij}^{s}$, which reflects the nonlocal
frustrations of the spin background on the kinetic energy of the charge
degrees of freedom. \emph{Vice versa }the spinons also perceive the doped
holes as $\pi $ flux quanta through $A_{ij}^{h},$ which represents the
dynamic frustrations of the doped holes on the spin degrees of freedom.

It is instructive to reformulate the above phase string model in the
following path-integral formalism\cite{KQW05}

\begin{equation}
Z=\int DhDb_{\uparrow }Db_{\downarrow }DA^{s}DA^{h}\exp \left(
-\int_{0}^{\beta }d\tau \int d^{2}\mathbf{r}L_{\mathrm{string}}\right)
\label{Z}
\end{equation}%
in which the Euclidean Lagrangian of the phase string model is given by%
\begin{equation}
L_{\mathrm{string}}=L_{h}+L_{s}+L_{CS}  \label{Lstring}
\end{equation}%
where
\begin{eqnarray}
L_{h} &=&\sum_{I}h_{I}^{\dagger }\left[ \partial _{\tau }-iA_{0}^{s}(I)%
\right] h_{I}-t_{h}\sum_{\left\langle IJ\right\rangle }\left(
e^{iA_{IJ}^{s}}h_{I}^{\dagger }h_{J}+c.c.\right)  \label{Lh} \\
L_{s} &=&\sum_{i\sigma }b_{i\sigma }^{\dagger }\left[ \partial _{\tau
}-i\sigma A_{0}^{h}(i)\right] b_{i\sigma }-J_{s}\sum_{\left\langle
ij\right\rangle \sigma }\left( e^{i\sigma A_{ij}^{h}}b_{i\sigma }^{\dagger
}b_{j-\sigma }^{\dagger }+c.c.\right)  \label{Ls} \\
L_{CS} &=&\frac{i}{\pi }\sum_{I}\epsilon ^{\mu \nu \lambda }A_{\mu
}^{s}(I)\partial _{\nu }A_{\lambda }^{h}(i)  \label{Lagrangian-CS}
\end{eqnarray}%
For simplicity, two chemical potential terms enforcing the global
constraints (\ref{global-1}) and (\ref{global-2}) are not included in $L_{%
\mathrm{string}}$.

In such a Lagrangian formalism, two matter fields, bosonic spinons and
holons, are \emph{minimally} coupled to the $U(1)$ gauge fields, $A^{s}$ and
$A^{h}$, whose gauge structure is decided by the mutual-Chern-Simons term $%
L_{CS}$ in (\ref{Lagrangian-CS}), in \emph{replacement} of the original
topological constraints (\ref{cond1}) and (\ref{cond2}). So the phase string
theory is also known as the mutual Chern-Simons theory. The time-reversal,
parity, and spin rotational symmetries can be explicitly shown to be
preserved.\cite{KQW05}

Note that the original constraints (\ref{cond1}) and (\ref{cond2}) can be
obtained by the equations of motion for the temporal components $A_{0}^{h}$
and $A_{0}^{s}$:
\begin{eqnarray}
\frac{\partial L}{\partial A_{0}^{s}(I)} &=&0\Rightarrow \epsilon ^{\alpha
\beta }\Delta _{\alpha }A_{\beta }^{h}(i)=\pi n_{I}^{h}  \label{constraint1}
\\
\frac{\partial L}{\partial A_{0}^{h}(i)} &=&0\Rightarrow \epsilon ^{\alpha
\beta }\Delta _{\alpha }A_{\beta }^{s}(I)=\pi \sum_{\sigma }\sigma
n_{i\sigma }^{b}  \label{constraint2}
\end{eqnarray}%
with $\Delta _{\alpha }A_{\beta }^{h}(i)\equiv A_{\beta }^{h}(i+\hat{\alpha}%
)-A_{\beta }^{h}(i)$ and $\Delta _{\alpha }A_{\beta }^{s}(I)\equiv A_{\beta
}^{s}(I)-A_{\beta }^{h}(I-\hat{\alpha}).$ Here the indices $\alpha $ and $%
\beta $ are used to denote the spatial components ($\alpha ,\beta =x$, $y$),
and the lattice gauge fields $A_{IJ}^{s}\equiv A_{\alpha }^{s}(I)$ ($J=I-%
\hat{\alpha}$) and $A_{ij}^{h}\equiv A_{\alpha }^{h}(j)$ ($i=j+\hat{\alpha}$%
). The lattice sites $I$ and $i$ refer to two sets of \textquotedblleft
dual\textquotedblright\ lattices where the holons and spinons live on,
respectively, which is a technical way to \textquotedblleft
regulate\textquotedblright\ the theory at the short-distance, which is not
expected to change the low-energy physics.\cite{KQW05}

\subsection{Bosonic RVB order parameter}

To justify the above phase string model, let us first consider the
superexchange term $H_{J}$ in Eq. (\ref{hj}).

$H_{J}$ is expressed in terms of the RVB operator $\hat{\Delta}_{ij}^{s}$
which is invariant under the gauge transformation (\ref{u(1)2}). It is
natural to define the bosonic RVB order parameter\cite{WST98}
\begin{equation}
\Delta ^{s}\equiv \left\langle \hat{\Delta}_{ij}^{s}\right\rangle _{nn}
\label{order}
\end{equation}%
for nn sites. At half filling, $\Delta ^{s}$ reduces to the well-known
Schwinger-boson mean-field order parameter\cite{AA} $\Delta ^{\mathrm{SB}%
}=\left\langle \sum_{\sigma }b_{i\sigma }b_{j-\sigma }\right\rangle
_{nn}\equiv \left\langle \hat{\Delta}_{ij}^{\mathrm{SB}}\right\rangle _{nn}$
as $A_{ij}^{h}$ $=0$. Since $\Delta ^{s}\neq 0$ up to a temperature $\sim
J/k_{\mathrm{B}}$ at half-filling, $\Delta ^{s}$ defined in Eq. (\ref{order}%
) is expected to survive and persist into a \emph{finite} doping and lower
temperature regime.

It is interesting to distinguish the Schwinger-boson order parameter and the
bosonic RVB order parameter at \emph{finite} doping. By using the
aforementioned unitary transformation\cite{WSCT97} $\hat{\Delta}_{ij}^{%
\mathrm{SB}}\longrightarrow e^{i\hat{\Theta}}\hat{\Delta}_{ij}^{\mathrm{SB}%
}e^{-i\hat{\Theta}}$, the Schwinger-boson order parameter can be expressed
in the phase string formalism as

\begin{equation}
\hat{\Delta}_{ij}^{\mathrm{SB}}=e^{i(1/2)\sum_{l}\left[ \theta
_{i}(l)+\theta _{j}(l)\right] n_{l}^{h}}\hat{\Delta}_{ij}^{s}  \label{dscb}
\end{equation}%
with $\theta _{i}(l)\equiv \mbox{Im ln
$(z_i-z_l)$}$. So $\hat{\Delta}_{ij}^{\mathrm{SB}}$ and $\hat{\Delta}%
_{ij}^{s}$ differ by a phase factor which is composed of $2\pi $ vortices
with the cores located at the hole sites, $l$'s, with $n_{l}^{h}=1$. Namely,
each doped hole will induce a $2\pi $ vortex in the Schwinger-boson order
parameter. The general topological vortex excitation in the Schwinger-boson
mean-field state has been previously discussed by Ng.\cite{Ng} The case that
a doped hole is bound to the vortex core has been discussed as one of
several possibilities there. It turns out that since the \emph{bare} hole
will induce a highly frustrated phase string effect, its binding with a
topological vortex will be stabilized, as the composite object can
effectively \emph{erase} the singular effect and is thus in favor of the
kinetic energy. Such a composite object is nothing but the bosonic holon in
the present phase string formalism. Generally the bosonic RVB state with $%
\Delta ^{s}\neq 0$ means that the original Schwinger-boson order parameter
is \emph{phase disordered} with $\Delta ^{\mathrm{SB}}=0$ according to Eq. (%
\ref{dscb}), unless the holons are localized which corresponds to a
low-doping AF state to be discussed in Sec. 4.5.

Based on Eq. (\ref{order}), a \textquotedblleft
mean-field\textquotedblright\ version of the superexchange Hamiltonian may
be rewritten as\cite{WST98}
\begin{equation}
H_{J}\rightarrow -J_{s}\sum_{\langle ij\rangle \sigma }\left( e^{i\sigma
A_{ij}^{h}}\right) b_{i\sigma }^{\dagger }b_{j-\sigma }^{\dagger
}+h.c.~+\lambda \sum_{i}\left( \sum_{\sigma }b_{i\sigma }^{\dagger
}b_{i\sigma }-1+\delta \right) ~  \label{hs0}
\end{equation}%
where
\begin{equation}
J_{s}\equiv J_{\mathrm{eff}}\Delta ^{s}/2
\end{equation}%
where $J_{\mathrm{eff}}\sim J$ is a renormalized superexchange coupling to
be further specified below. The Lagrangian multiplier $\lambda $ is
introduced to implement the condition (\ref{global-2}).

In the limit where the gauge field $A_{ij}^{h}$ can be treated as a static
field, \emph{e.g.,} in the holon condensed regime, Eq. (\ref{hs0}) can be
straightforwardly diagonalized by a Bogoliubov transformation\cite{WST98}
\begin{equation}
b_{i\sigma }=\sum_{m}w_{m\sigma }(i)\left( u_{m}\gamma _{m\sigma
}-v_{m}\gamma _{m-\sigma }^{\dagger }\right) ~,  \label{bogo}
\end{equation}%
as $H_{s}=\sum_{m\sigma }E_{m}\gamma _{m\sigma }^{\dagger }\gamma _{m\sigma
}+\mathrm{const.}$, where $u_{m}=\frac{1}{\sqrt{2}}\left( \frac{\lambda }{%
E_{m}}+1\right) ^{1/2},$ $v_{m}=\frac{1}{\sqrt{2}}\left( \frac{\lambda }{%
E_{m}}-1\right) ^{1/2}\mathrm{sgn}(\xi _{m}),$ and $E_{m}=\sqrt{\lambda
^{2}-\xi _{m}^{2}}.$ The Lagrangian multiplier $\lambda $ is determined by
enforcing $\sum_{i}\sum_{\sigma }\left\langle b_{i\sigma }^{\dagger
}b_{i\sigma }\right\rangle =(1-\delta )N$. The wave function $w_{m\sigma }$
and the spectrum $\xi _{m}$ are determined by the following eigen equation,
\begin{equation}
\xi _{m}w_{m\sigma }(i)=-J_{s}\sum_{j=nn(i)}e^{i\sigma {A}%
_{ij}^{h}}w_{m\sigma }(j)~.  \label{weq}
\end{equation}%
Self-consistently, an another gauge-invariant mean-field order parameter%
\begin{equation}
\left\langle \left( e^{i\sigma A_{ij}^{h}}\right) b_{i\sigma }^{\dagger
}b_{j\sigma }\right\rangle _{nn}\equiv 0  \label{bhop}
\end{equation}%
in this mean-field scheme.
\begin{figure}[tbph]
\begin{center}
\includegraphics[width=3in]{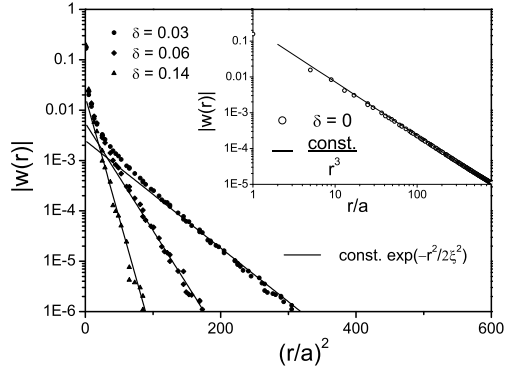}
\end{center}
\caption{The bosonic RVB amplitudes, $|W_{ij}|$, at three different hole
concentrations. The inset shows the result for half filling.}
\label{Wij}
\end{figure}

The corresponding ground state is given by\cite{WZM05}
\begin{equation}
|\mathrm{RVB}\rangle _{\mathrm{MF}}=\exp \left( \sum_{ij}W_{ij}b_{i\uparrow
}^{\dagger }b_{j\downarrow }^{\dagger }\right) |0\rangle ~,  \label{phirvb}
\end{equation}%
where the RVB amplitude $W_{ij}\equiv -\sum_{m}\frac{v_{m}}{u_{m}}w_{m\sigma
}^{\ast }(i)w_{m\sigma }(j).$ It can be further shown\cite{WZM05} that $%
W_{ij}\neq 0,$ \textrm{only if} $i,j\in $ \textrm{different sublattices},%
\emph{\ i.e.}, the RVB amplitude $W_{ij}$ only connects $\uparrow $ and $%
\downarrow $ spins on opposite sublattices. The calculated $\left\vert
W_{ij}\right\vert $ is plotted in Fig. \ref{Wij} for $i$ and $j$ belong to
different sublattices, which shows that while $W_{ij}\propto |\mathbf{r}%
_{ij}|^{-3}>0$, for $\left\vert \mathbf{r}_{ij}\right\vert \gg a$ at
half-filling, its absolute value becomes exponentially decay at finite
doping:\cite{WZM05}
\begin{equation}
\left\vert W_{ij}\right\vert \propto e^{-\frac{|\mathbf{r}_{ij}|^{2}}{2\xi
^{2}}}  \label{wij1}
\end{equation}%
with a characteristic length scale which is to be identified as the spin
correlation length later
\begin{equation}
\xi =a\sqrt{\frac{2}{\pi \delta }}=\sqrt{2}a_{c},  \label{xi}
\end{equation}%
where $a_{c}=a/\sqrt{\pi \delta }$ is the typical \textquotedblleft
cyclotron\textquotedblright\ radius decided by Eq. (\ref{weq}).

At half-filling, the ground state $|\mathrm{RVB}\rangle _{\mathrm{MF}}$
qualitatively describes the AF correlations fairly well.\cite{AA} In
particular, one may define%
\begin{equation}
|\mathrm{RVB}\rangle \equiv \hat{P}_{s}|\mathrm{RVB}\rangle _{\mathrm{MF}}
\label{rvb}
\end{equation}%
where $\hat{P}_{s}$ denotes the projection to a single occupancy with each
lattice site occupied by one spin. $|\mathrm{RVB}\rangle $ is equivalent to
Eq. (\ref{lda}) with $W_{ij}$ shown in Fig. \ref{Wij}. Based on Eq. (\ref%
{rvb}), a numerical calculation\cite{WZM05} gives rise to $\left\langle
\mathbf{S}_{i}\cdot \mathbf{S}_{j}\right\rangle _{nn}=-0.3344(2)$ and a
staggered magnetization $m=0.296(2)$ with the maximum sample size $64\times
64$. These results are essentially the same as the best variational result%
\cite{lda} and also compare extremely well with exact numerical results of $%
\left\langle \mathbf{S}_{i}\cdot \mathbf{S}_{j}\right\rangle _{nn}=-0.3346$
and $m=0.31$.

\subsection{Ground-state wave function}

At finite doping, it is easy to see that a simple ansatz for the full ground
state: $|\Psi _{b}\rangle \sim h_{l_{1}}^{\dagger }h_{l_{2}}^{\dagger }...|%
\mathrm{RVB}\rangle $ does not work here because Eq. (\ref{bhop}) would lead
to vanishing $\left\langle H_{t}\right\rangle $.

According to $H_{t}$ in Eq. (\ref{ht}), the hopping of a holon should be
always accompanied by a spinon \textquotedblleft backflow\textquotedblright\
or vice versa. But this is not properly \textquotedblleft
registered\textquotedblright\ in utilizing Eq. (\ref{bhop}). It implies that
the correct way to add holes in the ground state should be\cite{WZM05} $%
|\Psi _{G}\rangle \sim ...\left( h_{l_{1}}^{\dagger }b_{l_{1}\sigma
_{1}}\right) \left( h_{l_{2}}^{\dagger }b_{l_{2}\sigma _{2}}\right) ...|%
\mathrm{RVB}\rangle $ or more precisely

\begin{equation}
|\Psi _{G}\rangle =\sum_{\{l_{h}\}}\varphi _{h}(\{l_{h}\})\left(
h_{l_{1}}^{\dagger }h_{l_{2}}^{\dagger }...\right) \otimes \Pi \left(
\left\{ l_{h}\right\} \right) |\mathrm{RVB}\rangle ,  \label{bgs}
\end{equation}%
where $\varphi _{h}$ describes the bosonic holon wave function, while the
\textquotedblleft spinon backflow\textquotedblright\ operator is given by
\begin{equation}
\Pi \left( \left\{ l_{h}\right\} \right) =\sum_{\left\{ \sigma _{h}\right\}
}Z\left( \left\{ l_{h}\right\} ,\left\{ \sigma _{h}\right\} \right)
b_{l_{1}\sigma _{1}}b_{l_{2}\sigma _{2}}...  \label{pi}
\end{equation}%
where the summation over the spin index $\sigma _{h}$ is under the
constraint $1/2\sum_{h}\sigma _{h}=S_{z}$ (note that $S_{z}=0$ in $|\mathrm{%
RVB}\rangle )$. Here $|\mathrm{RVB}\rangle $ is understood as describing the
\textquotedblleft half-filling\textquotedblright\ spin background even at
finite doping. Obviously the no double occupancy constraint is automatically
satisfied by $|\Psi _{G}\rangle $ defined in Eq. (\ref{bgs}) at arbitrary
doping.

Then, in accordance with the gauge invariance under (\ref{u(1)1}) and (\ref%
{u(1)2}), in minimizing $\langle \Psi _{G}|H_{t}|\Psi _{G}\rangle $, $%
\varphi _{h}$ will be determined as the ground state of the effective
hopping Hamiltonian $H_{h}$ in Eq. (\ref{hh}), while the renormalized
hopping integral $t_{h}$ is given by

\begin{equation}
t_{h}=\langle \mathrm{RVB}|\Pi ^{\dagger }\sum_{\sigma }b_{j\sigma
}^{\dagger }b_{i\sigma }e^{-i\left( \phi _{ij}^{0}+\sigma A_{ij}^{h}\right)
}\Pi |\mathrm{RVB}\rangle .  \label{th}
\end{equation}%
In the dilute hole limit where the correlations between the
\textquotedblleft backflow spinons\textquotedblright\ are negligible, $%
Z\left( \left\{ l_{h}\right\} ,\left\{ \sigma _{h}\right\} \right) $ may be
reduced to a product of the single-spinon wave function $Z_{\sigma }(l),$
which can be variationally determined by optimizing $t_{h}$ as the maximal
eigen value of the following eigen equation\cite{WZM05}
\begin{equation}
(-t_{h})Z_{\sigma }(i)=-\frac{\tilde{t}}{4}\sum_{j=nn(i)}e^{-i\phi
_{ij}^{0}-i\sigma {A}_{ij}^{h}}Z_{\sigma }(j)~  \label{th1}
\end{equation}%
with $\tilde{t}=\left( \frac{\bar{n}^{b}}{2}+\frac{\left\vert \Delta
^{s}\right\vert ^{2}}{2\bar{n}^{b}}\right) t,$ $\bar{n}^{b}=1-\delta .$
Numerically $t_{h}$ thus determined is weakly doping dependent with $%
\left\vert t_{h}\right\vert \simeq 0.68t\sim O(t)$.\cite{WZM05}

Self-consistently let us go back to check the superexchange energy since $|%
\mathrm{RVB}\rangle $ is now modified by $\Pi |\mathrm{RVB}\rangle $ for the
spin degrees of freedom:

\begin{eqnarray}
\langle \Psi _{G}|H_{J}|\Psi _{G}\rangle &=&-\frac{J}{2}\sum_{\langle
ij\rangle }\sum_{\{l_{h}\}\neq i,j}\left\vert \varphi _{h}\right\vert
^{2}\langle \mathrm{RVB}|\Pi ^{\dagger }\left( \hat{\Delta}_{ij}^{s}\right)
^{\dagger }\hat{\Delta}_{ij}^{s}~\Pi |\mathrm{RVB}\rangle  \nonumber \\
&\simeq &-\frac{J_{\mathrm{eff}}}{2}\sum_{\langle ij\rangle }\langle \mathrm{%
RVB}|\left( \hat{\Delta}_{ij}^{s}\right) ^{\dagger }\hat{\Delta}_{ij}^{s}~|%
\mathrm{RVB}\rangle  \label{hj1}
\end{eqnarray}%
Note that $\sum_{\{l_{h}\}\neq i,j}\left\vert \varphi _{h}\right\vert
^{2}=1-2\delta +O(\delta ^{2})$. And $\langle \mathrm{RVB}|\Pi ^{\dagger
}\left( \hat{\Delta}_{ij}^{s}\right) ^{\dagger }\hat{\Delta}_{ij}^{s}~\Pi |%
\mathrm{RVB}\rangle \simeq f(\delta )\langle \mathrm{RVB}|\left( \hat{\Delta}%
_{ij}^{s}\right) ^{\dagger }\hat{\Delta}_{ij}^{s}~|\mathrm{RVB}\rangle $
with $f(\delta )=1-2(g-1)\delta +O(\delta ^{2}),$ ($g>1$), such that
\begin{equation}
J_{\mathrm{eff}}=J(1-2g\delta +O(\delta ^{2}))
\end{equation}%
at small doping. Here $g$ has been empirically determined\cite{GW05} by $g=2$
by comparing with the experimental measurements. The mean-field treatment of
the last line in Eq. (\ref{hj1}) leads to the effective spinon Hamiltonian (%
\ref{hs}).

Generally speaking, in order to minimize $\langle \Psi _{G}|H_{J}|\Psi
_{G}\rangle ,$ the \textquotedblleft backflow spinons\textquotedblright\ in $%
\Pi $ are better paired up\cite{WZM05}

\begin{equation}
\Pi \left( \left\{ l_{h}\right\} \right) \propto \exp \left[
\sum_{ll^{\prime }\in \{l_{h}\},\sigma }G_{ll^{^{\prime }}}^{\sigma
}b_{l\sigma }b_{l^{\prime }-\sigma }\right] |\mathrm{RVB}\rangle ~
\label{psib1}
\end{equation}%
with $G_{ll^{^{\prime }}}^{\sigma }=$ $Z_{\sigma }(l)g(l-l^{\prime
})Z_{-\sigma }(l^{\prime })$, where $g(l-l^{\prime })$ denotes the pairing
amplitude between the two \textquotedblleft backflow
spinons\textquotedblright , which is no longer restricted to the pairing
between two opposite sublattices. The paring $g(l-l^{\prime })$ is expected
to reduce the hopping integral $t_{h}$ and enhance $J_{\mathrm{eff}},$ but
the detailed values of them will not affect the general properties of the
phase string model discussed below.

\section{PHYSICAL CONSEQUENCES}

The minimal phase string model is composed of Eqs. (\ref{hh}) and (\ref{hs}%
). In the following we shall see that such a simple model will possess a
rich phase diagram unifying the AF state, the SC phase, the pseudogap regime
including both the upper and lower pseudogap phases, as well as a
high-temperature \textquotedblleft normal state\textquotedblright . The
richness of this model can be attributed to the unconventional competition
between the charge and spin degrees of freedom via the mutual Chern-Simons
gauge structure.

\subsection{Superconducting (SC)\ phase}

The SC state is a simple self-consistent solution of the phase string model
at finite doping.\cite{WST98} First, the bosonic holons will experience a
Bose condensation at $T=0$ if $A_{ij}^{e}=A_{ij}^{s}=0$ in $H_{h}$. Once the
holons are condensed, the gauge field $A_{ij}^{h}$ will reduce to a
non-dynamic $\bar{A}_{ij}^{h}$ to describe a uniform flux of strength
\begin{equation}
\sum\nolimits_{\square }\bar{A}_{ij}^{h}=\pi \delta  \label{fluxh}
\end{equation}%
per plaquette. Then, according to $H_{s}$, a gap will open up in the spinon
spectrum, such that the fluctuations of $A_{ij}^{s}$ get gapped, which in
return self-consistently ensures the holon condensation in $H_{h}$.

With the holon condensation $\left\langle h_{i}^{\dagger }\right\rangle \neq
0$, the amplitude of the SC order parameter, Eq. (\ref{sca}), becomes finite:

\begin{equation}
\Delta _{ij}^{0}\equiv \left\langle \hat{\Delta}_{ij}^{0}\right\rangle
\propto \left\langle h_{i}^{\dagger }\right\rangle \left\langle
h_{j}^{\dagger }\right\rangle \Delta _{ij}^{s}  \label{d0}
\end{equation}%
and in the ground state, the phase coherence
\begin{equation}
\left\langle e^{-i(1/2)\left( \Phi _{i}^{s}+\Phi _{j}^{s}\right)
}\right\rangle \neq 0  \label{pcoh}
\end{equation}%
can be realized because of a finite-range RVB pairing of spinons with a
finite excitation energy gap (\emph{cf}. Sec. 4.1.6.). Then the
superconducting order parameter defined in Eq. (\ref{sco}) gains a finite
mean value%
\begin{equation}
\left\langle \hat{\Delta}_{ij}^{\mathrm{SC}}\right\rangle \neq 0.
\label{dsc}
\end{equation}%
Note that the phase factor $e^{-i(1/2)\left( \Phi _{i}^{s}+\Phi
_{^{j}}^{s}\right) }$ will also decide the d-wave symmetry of $\hat{\Delta}%
_{ij}^{\mathrm{SC}}$ ($\Delta _{ij}^{0}$ is s-wave-like in general).\cite%
{ZMW03}

\subsubsection{Ground-state wave function}

The holon condensation as the solution of Eq. (\ref{hh}) at $A^{e}=0$ may be
approximately treated as an ideal one with

\begin{equation}
\varphi _{h}(l_{1},l_{2,}...,l_{N_{h}})=\mathrm{const.}
\end{equation}%
Then the corresponding ground state of Eqs. (\ref{bgs}) and (\ref{psib1}) is
simplified to\cite{WZM05}

\begin{eqnarray}
|\Psi _{G}\rangle _{\mathrm{SC}} &=&\mathrm{const.}\left[ \hat{D}\right]
^{N_{h}/2}|\mathrm{RVB}\rangle  \nonumber \\
&=&\hat{P}_{N_{h}}\exp \left[ \hat{D}\right] |\mathrm{RVB}\rangle ~
\label{psib2}
\end{eqnarray}%
in which $\hat{P}_{N_{h}}$ denotes a projection onto a $N_{h}$-hole state and%
\begin{equation}
\hat{D}=\sum_{ij\sigma }G_{ij}^{\sigma }\left( h_{i}^{\dagger }b_{i\sigma
}\right) \left( h_{j}^{\dagger }b_{j-\sigma }\right) .  \label{D}
\end{equation}

Equation (\ref{psib2}) implies $\left\langle \hat{D}\right\rangle \neq 0.$
Then, in terms of Eqs. (\ref{sca}) and (\ref{sco})\cite{WZM05}\
\begin{eqnarray}
\left\langle \hat{D}\right\rangle &\simeq &\sum_{ij}\tilde{G}_{ij}\frac{%
\left\langle \hat{\Delta}_{ij}^{0}\right\rangle }{2}  \nonumber \\
&=&\frac{1}{2}\sum_{ij}\tilde{G}_{ij}\left\langle e^{-i(1/2)\left( \Phi
_{i}^{s}+\Phi _{j}^{s}\right) }\hat{\Delta}_{ij}^{\mathrm{SC}}\right\rangle
\neq 0  \label{D2}
\end{eqnarray}%
where $\tilde{G}_{ij}=g(i-j)\sum_{\sigma }Z_{\sigma }(i)Z_{-\sigma
}(j)(-1)^{j}e^{i\Phi _{j}^{0}+i\phi _{ij}^{0}+i\sigma \bar{A}_{ij}^{h}}$ is
a s-wave constant\cite{WZM05} based on Eq. (\ref{th1}). Thus, it confirms
again that the SC off-diagonal-long-range-order (ODLRO) [Eq. (\ref{dsc})] is
established once the phase coherence (\ref{pcoh}) is realized in the ground
state.

\subsubsection{Generalized Ginzburg-Landau (GL) equation}

In the SC state, one may treat the charge condensate in terms of a slowly
varying continuous field, $\left\langle h_{i}\right\rangle \rightarrow \psi
_{h}(\mathbf{r}_{i})$. Note that in the continuum limit Eq. (\ref{hh}) can
be rewritten as
\begin{equation}
H_{h}={\frac{1}{2m_{h}}}\int d^{2}\mathbf{r}~h^{\dagger }(\mathbf{r})\left(
-i\nabla -\mathbf{A}^{s}-e\mathbf{A}^{e}\right) ^{2}h(\mathbf{r})~
\label{hholon}
\end{equation}%
where $m_{h}$ $=(2t_{h}a^{2})^{-1}$ and $\mathbf{A}^{s}$ is the continuum
version of $A_{ij}^{s}=\mathbf{r}_{ij}\cdot \mathbf{A}^{s}$ given by%
\begin{equation}
\mathbf{A}^{s}(\mathbf{r})=\frac{1}{2}\int d^{2}\mathbf{r}^{\prime }~\frac{%
\hat{\mathbf{z}}\times (\mathbf{r}-\mathbf{r}^{\prime })}{|\mathbf{r}-%
\mathbf{r}^{\prime }|^{2}}\left[ n_{\uparrow }^{b}(\mathbf{r}^{\prime
})-n_{\downarrow }^{b}(\mathbf{r}^{\prime })\right] ~  \label{asc}
\end{equation}%
with $n_{\sigma }^{b}(\mathbf{r}_{i})\equiv n_{i\sigma }^{b}/a^{2}$.

By noting that the holons here are hard-core bosons with a repulsive
short-range interaction, one may generally write down the corresponding GL
free energy $F_{h}=\int d^{2}\mathbf{r}$ $f_{h}$ where\cite{muthu2002}
\begin{equation}
f_{h}=f_{h}^{0}+\alpha \left\vert \psi _{h}\right\vert ^{2}+\frac{\eta }{2}%
|\psi _{h}|^{4}+\frac{1}{2m_{h}}\psi _{h}^{\ast }\left( -i\nabla -\mathbf{A}%
^{s}-e\mathbf{A}^{e}\right) ^{2}\psi _{h}  \label{fh}
\end{equation}%
with $f_{h}^{0}$ denoting the \textquotedblleft normal
state\textquotedblright\ free energy density. \ And the \textquotedblleft
supercurrent\textquotedblright\ density is given by
\begin{equation}
\mathbf{J}(\mathbf{r})=-\frac{i}{2m_{h}}\left[ \psi _{h}^{\ast }(\mathbf{r}%
)\nabla \psi _{h}(\mathbf{r})-\nabla \psi _{h}^{\ast }(\mathbf{r})\psi _{h}(%
\mathbf{r})\right] -\frac{\mathbf{A}^{s}+e\mathbf{A}^{e}}{m_{h}}\psi
_{h}^{\ast }(\mathbf{r})\psi _{h}(\mathbf{r})~.  \label{j0}
\end{equation}

These equations are similar to an ordinary GL theory describing a charge $+e$
Bose condensate coupled to an external electromagnetic field $\mathbf{A}^{e}$%
, \emph{except} that $\psi _{h}$ is further coupled to the spin degrees of
freedom through the vector potential $\mathbf{A}^{s}$. It means that each
isolated spin (spinon) will register as a $\pm \pi $ flux tube in Eq. (\ref%
{fh}) to exert frustration effect on the charge condensate. Thus, such a
generalized GL must be \emph{coupled} to the spinon Hamiltonian $H_{s}$ to
govern the basic physics in the SC state.
\begin{figure}[tbph]
\begin{center}
\includegraphics[width=3in]
{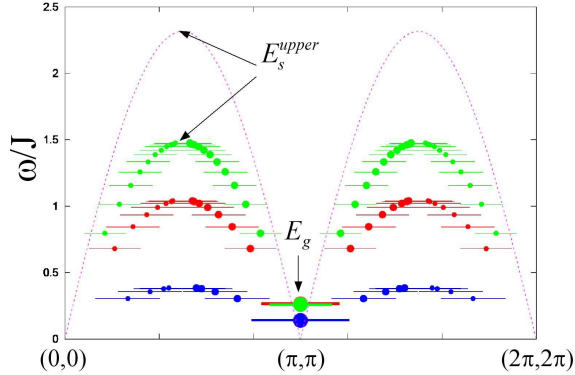}
\end{center}
\caption{Basic energy scales, $E_{s}^{\mathrm{upper}}$ and $E_{g}$,
determined by the dynamic spin susceptibility function $\protect\chi %
^{\prime \prime }(\mathbf{Q},\protect\omega )$ at $T=0$. The peak positions
of $\protect\chi ^{\prime \prime }$ at $\protect\delta =0$ is shown in the
energy and momentum (along the $Q_{x}=Q_{y}$ axis) by the dotted curve,
which tracks the spin wave dispersion with $E_{s}^{\mathrm{upper}}\simeq
2.3J $ and $E_{g}=0$. The upper-bound energy $E_{s}^{\mathrm{upper}}$
monotonically decreases with increasing doping from $\ 0.05$, $0.125$, to $%
0.2$. $E_{g}$ denotes the resonancelike peak energy at $\mathbf{Q}_{\mathrm{%
AF}}=(\protect\pi ,\protect\pi )$, which emerges in the SC state. Note that
the finite horizontal bars at finite doping indicate the momentum widths for
these non-propagating modes.\protect\cite{CW05}}
\label{dispersion}
\end{figure}

\begin{figure}[tbp]
\begin{center}
\includegraphics[width=2.5in]{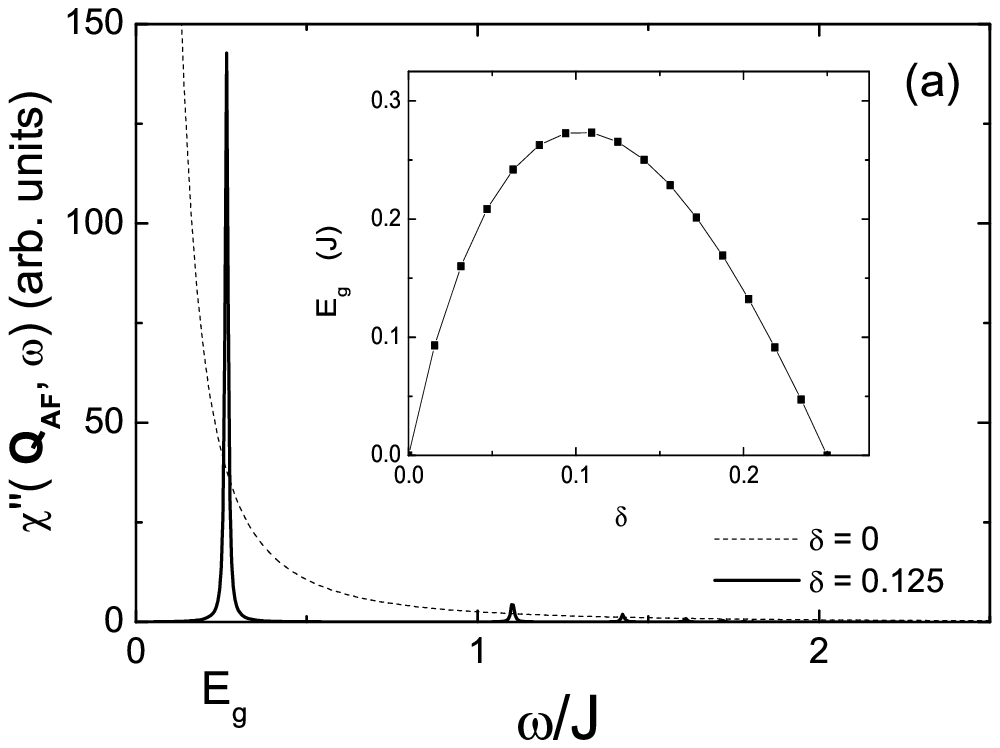} %
\includegraphics[width=2.3in]{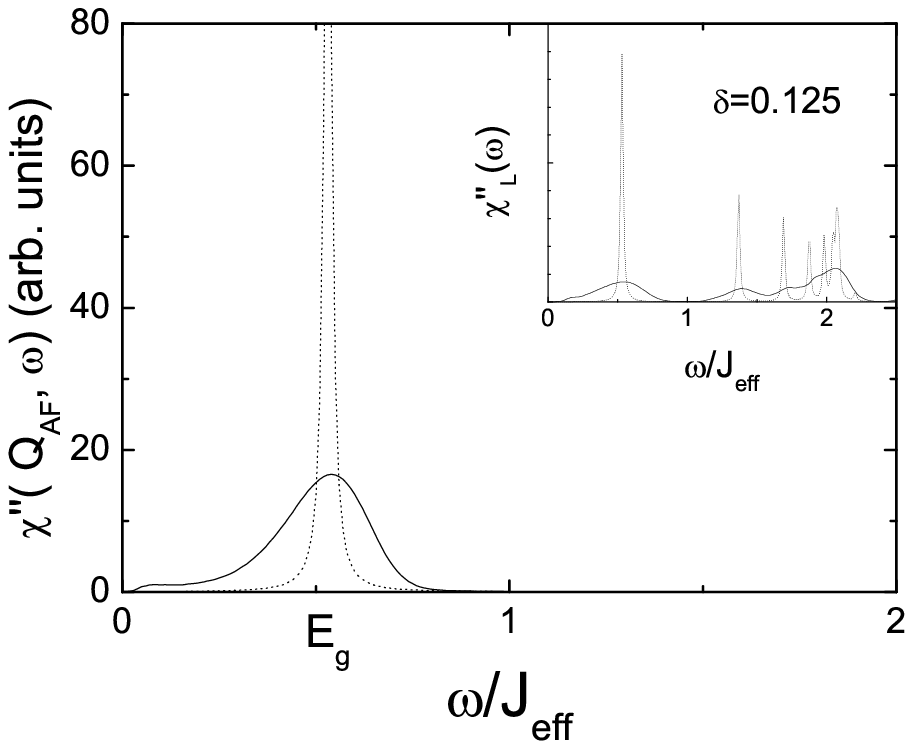}
\end{center}
\caption{Left panel: $\protect\chi ^{\prime \prime }(\mathbf{Q}_{\mathrm{AF}%
},\protect\omega )$ shows a resonance peak at energy $E_{g}$ [$\mathbf{Q}_{%
\mathrm{AF}}=(\protect\pi ,\protect\pi )]$ at $\protect\delta =0.125$.
Inset: the evolution of $E_{g}$ as a function of $\protect\delta $. Right
panel: $\protect\chi ^{\prime \prime }(\mathbf{Q}_{\mathrm{AF}},\protect%
\omega )$ with incorporating the fluctuational effect induced by the charge
inhomogeneity. Inset: the local susceptibility $\protect\chi _{L}^{\prime
\prime }(\protect\omega )$ in the same situation. [From Ref. \protect\cite%
{CW05}]}
\label{res}
\end{figure}
\begin{figure}[tbph]
\begin{center}
\includegraphics[width=3.3in]{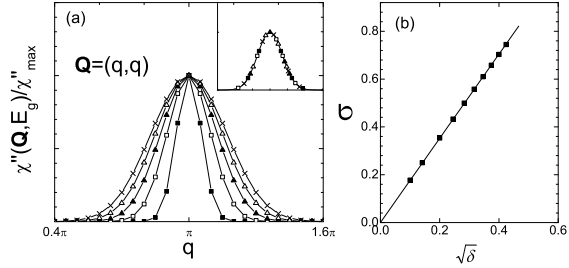}
\end{center}
\caption{(a) Momentum distribution of $\protect\chi ^{\prime \prime }(%
\mathbf{Q},E_{g})$, scanned along the diagonal direction $\mathbf{Q}=(q,q)$
at various hole concentrations. The intensities are normalized at the
maximums. Inset shows that the data in the main panel can be well fit into a
Gaussian function $\exp (-(\mathbf{Q}-\mathbf{Q}_{\mathrm{AF}})^{2}/2\protect%
\sigma ^{2})$, with $\protect\sigma =\protect\sqrt{2}a/\protect\xi =$ $%
\protect\sqrt{\protect\pi \protect\delta }$, as shown in (b). [From Ref.
\protect\cite{CW05}]}
\label{Q}
\end{figure}

\subsubsection{Non-BCS-type elementary excitation: $S=1$ spin excitation}

As outlined in Sec. 3.3, the spinon Hamiltonian $H_{s}$ can be diagonalized
under the condition (\ref{fluxh}). Figure \ref{dispersion} shows the
dispersive behavior of the $S=1$ spin excitation based on the peak position
of the spin dynamic susceptibility $\chi ^{\prime \prime }(\mathbf{Q},\omega
)$ at different doping concentrations, which clearly depicts how the spin
excitation evolves from the spin-wave picture at half-filling (dotted curve)
to the non-propagating modes (solid bars) in the SC state.\cite{CW05}

A clear spin gap is opened up at $E_{g}$ in the SC state with the gapless
spin wave replaced by a resonancelike mode (Fig. \ref{res}) around the AF
wave vector $\mathbf{Q}_{\mathrm{AF}}=(\pi ,\pi ),$ whose doping dependence
is also shown in the inset of the left panel in Fig. \ref{res}. Such a
resonance mode has a finite width in momentum which implies a finite spin
correlation length $\xi $ [Eq. (\ref{xi})] as shown in Fig. \ref{Q}.
Furthermore, the spatial charge inhomogeneity can affect the width of the
resonance peak via $A^{h}$ as shown\cite{CW05} in the right panel of Fig. %
\ref{res}, in which the inset illustrate the local spin susceptibility $\chi
_{L}^{\prime \prime }(\omega )=\left( 1/N\right) \sum\nolimits_{\mathbf{Q}}$
$\chi ^{\prime \prime }(\mathbf{Q},\omega )$. Finally it is noted that the
envelop of the high energy $S=1$ excitation still roughly tracks the spin
wave with a softened upper bound energy $E_{s}^{\mathrm{upper}}$, which
decreases monotonically with doping [Fig. \ref{dispersion}].

\subsubsection{Non-BCS-type topological excitation: Spinon vortex}

In the above we have examined the $S=1$ spin excitation which is composed of
a pair of $S=1/2$ spinons according to Eq. (\ref{s+}). However, a single $%
S=1/2$ spinon excitation will be \textquotedblleft
confined\textquotedblright\ in the SC state, \emph{i.e., }will not appear in
the finite energy spectrum. In this sense, the above $S=1$ excitations are
true elementary ones, which do not fractionalize. We shall elaborate this as
follows.

It is convenient to rewrite the nn SC order parameter as the mean value of (%
\ref{sco}) in the continuum version (without considering the d-wave symmetry
of the relative coordinate for simplicity):\cite{muthu2002}
\begin{equation}
\Delta ^{\mathrm{SC}}=\Delta ^{0}\left\langle e^{i\Phi ^{s}(\mathbf{r}%
)}\right\rangle ~  \label{psisc}
\end{equation}%
where the amplitude
\begin{equation}
\Delta ^{0}=\Delta ^{s}\left( \psi _{h}^{\ast }\right) ^{2}
\label{amplitude}
\end{equation}%
and the phase
\begin{equation}
\Phi ^{s}(\mathbf{r})=\int d^{2}\mathbf{r}^{\prime }~\mathrm{Im~ln}\left[
z-z^{\prime }\right] ~\left[ n_{\uparrow }^{b}(\mathbf{r}^{\prime
})-n_{\downarrow }^{b}(\mathbf{r}^{\prime })\right] ~.  \label{phis0}
\end{equation}%
\

\begin{figure}[tbp]
\begin{center}
\includegraphics[width=2.3in]{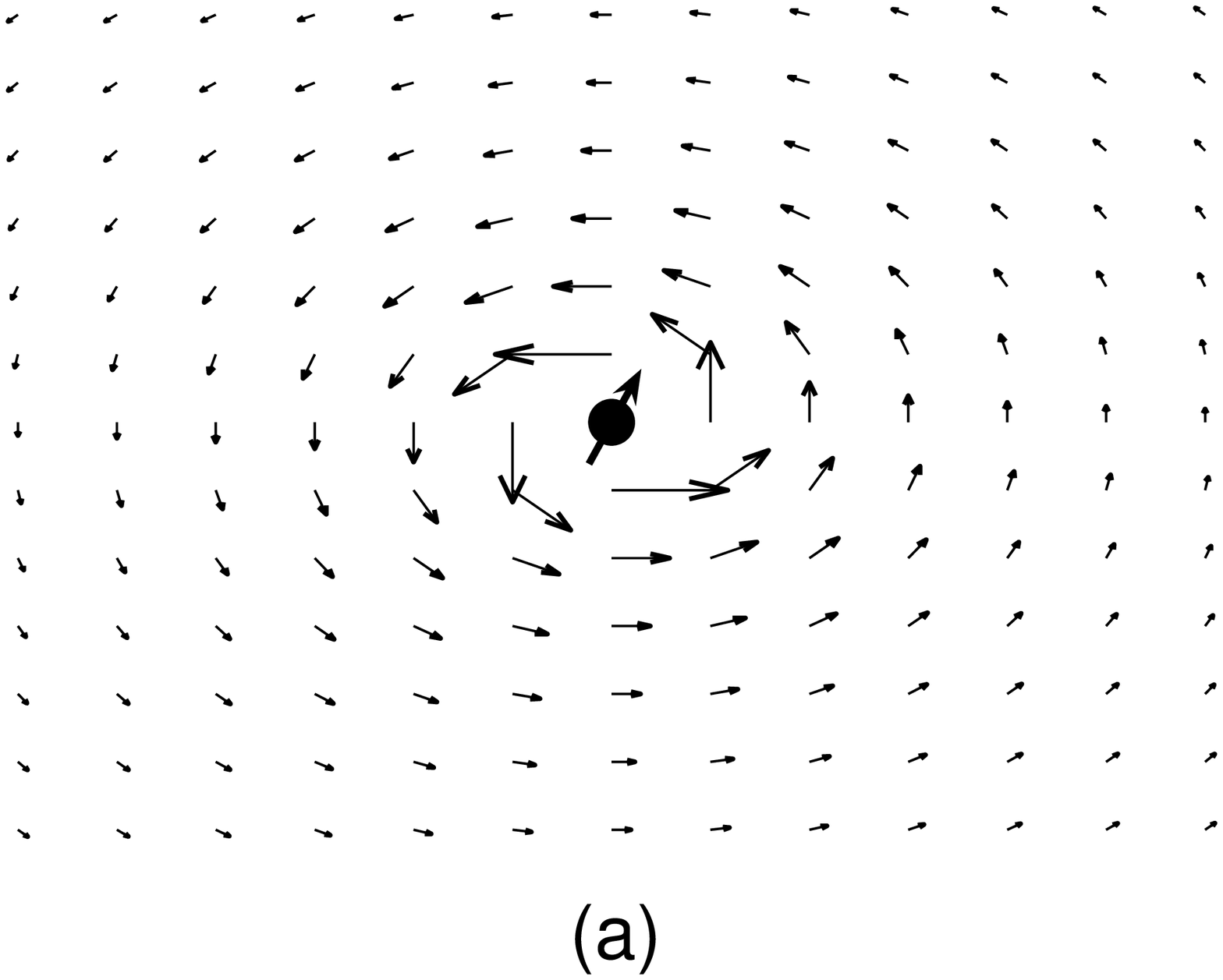} %
\includegraphics[width=2.3in]{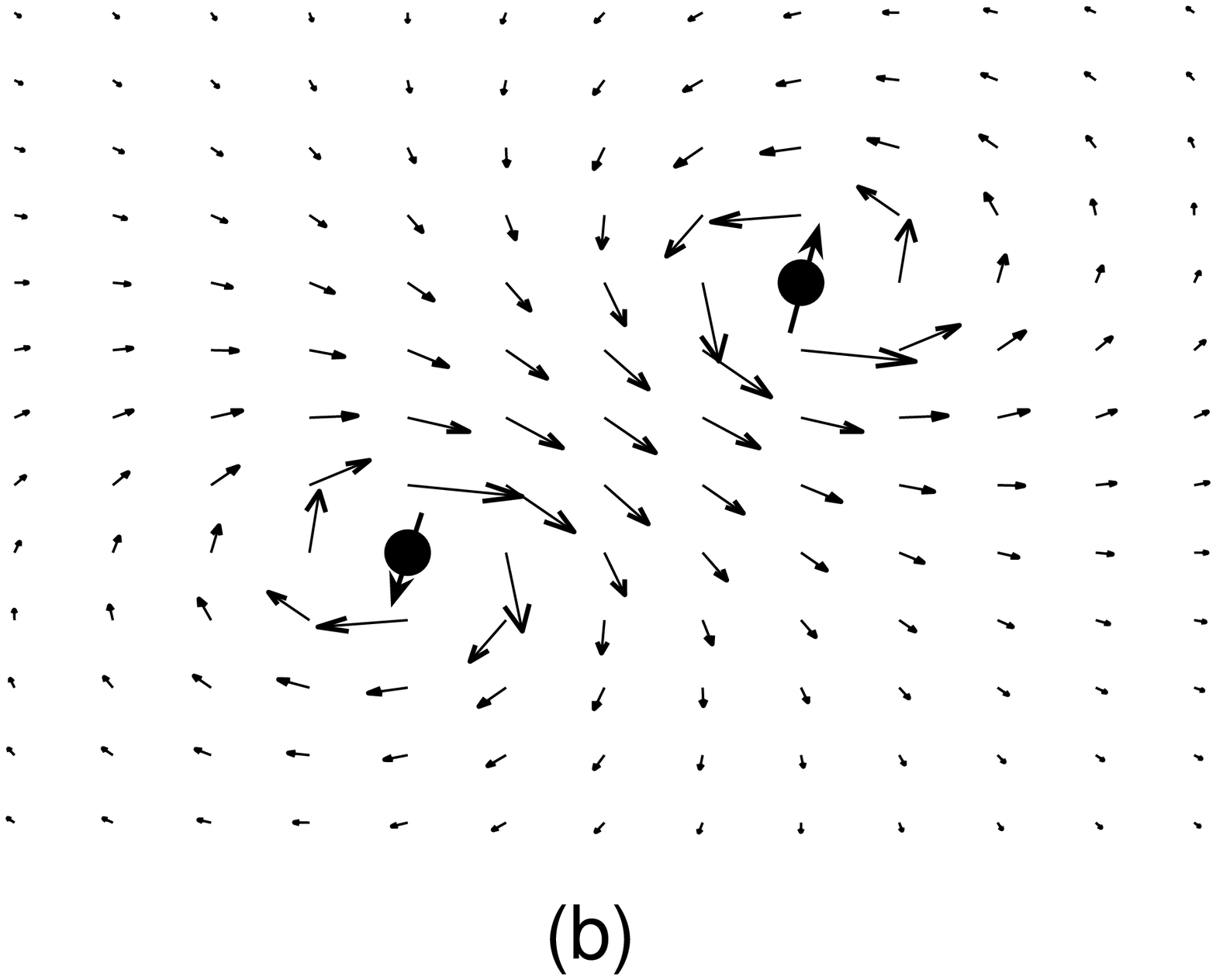}
\end{center}
\caption{ (a) An isolated spinon is always associated with a $2\protect\pi $
vortex in the phase of the SC order parameter. (b) A pair of
vortex-antivortex with spinons located at the cores. }
\label{topo}
\end{figure}

From Eq. (\ref{phis0}), it is clear that $\Phi ^{s}$ describes $2\pi $ phase
vortices whose cores are centered at spinons: \emph{i.e.,} $\Phi
^{s}\rightarrow \Phi ^{s}\pm 2\pi $ or $\Delta ^{\mathrm{SC}}\rightarrow
\Delta ^{\mathrm{SC}}e^{\pm 2\pi i}$ each time as the coordinate $\mathbf{r}$
continuously winds around a \emph{spinon} once according to Eq. (\ref{phis0}%
). In other words, a spinon is always associated with a $2\pi $ vortex in $%
\Delta ^{\mathrm{SC}}$, known as a \emph{spinon-vortex composite}\cite%
{muthu2002,WM02} which is schematically illustrated in Fig. \ref{topo}(a). A
spinon-vortex and -antivortex pair of a finite range of separation will
result in the cancellation of the phase $\Phi ^{s}$ at a large length scale
as shown in Fig. \ref{topo}(b).

By writing%
\begin{equation}
\psi _{h}=\sqrt{\rho _{h}}e^{i\phi _{h}}  \label{psih}
\end{equation}%
a London equation for the supercurrent based on Eq. (\ref{j0}) is given by%
\cite{muthu2002}%
\begin{equation}
\mathbf{J}(\mathbf{r})=\frac{\rho _{h}}{m_{h}}\left[ \nabla \phi _{h}-%
\mathbf{A}^{s}-e\mathbf{A}^{e}\right]  \label{j2}
\end{equation}%
Since each unpaired spinon will contribute to $\oint\nolimits_{c}d\mathbf{r}%
\cdot \mathbf{A}^{s}=\pm \pi $ in terms Eq. (\ref{asc}) if the loop $c$
encloses such a spinon, a \emph{minimal }supercurrent vortex centered around
it is then given by
\begin{equation}
\oint\nolimits_{c}d\mathbf{r}\cdot \mathbf{J}(\mathbf{r})=\pm \pi \frac{\rho
_{h}}{m_{h}}  \label{cvortex}
\end{equation}%
at $\mathbf{A}^{e}=0$ according to Eq. (\ref{j2}).

For a single spinon vortex centered at the origin, we have $\displaystyle%
\mathbf{A}^{s}(\mathbf{r})={\frac{1}{2}}{\frac{\hat{\mathbf{z}}\times
\mathbf{r}}{r^{2}}}$, for distances $r>>a_{c}\sim \xi $, the size of the
vortex core. Using $\mathbf{J}=-{\frac{\rho _{h}}{m_{h}}}\mathbf{A}^{s}$,
one can estimate the energy cost of a spinon-induced vortex current based on
Eq. (\ref{hholon}) by\cite{muthu2002}
\begin{eqnarray}
E_{v} &=&-\int d^{2}\mathbf{r}\text{ }\mathbf{A}^{s}\cdot \mathbf{J}-\int
d^{2}\mathbf{r}\text{{}}\rho _{h}{\frac{(\mathbf{A}^{s})^{2}}{2m_{h}}}
\nonumber \\
&=&{\frac{\rho _{h}}{2m_{h}}}\int d^{2}\mathbf{r}(\mathbf{A}^{s})^{2} \\
&=&{\frac{\pi \rho _{h}}{4m_{h}}}\int dr{\frac{1}{r}}\propto \mathrm{ln}{%
\frac{L}{a_{c}}}~,  \label{confinement}
\end{eqnarray}%
where $L$ is the size of the sample. Thus one concludes that a single $S=1/2$
spinon excitation is forbidden owing to a logarithmically diverging energy.%
\cite{muthu2002}

\subsubsection{Topological defects: Flux quantization and Zn impurity}

The phase string model predicts that an isolated spinon excitation is a
topological vortex which cannot live alone in the bulk of the SC state. So
there is no electron fractionalization at low energy and long distance.
However, in the presence of two kinds of special defects in the
superconductor, a single spinon excitation can be naturally induced as a
unique prediction of the model.

\paragraph{Flux quantization}

\begin{figure}[tbp]
\begin{center}
\includegraphics[width=3.5in]{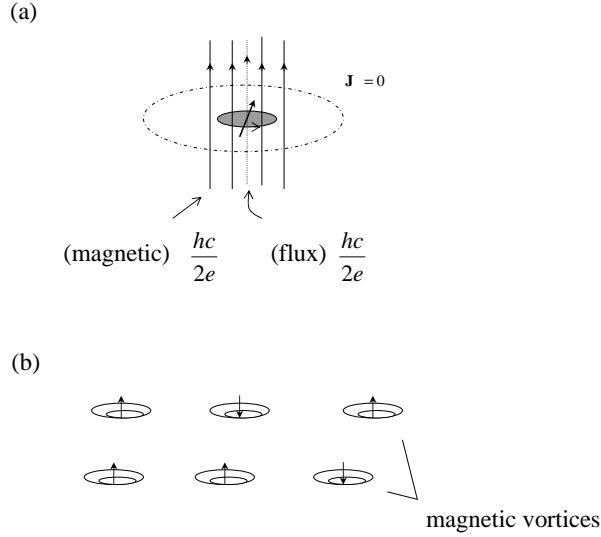}
\end{center}
\caption{Flux quantization occurs at $hc/2e$, with a bosonic $S=1/2$ spinon
trapped inside the core.}
\label{fq}
\end{figure}

In the presence of magnetic field, using Eq. (\ref{j2}) we get
\begin{equation}
{\frac{m_{h}}{\rho _{h}}}\oint_{c}\mathbf{J}(\mathbf{r})\cdot d\mathbf{r}%
=2\pi n-\oint_{c}d\mathbf{r}\cdot \left( \mathbf{A}^{s}+e\mathbf{A}%
^{e}\right) ~,
\end{equation}%
where the integral is over a closed loop $\ c$ and $n$, an integer. Now
suppose that the integration is carried over a loop that is far away from
the magnetic vortex core, where $\mathbf{J}=0$. Then one arrives at
\begin{equation}
\left( 2\pi n-e\oint_{c}d\mathbf{r}\cdot \mathbf{A}^{e}\right) -\oint_{c}d%
\mathbf{r}\cdot \mathbf{A}^{s}=0~.  \label{fq1}
\end{equation}%
If $\mathbf{A}^{s}=0$, the magnetic flux will be quantized at $2\pi n$ in
units of $\hbar c/e$; \emph{i.e.}, the minimal flux quantum in this case is $%
hc/e\equiv \Phi _{0}$, as expected for a charge $e$ Bose system. However,
the presence of $\mathbf{A}^{s}$ changes the quantization condition
radically. Suppose there is one excited spinon trapped in the core of a
magnetic fluxoid [Fig. \ref{fq}(a)]. Then, from Eq. (\ref{fq1}), we obtain
the minimal flux quantization condition\cite{muthu2002}
\begin{equation}
\oint_{c}d\mathbf{r}\cdot \mathbf{A}^{e}=\pm \pi ~,  \label{fq2}
\end{equation}%
which is precisely the quantization condition at $\phi _{0}=\Phi
_{0}/2=hc/2e $. As the holons do not distinguish between internal
(fictitious) and external (magnetic) flux in (\ref{fh}), they still perceive
a total flux quantized at $\Phi _{0}$ [see Fig. \ref{fq}(a)], even though
the true magnetic flux quantum is $\phi _{0}$.

Therefore, an excited spinon or free $S=1/2$ moment will be
\textquotedblleft nucleated\textquotedblright\ inside the core of a magnetic
vortex, which plays a crucial role to ensure the flux quantization at $\phi
_{0}$.\cite{muthu2002} But normally the Kondo screening effect due to the
coupling to the background quasiparticles may complicate the analysis of
possible experimental behavior of such a \emph{free} moment at very low
temperatures.

\paragraph{Zn impurity}

\begin{figure}[tbph]
\begin{center}
\includegraphics[width=2in]{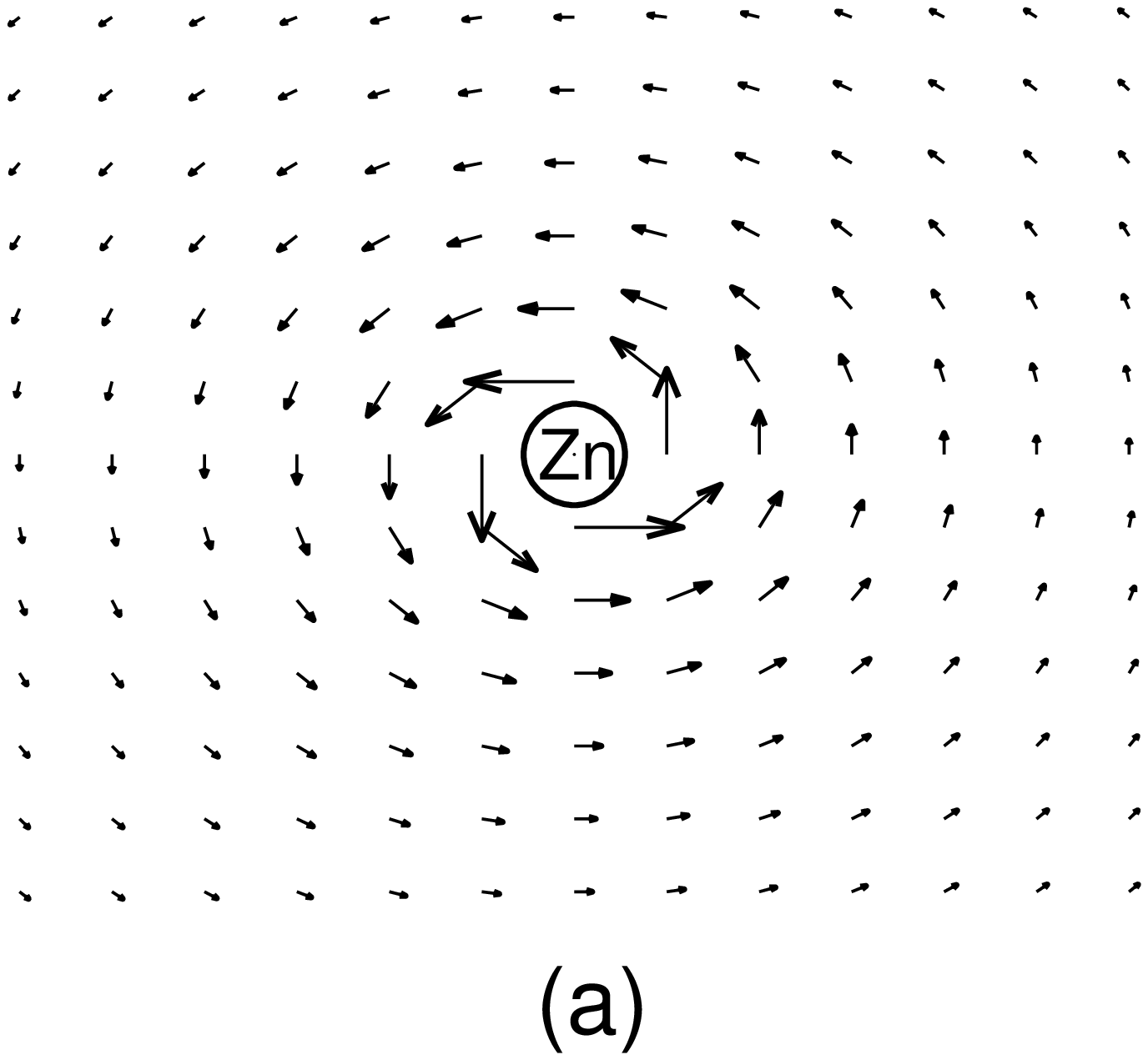} \includegraphics[width=2in]{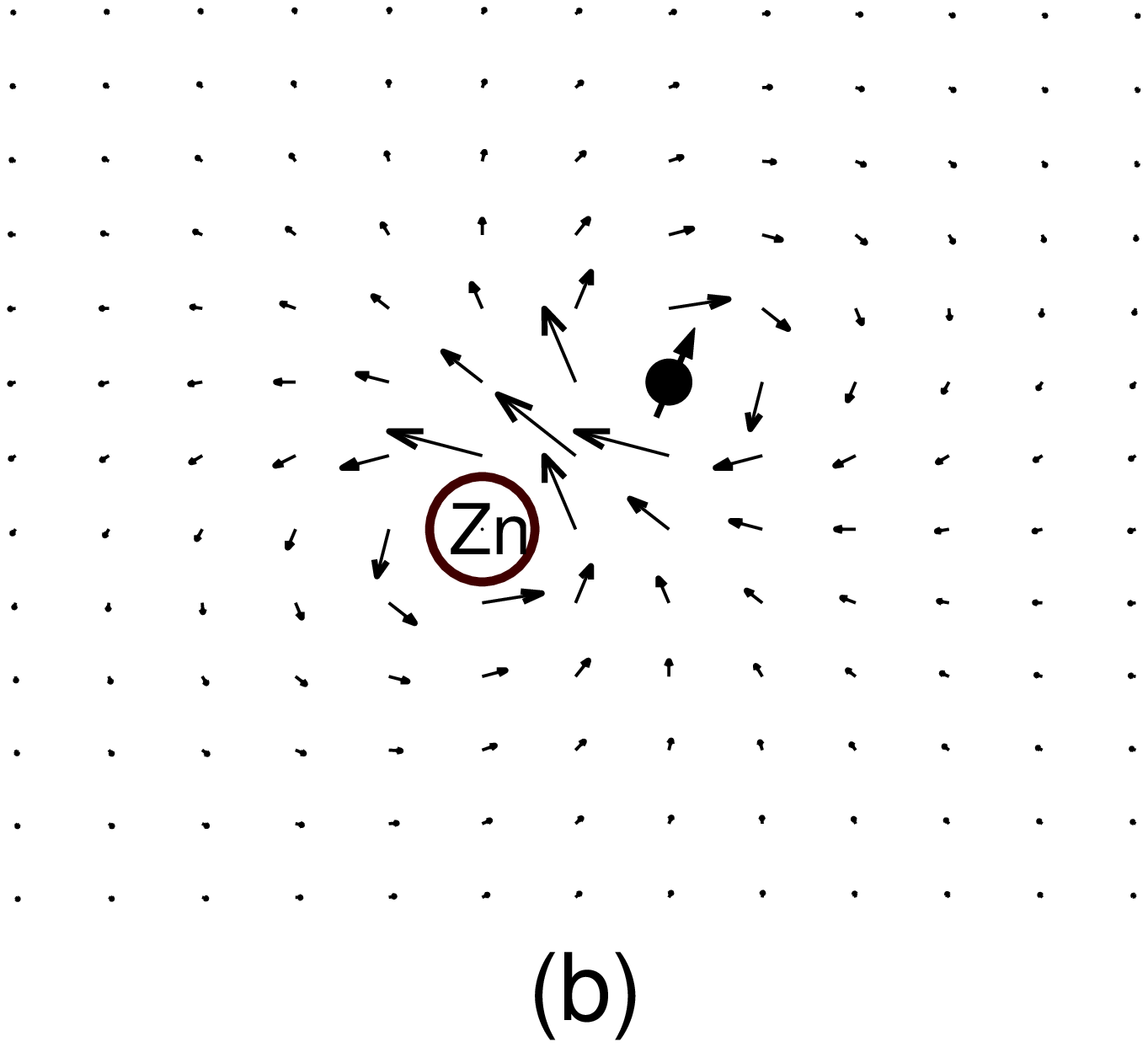}
\end{center}
\caption{(a) A vacancy (zinc impurity) always induces a vortex-like
supercurrent response in the SC phase due to the phase string effect. (b) To
compensate such a vortex effect, a spinon, which carries an antivortex, has
to be trapped around the zinc impurity, giving rise to a local $S=1/2$
moment. }
\label{Topo}
\end{figure}

In the $t$-$J$ model, a zinc impurity is treated as an \emph{empty} site
with excluding the occupation of any electrons. If one reformulates the
model, with the defect, in the phase string representation outlined in Secs.
2 and 3, it is found\cite{QW05} that such a \textquotedblleft
vacancy\textquotedblright\ site will behave like a topological defect which
induces a vortex current in the resulting phase string model, as shown in
Fig. \ref{Topo}(a). A heuristic understanding is to imagine exciting a
spinon in the pure system at a given site and then freezing its spin such
that the superexchange coupling with the surrounding spins is effectively
cutoff. Nor a holon can hop to this site such that the effect of a zinc
impurity is effectively created, which is nothing but a vortex in Fig. \ref%
{Topo}(a).

Now it is natural to see why a zinc impurity will generally induce a spin-$%
1/2$ around it in the SC state: Such a Zn-vortex would cost a
logarithmically divergent energy and thus must be \textquotedblleft
screened\textquotedblright\ by nucleating a neutral $S=1/2$ spinon which
carries an antivortex and is bound to the latter, as shown in Fig. \ref{Topo}%
(b).

\begin{figure}[tbp]
\begin{center}
\includegraphics[width=2.2in] {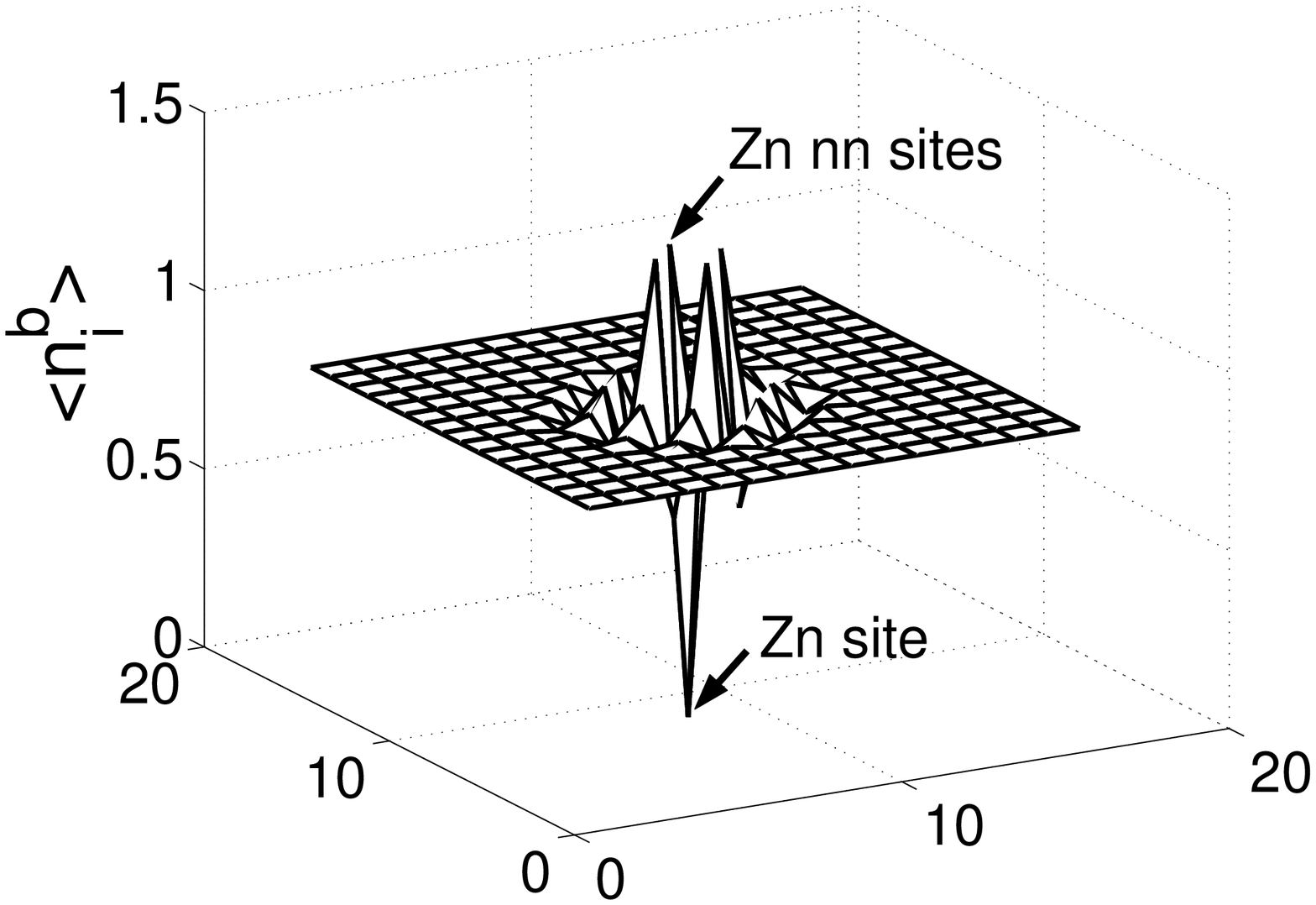}
\includegraphics[width=2.2in]
{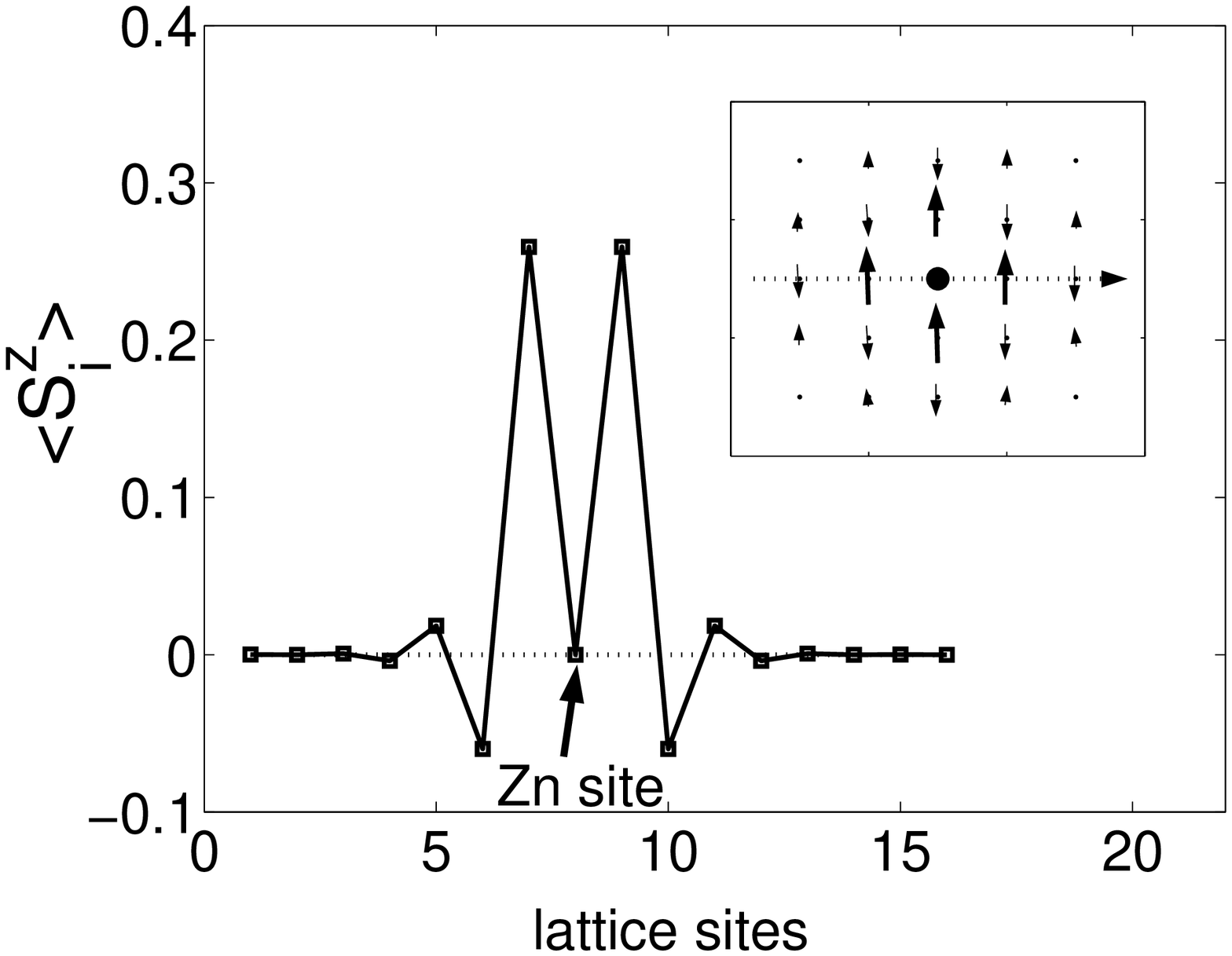}
\end{center}
\caption{Left panel: Spinon density distribution $\left\langle
n_{i}^{b}\right\rangle =\left\langle \sum_{\protect\sigma }b_{i\protect%
\sigma }^{\dagger }b_{i\protect\sigma }\right\rangle $ around the zinc
impurity at $\protect\delta =0.125.$ Right panel: The distribution of $%
\left\langle S_{i}^{z}\right\rangle $ near the zinc impurity, with the scan
along the dashed direction shown in the inset, where the zinc site is marked
by the filled circle. [From Ref. \protect\cite{QW05}]}
\label{Sdis}
\end{figure}

\begin{figure}[tbp]
\begin{center}
\includegraphics[width=2.2in] {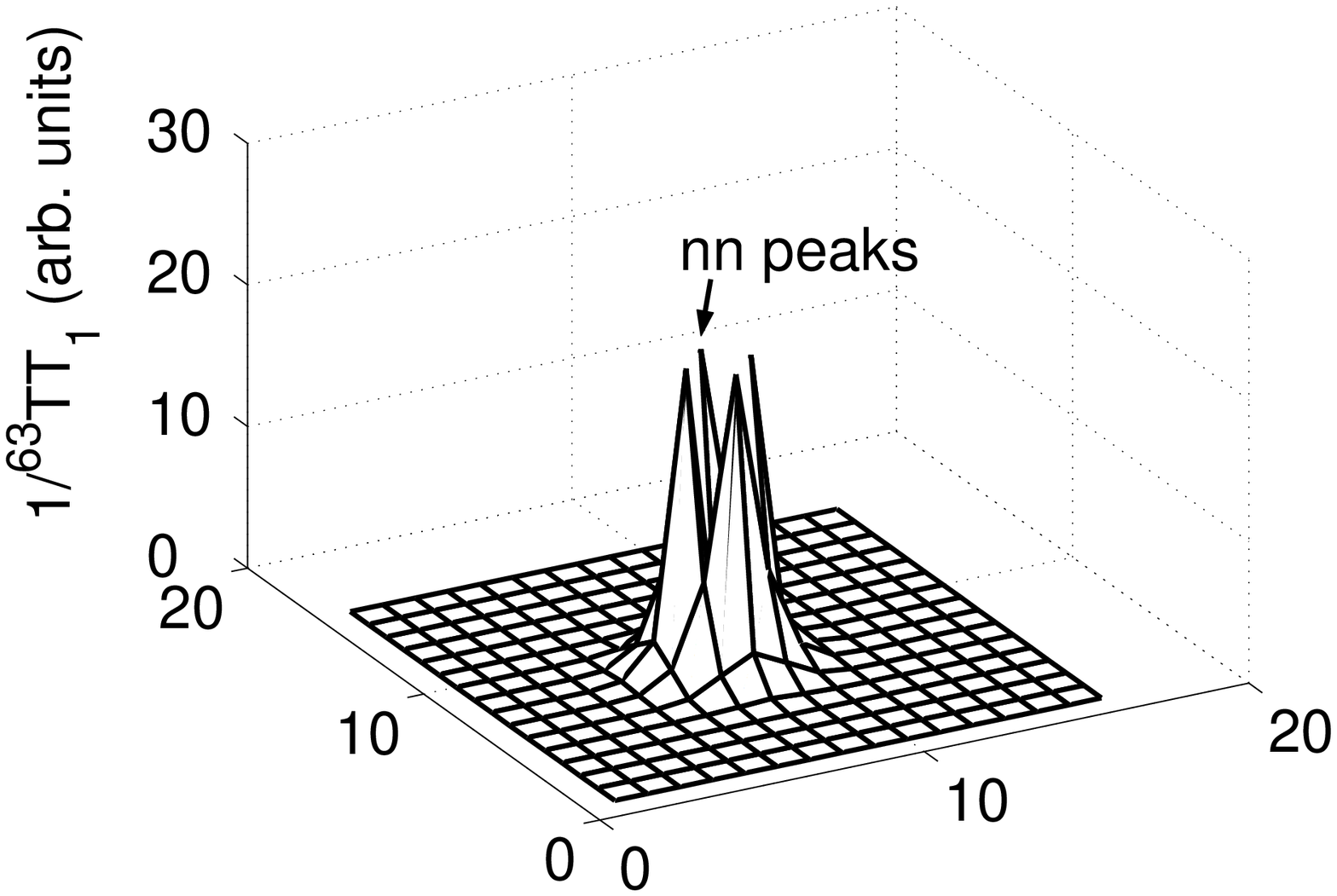}
\includegraphics[width=2.2in]
{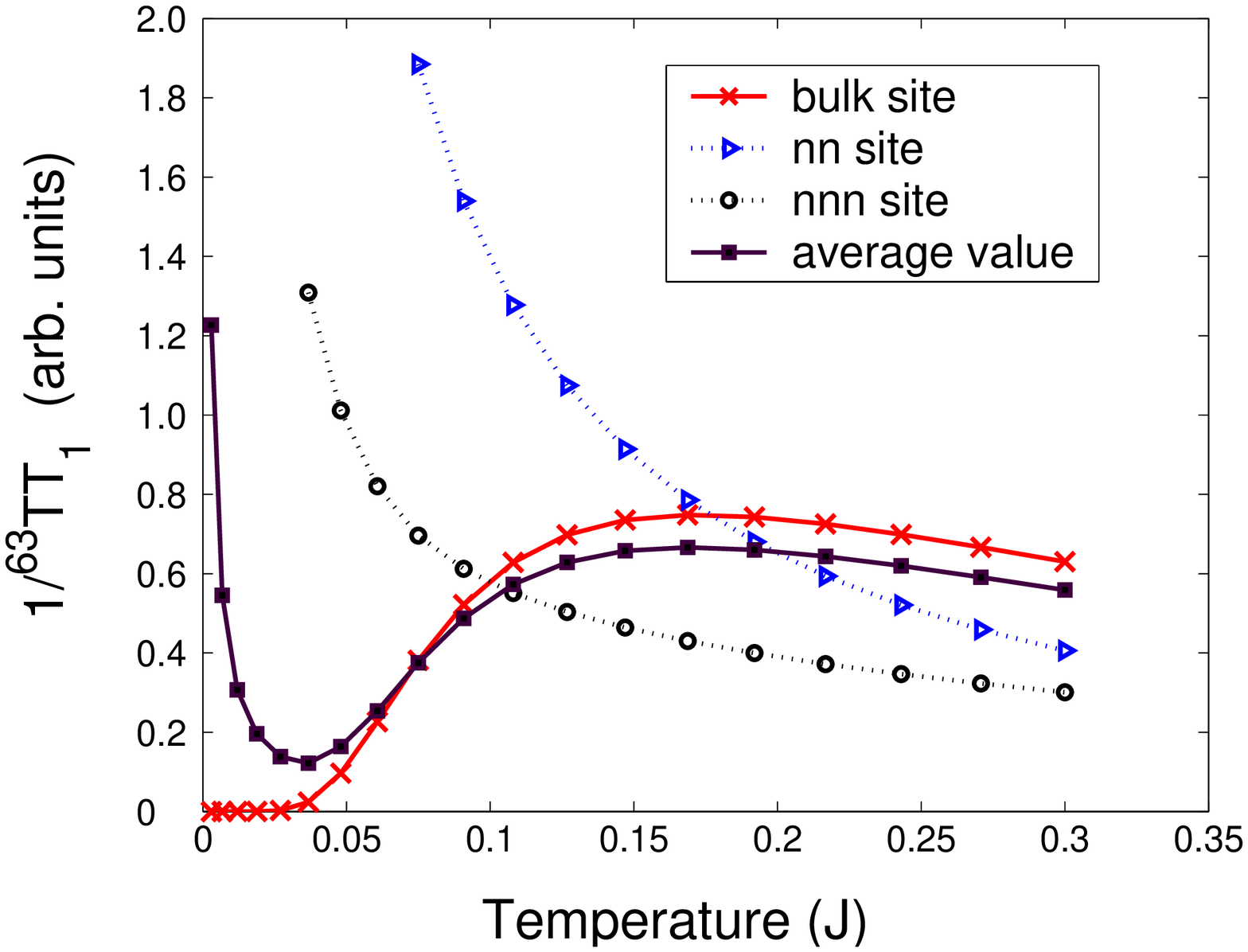}
\end{center}
\caption{Left panel: Distribution of the contributions to $1/^{63}T_{1}T$
from individual sites near the zinc impurity, at temperature $T=0.0067J$.
Right panel: $1/^{63}T_{1}T$ vs. T at different sites. Solid curve with
crosses: from the site far from the zinc impurity; Dashed curve with
triangles: the nn site near the zinc; Dashed curve with circles: the next
nearest neighbor (nnn) site near the zinc; Solid curve with squares: average
over all sites in a $16\times 16$ lattice with one zinc. [From Ref.
\protect\cite{QW05}]}
\label{1ovTT1}
\end{figure}

\begin{figure}[tbp]
\begin{center}
\emph{\includegraphics[width=2.2in] {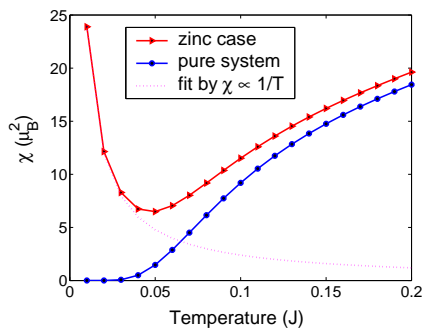}}
\end{center}
\caption{Uniform spin susceptibility in the pure system is shown by the
solid curve with full circles; The case with one zinc is illustrated by the
solid curve with triangles; The dashed curve is a fit by$\ \protect\chi %
=0.2390/T$. [From Ref. \protect\cite{QW05}]}
\label{Unif}
\end{figure}

Once the topological origin of the $S=1/2$ moment around a zinc impurity is
established, a simple effective description of the system with one zinc
impurity can be developed\cite{QW05} based on a \textquotedblleft sudden
approximation\textquotedblright\ using the phase string model. The physical
consequences, as clearly illustrated in Figs. \ref{Sdis}, \ref{1ovTT1}, and %
\ref{Unif}, are that the $S=1/2$ \emph{free }moment induced by the zinc
impurity distributes in an AF staggered pattern around the \textquotedblleft
vacancy\textquotedblright\ site.

\subsubsection{\protect\bigskip Superconducting transition}

We have seen that a single excited spinon is a topological vortex. The
interaction between spinon-vortices can be obtained by substituting Eq. (\ref%
{psih}) into Eq. (\ref{fh}) and carrying out the area integration\cite{WM02}
\begin{equation}
{F}_{h}=\int \int d^{2}\mathbf{r}_{1}d^{2}\mathbf{r}_{2}\left[
\sum\nolimits_{\alpha }\alpha n_{\alpha }^{b}(\mathbf{r}_{1})\right] V(%
\mathbf{r}_{12})\left[ \sum\nolimits_{\beta }\beta n_{\beta }^{b}(\mathbf{r}%
_{2})\right] +\mathrm{const.}  \label{hv}
\end{equation}%
in which $\alpha ,\beta =\pm $ refer to the signs of vorticities carried by
spinons and
\begin{equation}
V(\mathbf{r}_{12})=-\frac{\pi \rho _{h}}{{4m_{h}}}\ln \frac{|\mathbf{r}_{1}-%
\mathbf{r}_{2}|}{r_{c}}
\end{equation}%
with $r_{c}\sim a$. Eq. (\ref{hv}) is similar to the XY model, except that $%
\mathbf{A}^{s}$ introduces $\pi $ instead $2\pi $ vortices and the vortex
cores are attached to spinons which have their own dynamics as governed by $%
H_{s}$ with an intrinsic quantum length scale $a_{c}$.

These spinon-vortex composites form vortex-antivortex bound pairs at low
temperature [\emph{cf.} Fig. \ref{topo}(b)]. Such a binding directly results
in a phase coherence%
\begin{equation}
\left\langle e^{i\Phi ^{s}(\mathbf{r})}\right\rangle \neq 0  \label{pc}
\end{equation}%
of the SC order parameter (\ref{psisc}). A detailed renormalization group
analysis leads to the following $T_{c}$ relation\cite{SWT03}
\begin{equation}
T_{c}\simeq \frac{E_{g}}{4k_{\mathrm{B}}}  \label{tc}
\end{equation}%
which connects the phase coherence temperature with the spin resonancelike
energy $E_{g}$ (\emph{cf.} Fig. \ref{res}).

The superconducting phase coherence (\ref{pc}) implies that spinons are
\emph{confined} in the bulk where a single spinon-vortex excitation costs a
logarithmically divergent energy. In this case, a finite energy elementary
excitation is an $S=1$ spin excitation composed of pairs of spinon-vortices,
whose minimal excitation corresponds to the spin resonancelike mode with $%
E_{g}=2E_{s}\sim \delta J$ with $E_{s}=(E_{m})_{\mathrm{\min }}$ denoting
the lowest level in the spinon spectrum in terms of $H_{s}$.

\subsubsection{Emergence of the nodal quasiparticle}

So far we have surveyed some novel properties of the SC state, which
generally are non-BCS like. They involve either \textquotedblleft high
energy\textquotedblright\ (\emph{e.g.}, spin excitations above the spin gap $%
E_{g}$) or short-distance (\emph{e.g.,} the vortex core at a length scale $%
\sim \xi $) physics. However, in a sufficiently long wavelength and low
energy regime, a typical d-wave BCS superconductor will eventually emerge,%
\cite{WST00} where the physical properties will be essentially dominated by
the conventional nodal quasiparticle excitations.

A quasiparticle excitation is created by the electron $c$-operator.\cite%
{WST00} According to the definition (\ref{mutual}), it is composed of a pair
of holon and spinon together with a phase vortex factor $e^{i\hat{\Theta}}$
as a \textquotedblleft bound state\textquotedblright . Such a low-lying
excitation is a \textquotedblleft collective mode\textquotedblright ,\cite%
{WST00} which will be independent of the \textquotedblleft
mean-field\textquotedblright\ phase string model $H_{\mathrm{string}}$ below
the spinon gap $E_{s}=E_{g}/2$. Besides the spinon gap in $H_{s}$, the
gapless \textquotedblleft phonon mode\textquotedblright\ for the condensed
holons in $H_{h}$ will be turned into the plasma mode by coupling to the
external electromagnetic field --- the Anderson-Higgs mechanism. So a
low-lying quasiparticle should remain coherent since there are neither other
low-lying states for it to decay into nor to scatter with.

To further see how a quasiparticle is compatible with $H_{\mathrm{string}}$,
one may examine the behavior of its three components. It has been seen that
an isolated spinon excitation will cost a logarithmically divergent energy
via the gauge field $A_{ij}^{s}$. Similarly a single localized holon excited
from the condensate will also create a divergent energy due to the global
spinon response via the gauge field $A_{ij}^{h}$ in $H_{s}$.\cite{ZMW03}
Furthermore, a divergent energy is also associated with the creation of
vortices by $e^{i\hat{\Theta}}$.\cite{ZMW03} However, as the bound state of
these three fields, a quasiparticle excitation will no longer invite any
nonlocal response from $H_{\mathrm{string}}$ neither from the holons nor the
spinons.\cite{WST00,ZMW03} In other words, the holon and spinon are confined
inside a quasiparticle, making it a stable \textquotedblleft
local\textquotedblright\ excitation with a \emph{finite} excitation energy.
The confining potential is logarithmic. \

The existence of the low-lying quasiparticle excitation may be also
understood based on the ground state (\ref{psib2}),\cite{WZM05} which can be
considered as the Bose condensation of the bosonic field $\hat{D}$ on the
RVB background $|\mathrm{RVB}\rangle $ with $\left\langle \hat{D}%
\right\rangle \neq 0$. Then a low-lying excitation may be constructed based
on $\hat{D}|\Psi _{G}\rangle _{\mathrm{SC}}$, with some \emph{smooth change}
in the $\hat{D}$. Generally with achieving the phase coherence (\ref{pcoh}),
one has
\begin{equation}
\hat{D}|\Psi _{G}\rangle _{\mathrm{SC}}\simeq \sum_{ij}\tilde{g}%
_{ij}\sum_{\sigma }\sigma c_{i\sigma }c_{j-\sigma }|\Psi _{G}\rangle _{%
\mathrm{SC}}  \label{dd3}
\end{equation}%
with
\begin{equation}
\tilde{g}_{ij}=\frac{1}{2}\tilde{G}_{ij}\left\langle e^{-i(1/2)\left( \Phi
_{i}^{s}+\Phi _{j}^{s}\right) }\right\rangle
\end{equation}%
according to the discussion in Sec. 4.1.1. Therefore, such type of low-lying
excitations, constructed with a smooth change in $\tilde{g}_{ij}$, can
always be described in terms of a pair of fermionic electronic
(quasiparticle) excitations. In particular, due to the d-wave nature of $%
\tilde{g}_{ij}$, the separation of the quasiparticles along the nodal line
may be infinitely large. As \textquotedblleft collective
modes\textquotedblright ,\ the quasiparticles are not directly described by $%
H_{\mathrm{string}}$, consistent with the fact that $H_{\mathrm{string}}$
determines $\varphi _{h}$ and $|\mathrm{RVB}\rangle $ in Eq. (\ref{bgs}),
but not the spinon backflow wave function $Z$ in $\Pi $.

\begin{figure}[tbp]
\begin{center}
\includegraphics[width=3in]{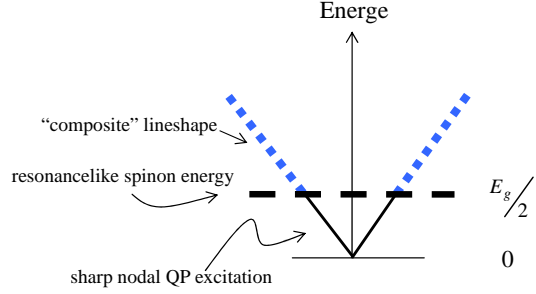}
\end{center}
\caption{{}A schematic nodal quasiparticle dispersion in the SC phase. The
lineshape below the characteristic spinon energy $E_{s}=E_{g}/2$ is very
sharp as energetically the quasiparticle cannot decay into a spinon and a
holon in the condensate. Above $E_{s}$, however, such a decay is
energetically allowed \emph{locally} with the spinon and holon remain
loosely confined at large distance. [From Ref. \protect\cite{WQ06}]}
\label{qp}
\end{figure}
\begin{figure}[tbph]
\begin{center}
\includegraphics[width=2in]{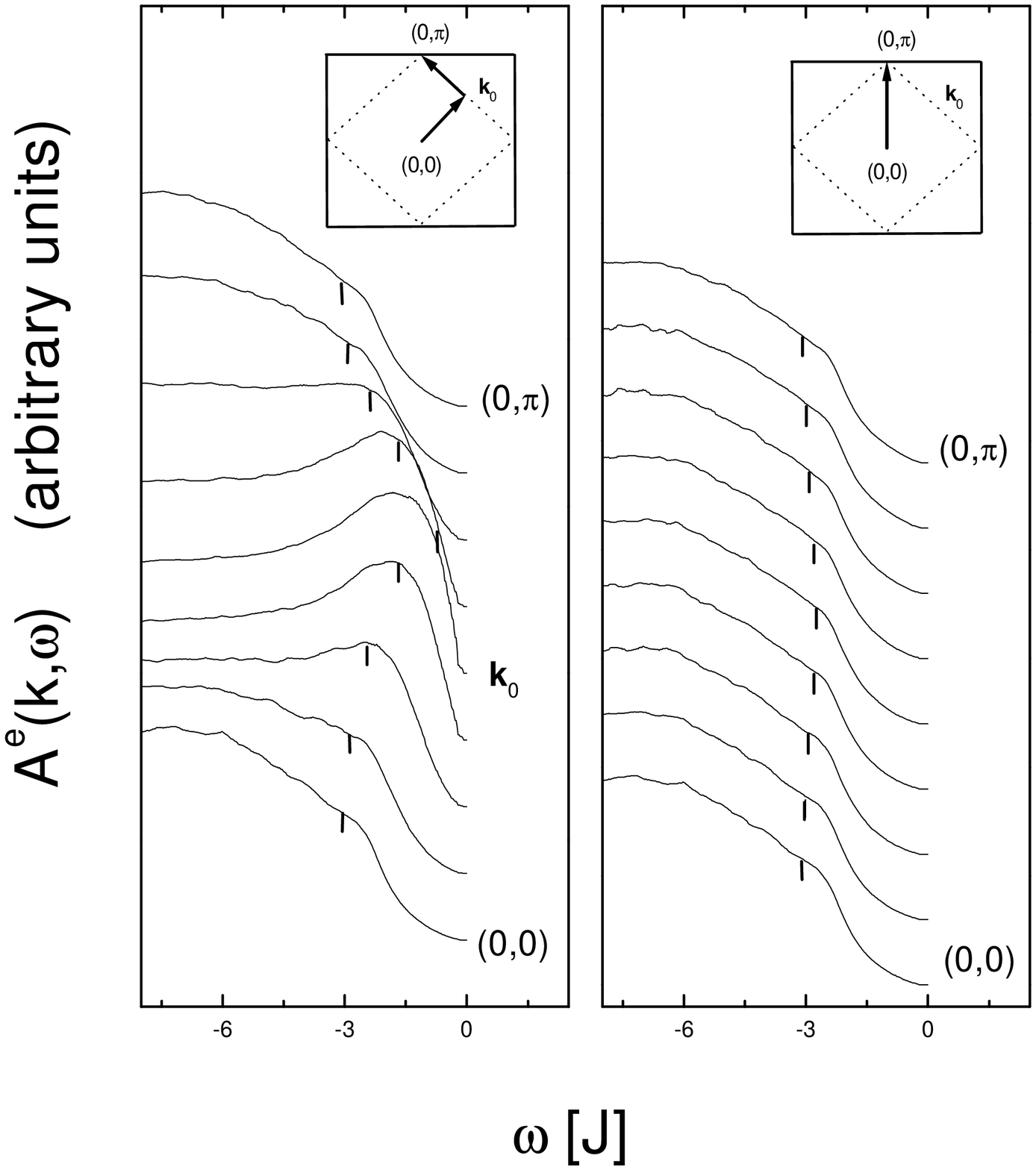}%
\includegraphics[width=2.1in]{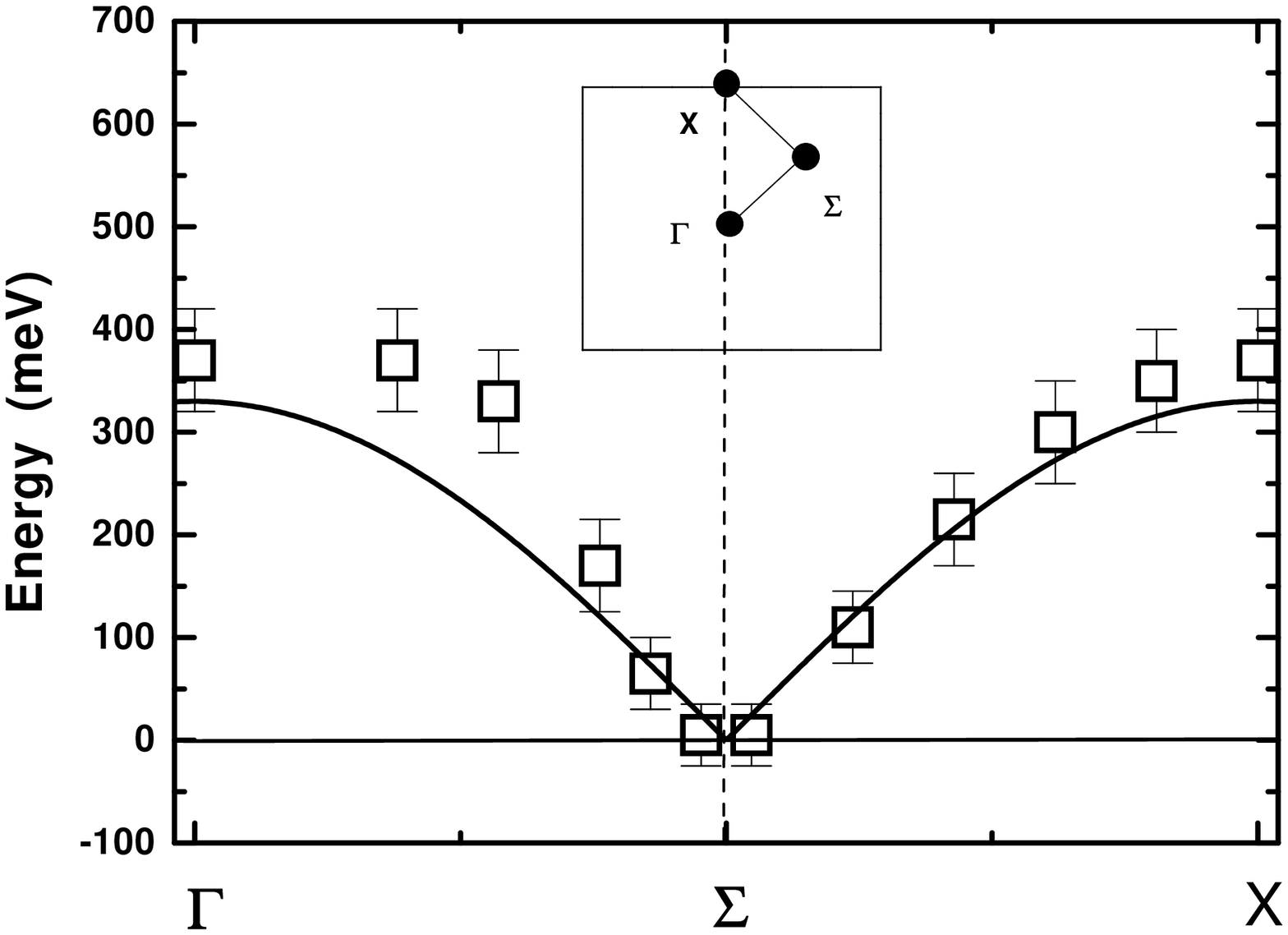}
\end{center}
\caption{Left: The single-electron spectral function in the one-hole case.
The small bars mark the spinon spectrum $E_{\mathbf{k}+\mathbf{k}_{0}}^{s}-%
\protect\mu _{0}$. Right: The \textquotedblleft
quasiparticle\textquotedblright\ spectrum determined by the ARPES (open
square) and by the spinon spectrum $E_{\mathbf{k}+\mathbf{k}_{0}}^{s}$
(solid curve). (From Ref. \protect\cite{WMST01}) }
\label{1hole}
\end{figure}

The effective Hamiltonian governing the motion of a quasiparticle excitation
can be derived based on the original $t$-$J$ model.\cite{WST00,GW08} The
renomalization effect from interacting with the background electrons will be
represented by the mean-field order parameter $\Delta _{ij}^{\mathrm{SC}}$,
etc. An equation of motion description of such a quasiparticle has been
developed,\cite{WST00,GW08} which shows the low-energy part of the spectral
function is similar to a sharp d-wave BCS one, but the high-energy part ($%
>E_{s}$) is nontrivial as the composite (spin-charge separation) feature
will show up, where the quasiparticle can decay into a pair of spinon and
holon \emph{locally} without costing much from the logarithmic confining
potential. Figure \ref{qp} illustrates the d-wave nodal quasiparticle
dispersion schematically, where the quasiparticle spectral function has a
sharp lineshape at
\begin{equation}
E_{\mathrm{qasiparticle}}<E_{\mathrm{spinon}}+E_{\mathrm{holon}}
\end{equation}%
Namely, the spinon gap $E_{s}$ (or spin gap $E_{g})$ will provide a
protection for the coherent quasiparticles in the SC state. Indeed, in the
half-filling limit, $E_{s}\rightarrow 0$ and the single-electron spectral
function has only an incoherent \textquotedblleft
composite\textquotedblright\ part\cite{WMST01} as shown in Fig. \ref{1hole}.
The more detailed results for the spectral function in the SC state will be
presented in a forthcoming paper.\cite{GW08}

\subsection{Lower Pseudogap Phase (LPP) [Spontaneous Vortex Phase (SVP)]}

Due to the composite structure of the SC order parameter $\Delta ^{\mathrm{SC%
}}\,$\ [Eq. (\ref{psisc})], a regime may exist at $T>T_{c}$ where $%
\left\langle e^{i\Phi ^{s}(\mathbf{r})}\right\rangle =0$ but the Cooper pair
amplitude $\Delta ^{0}$ remains finite. Such a regime is known as the
spontaneous vortex phase (SVP) or the lower pseudogap phase (LPP) of the
phase string model.\cite{WM02,WQ06}

The LPP is described by \emph{free} spinon vortices, which are thermally
excited and proliferate such that the phase $\Phi ^{s}$ gets disordered.
Namely, the LPP is an electron \textquotedblleft
fractionalized\textquotedblright\ state with the proliferation of unbinding
spinons, and the main distinction between the SC phase and LPP lies in the
phase (de)coherence or spinon (de)confinement.

In the LPP, a finite $\Delta ^{0}$ ensures that the spinon vortices are
still well defined. In fact, $\psi _{h}\neq 0$ means that the generalized GL
equations [Sec. 4.1.2.] are also applicable in the LPP. Since the holon
condensation persists in the LPP, the spin dynamics at $E>E_{g}$ should also
remain qualitatively similar to the SC phase [Sec. 4.1.3.].

\subsubsection{Phase diagram for the LPP}

\begin{figure}[tbp]
\begin{center}
\includegraphics[width=2.5in]{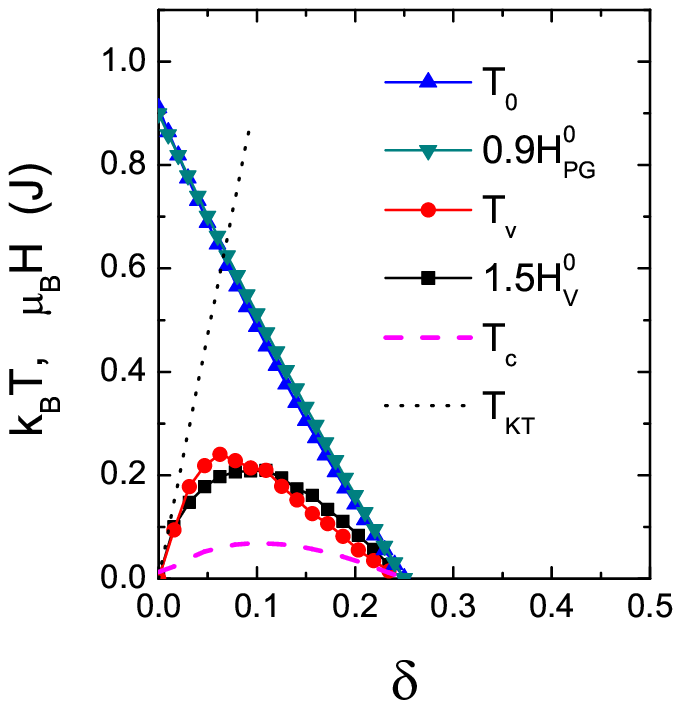}%
\includegraphics[width=2.5in]{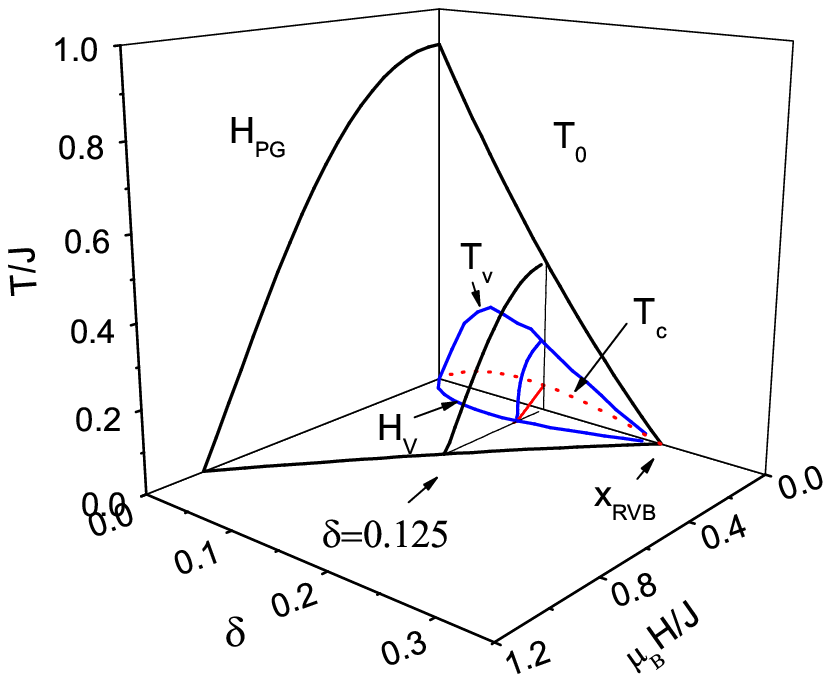}
\end{center}
\caption{Left panel: The characteristic temperature and magnetic field
scales which determine the phase diagram of the upper and lower pseudogap
and superconducting phases based on the phase string model. Here $T_{0}$ and
$H_{\mathrm{PG}}^{0}\equiv $ $H_{\mathrm{PG}}(T=0)$ for the UPP, $T_{v}$ and
$H_{v}^{0}\equiv H_{v}^{{}}(T=0)$ for the LPP, and $T_{c}$ for the SC phase.
Right panel: The phase diagram of the pseudogap regimes in the
three-dimensional space of magnetic field, doping, and temperature. [From
Ref. \protect\cite{WQ06}]}
\label{phase3D}
\end{figure}

$\Delta ^{0}$ is composed of the RVB pairing $\Delta ^{s}$ and holon
condensate $\psi _{h}^{\ast }$, which disappear at some $T_{0}$ and $T_{v}$,
respectively. We will see generally $T_{v}<T_{0}$, as $\psi _{h}\neq 0$ is
always underpinned by the spin singlet pairing. Thus $T_{v}$ will represent
the characteristic temperature for the LPP, whereas $T_{0}$ defines the
boundary of the so-called upper pseudogap phase (UPP) to be discussed in the
next section.

Note that without $\mathbf{A}^{s}$, the holon system would be a simple 2D
hard-core boson problem according to Eq. (\ref{hh}) or (\ref{hholon}), with
the Kosterlitz-Thouless (KT) transition temperature for the holon
condensation given by $T_{\mathrm{KT}}=\pi \delta \left( 2a^{2}m_{h}\right)
^{-1}$ as shown in Fig. \ref{phase3D} by a dotted line (with $t_{h}=3J$).

However, the frustration effect of $\mathbf{A}^{s}$ on the holon
condensation will play a crucial role here. The spinon-vortex density is
determined by $n_{v}=\sum_{m\sigma }\left\langle \gamma _{m\sigma }^{\dagger
}\gamma _{m\sigma }\right\rangle /N$. Due to the opening up of a spin gap $%
E_{g}$ in the holon condensation phase, $n_{v}$ is exponentially small for $%
T\ll E_{g}$. With increasing temperature, $n_{v}$ will monotonically
increase until reaching the maximal number $n_{v}^{\mathrm{max}}=1-\delta $
at $T=T_{0}$ where all the RVB pairs break up.
\begin{figure}[tbp]
\begin{center}
\includegraphics[width=2.5in]{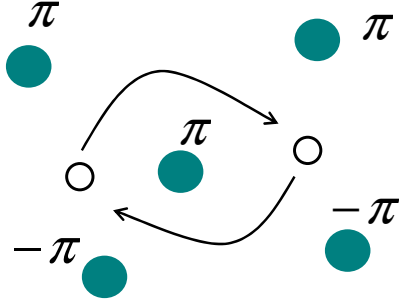}
\end{center}
\caption{Two holons (open circles) exchanging positions can pick up a minus
sign if an unpaired spinon (grey circle) is enclosed in between, which
carries a $\pm \protect\pi $ flux tube. So the phase coherence of bosonic
holons can be effectively destroyed to prevent the Bose condensation if
there is a sufficient number of free spinon excitations in the spin
background. }
\label{holonexchange}
\end{figure}

Since each free spinon carries a $\pi $ fluxoid as perceived by the holons,
the quantum phase coherence among bosonic holons will be violently
interrupted if on average there is an excited (unpaired) spinon sitting
between two neighboring holons (as illustrated by Fig. \ref{holonexchange}),
where an exchange between a pair of holons can gain a minus sign in the wave
function. In other words, the holon condensation must break down when the
vortex density $n_{v}$ is comparable to holon density $\delta $, far less
than $n_{v}^{\mathrm{max}}$ at low doping. Such a consideration provides an
estimate of the upper bound for $T_{v}$ as\cite{WQ06}
\begin{equation}
n_{v}=\delta .  \label{tv}
\end{equation}

Equation (\ref{tv}) can be also understood based on the \textquotedblleft
core touching\textquotedblright\ picture of spontaneously excited spinon
vortices. Note that the average distance between excited spinons may be
defined by $l_{s}\equiv 2a/\sqrt{\pi n_{v}}.$ Since the characteristic core
size of a spinon vortex is $a_{c}$, then one expects that the
\textquotedblleft supercurrents\textquotedblright\ carried by
spinon-vortices are totally destroyed when the sample is fully packed by the
\textquotedblleft cores\textquotedblright\ with $l_{s}=2a_{c}$ which also
results in Eq. (\ref{tv}).

\begin{figure}[tbp]
\begin{center}
\includegraphics[width=2.5in]{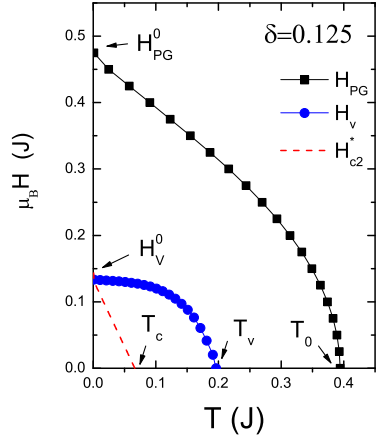}
\end{center}
\caption{The magnetic field -- temperature phase diagram of the pseudogap
phases at doping $\protect\delta =0.125$. [From Ref. \protect\cite{WQ06}]}
\label{HTPhD}
\end{figure}

The numerical result of the characteristic temperature $T_{v}$ in terms of
Eq. (\ref{tv}) is plotted in the left panel of Fig. \ref{phase3D} by the
filled circles. It shows that $T_{v}$ is nested below $T_{\mathrm{KT}}$ at
low doping and is dramatically reduced from $T_{\mathrm{KT}}$ with the
further increasing of doping due to the frustrations induced by spinon
excitations, which at larger doping remains nested below $T_{0}$ of the UPP
and eventually vanishes at $x_{\mathrm{RVB}}$ together with $T_{0}$ (see the
next section).

An external magnetic field will break up \emph{more }RVB pairs through the
Zeeman effect (\ref{zeeman}) at a given temperature. By considering the
Zeeman effect on the energy spectrum, $E_{m\sigma }=E_{m}-\sigma \mu _{%
\mathrm{B}}B$, the magnetic field dependence of $T_{v}=T_{v}(H)$ can be
further obtained from Eq. (\ref{tv}) (note that we do not distinguish the
magnetic field $H$ and induction $B$ as the magnetization is usually very
weak here). Or conversely, for each $T<T_{v}(0)$ there is a characteristic
field $H_{v}(T)$ at which the LPP phase is destroyed [Fig. \ref{HTPhD}]. $%
H_{v}^{0}\equiv H_{v}$($T=0)$ determined this way is shown in the left panel
of Fig. \ref{phase3D} by the filled squares.

For comparison, the SC temperature $T_{c}$ [Eq. (\ref{tc})] is shown as the
dashed curve in the left panel of Fig. \ref{phase3D}. Furthermore, in the
mixed state below $T_{c}$, by including the Zeeman energy, $E_{g}$ is
reduced to $E_{g}^{\ast }=E_{g}(B)-2\mu _{\mathrm{B}}B$ such that one can
estimate $T_{c}(B)$ by using a simple relation $T_{c}(B)\sim E_{g}^{\ast
}/4k_{\mathrm{B}}.$ Then in turn one may define $H_{c2}^{\ast }\equiv
B(T_{c}),$ which\ is shown in Fig. \ref{HTPhD} by a dashed curve. Note that $%
H_{c2}^{\ast }$ so defined will vanish at $T_{c},$ resembling the
conventional $H_{c2}$ in a BCS superconductor. However, since free spinon
vortices are generally present at $H>$ $H_{c2}^{\ast }$, $H_{c2}^{\ast }$ is
a crossover field which no longer has the same meaning as $H_{c2}$ in a
conventional BCS superconductor. Roughly speaking, the Abrikosov magnetic
vortices are expected to be present below $H_{c2}^{\ast }$ where the
spontaneous spinon-vortices, generated by the Zeeman effect, are still
loosely paired, whereas the vortex unbinding occurs above $H_{c2}^{\ast }$.
So $H_{c2}^{\ast }$ defines a crossover between two types of vortex regime.
The numerical result shows that $\mu _{\mathrm{B}}H_{c2}^{\ast }(0)\simeq
E_{g}(B=0,T=0)/2$, which results in $H_{c2}^{\ast }(0)\simeq 3T_{c}$ $\left(
\mathrm{Tesla/Kelvin}\right) ,$ according to Eq. (\ref{tc}).

\subsubsection{Nernst effect}

\begin{figure}[tbp]
\begin{center}
\includegraphics[width=4.5in]{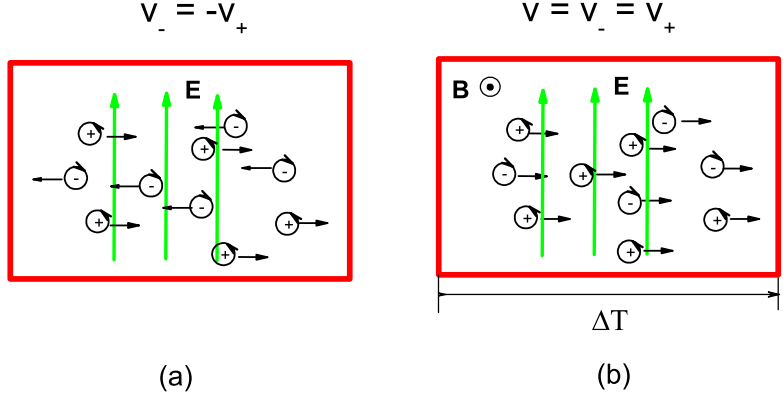}
\end{center}
\caption{Schematic picture of (a) the flux flowing under an electric field
in the LPP, which can lead to a flux flow resistivity as well as the spin
Hall effect, and (b) the flux flowing under a temperature gradient $\protect%
\nabla T$, which must be balanced by an electric field $E$ and thus leads to
the Nernst effect. }
\label{SchematicNernst}
\end{figure}

Base on the London-like equation (\ref{j2}), using the steady current
condition%
\begin{equation}
\partial _{t}\mathbf{J}=0  \label{j1}
\end{equation}%
and the electric field $\mathbf{E}=-\partial _{t}\mathbf{A}^{e}$ in the
transverse gauge, one finds\cite{WM02}
\begin{equation}
\mathbf{E}=\mathbf{\hat{z}}\times \phi _{0}\left( n_{v}^{+}\mathbf{v}%
_{+}-n_{v}^{-}\mathbf{v}_{-}\right) \text{ \ }  \label{ey}
\end{equation}%
where $n_{v}^{\pm }$ denotes the density of spinon vortices and antivortices
with drifting velocity $\mathbf{v}_{\pm }$ along a direction perpendicular
to the electric field. As illustrated by Fig. \ref{SchematicNernst}(a), the
electric field and the drifting of vortices and antivortices must be
balanced according to Eq. (\ref{ey}) in order to avoid the system being
accelerated indefinitely with $\partial _{t}\mathbf{J}\neq 0$. So the
applied electric field will drive the vortices and antivortices moving along
a perpendicular direction with opposite velocities: $\mathbf{v}_{+}=-\mathbf{%
v}_{-}$ if the vortices and antivortices are not polarized by the external
magnetic field, \emph{i.e., }$n_{v}^{+}$ $\mathbf{=}$ $n_{v}^{-}$.

The Nernst signal $e_{N}$ is the electric field measured along the $\hat{y}$%
-direction when spinon vortices and antivortices are both driven by a
temperature gradient in the \emph{same} direction along the $\hat{x}$%
-direction:%
\begin{equation}
e_{N}=\frac{E_{y}}{-\nabla _{x}T}\text{ \ .}  \label{nuv0}
\end{equation}%
Such a case is shown in Fig. \ref{SchematicNernst}(b), where the
spinon-vortices and -antivortices move along the\emph{\ }$\hat{x}$-direction
with $\mathbf{v}_{+}=\mathbf{v}_{-}=\mathbf{v}$. To have a finite $\mathbf{E}
$ in terms of Eq. (\ref{ey}), then the vortex density $n_{v}^{\pm }$ has to
be polarized by the external magnetic field $\mathbf{B=}B\mathbf{\mathbf{%
\hat{z}}}$ according to the \textquotedblleft neutrality\textquotedblright\
condition $B=\phi _{0}\left( n_{v}^{+}-n_{v}^{-}\right) $ such that
\begin{equation}
\mathbf{E}=\mathbf{B}\times \mathbf{v.}
\end{equation}

Suppose $s_{\phi }$ is the \emph{transport} entropy carried by a spinon
vortex and $\eta _{s}$ is its viscosity such that the drift velocity $%
\mathbf{v}^{s}$ can be decided by $s_{\phi }\nabla T=-\eta _{s}\mathbf{v.}$
Then one has $e_{N}=B\frac{s_{\phi }}{\eta _{s}}$. On the other hand, in the
absence of the temperature gradient, a charge current can also drive a
transverse motion of spinon vortices and antivortices along \emph{opposite}
directions, i.e., $\mathbf{v}_{\pm }=\pm \mathbf{v},$ such that an electric
field is generated along the current direction according to Eq. (\ref{ey}),
leading to a finite resistivity due to the presence of free vortices, which
is given by
\begin{equation}
\rho =\frac{n_{v}}{\eta _{s}}\phi _{0}^{2},  \label{rhov}
\end{equation}%
with $n_{v}\equiv n_{v}^{+}+n_{v}^{-}$. This formula is familiar in the
vortex flow regime of a conventional superconductor. Then, by eliminating $%
\eta _{s}$, one obtains\cite{WQ06}%
\begin{equation}
\alpha _{xy}\equiv \frac{e_{N}}{\rho }=\frac{Bs_{\phi }}{\phi _{0}^{2}n_{v}}.
\label{alpha}
\end{equation}%
\begin{figure}[tbp]
\begin{center}
\includegraphics[width=2.5in]{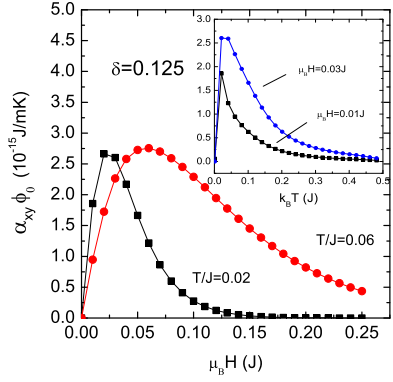}
\end{center}
\caption{{}The quantity $\protect\alpha _{xy}\protect\phi _{0}/d\equiv e_{N}%
\protect\phi _{0}/\protect\rho d$, which is related to the Nernst signal
without involving the viscosity coefficient, is shown as a function of the
magnetic field and temperature (the inset) at a given doping. Here $d=7.7%
\mathring{A}$ is the distance between two $\mathrm{CuO}_{2}$ layers. [From
Ref. \protect\cite{WQ06}]}
\label{figalpha}
\end{figure}

What really makes the Nernst transport unique in the present theory is that
the transport entropy $s_{\phi }$ here is associated with the spin degree of
freedom due to its free $S=1/2$ moment, instead of a normal core in a
conventional BCS superconductor, which is given by $s_{\phi }=k_{\mathrm{B}%
}\left\{ \ln \left[ 2\cosh \left( \beta \mu _{\mathrm{B}}B\right) \right]
-\beta \mu _{\mathrm{B}}B\tanh \left( \beta \mu _{\mathrm{B}}B\right)
\right\} .$ The temperature and magnetic field dependence of $\alpha _{xy}$
is shown in Fig. \ref{figalpha}.\cite{WQ06} The magnitude of such a quantity
is comparable to the experimental data, implying that the transport entropy
due to the free moment in a spinon-vortex is capable to produce the Nernst
signal observed experimentally.\cite{Ong}

\subsubsection{Spin Hall effect}

A unique prediction related to the spinon-vortex motion driven by an
external electric field is the existence of a conserved dissipationless spin
Hall current.\cite{KQW05} As shown in Fig. \ref{SchematicNernst}(a),
vortices can be driven by an in-plane electric field to move along the
transverse direction. Since each vortex carries a free moment, if these
moments are partially polarized, say, by an external magnetic field via the
Zeeman effect, then a spin Hall current can be generated along the vortex
motion direction. The spin Hall conductivity is determined as follows:\cite%
{she2005}

\begin{equation}
\sigma _{H}^{s}=\frac{\hbar \chi _{s}}{g\mu _{B}}\left( \frac{B}{n_{v}\phi
_{0}}\right) ^{2}  \label{sigmh}
\end{equation}%
which only depends on the intrinsic properties of the system like the
uniform spin susceptibility $\chi _{s},$ with the electron $g$-factor $%
g\simeq 2$. It is important to note that the external magnetic field $B$
applied perpendicular to the 2D plane reduces the spin rotational symmetry
of the system to the conservation of the $S^{z}$ component only, satisfying $%
\frac{\partial S^{z}}{\partial t}+\nabla \cdot \mathbf{J}^{s}=0.$ Thus the
polarized spin current $\mathbf{J}^{s}$ is still conserved and remains
dissipationless as the \emph{current} of its carriers -- vortices is \emph{%
dissipationless} in the LPP. In contrast, the charge current remains
dissipative as $\rho \neq 0$.

\subsubsection{Magnetization}

Inspired by the experiment,\cite{Ong} the diamagnetism has been also
studied. The total magnetization can be expressed as\cite{QW07}
\begin{equation}
M_{\mathrm{tot}}=M_{\mathrm{dia}}+M_{\mathrm{para}}
\end{equation}%
in which $M_{\mathrm{dia}}$ and $M_{\mathrm{para}}$ stand for the orbital
diamagnetism from the vortices and the paramagnetism from the Zeeman
coupling, respectively. Based on the mutual Chern-Simons formulation of the
phase string model outlined in Sec. 3.2, with a systematic description of
multi-spinon excitations and interaction between them, the nonlinear effect
of the magnetization vs. magnetic field can be effectively treated.\cite%
{QW07}

\begin{figure*}[tbp]
\begin{center}
\includegraphics[width=2in] {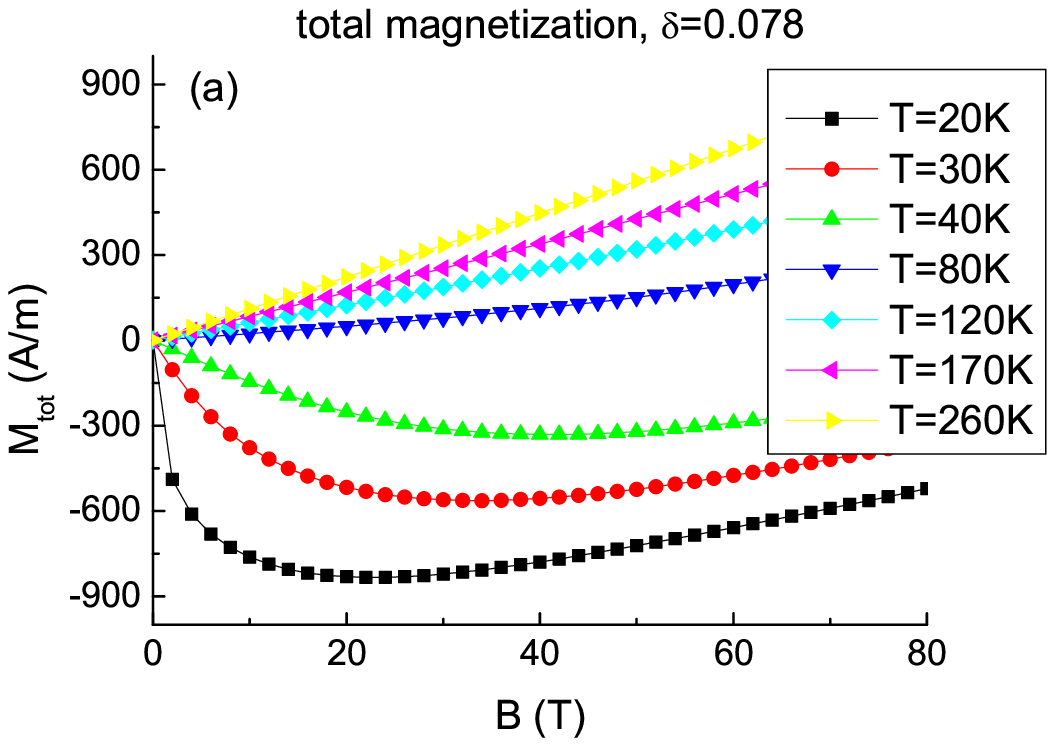}%
\includegraphics[width=2in]{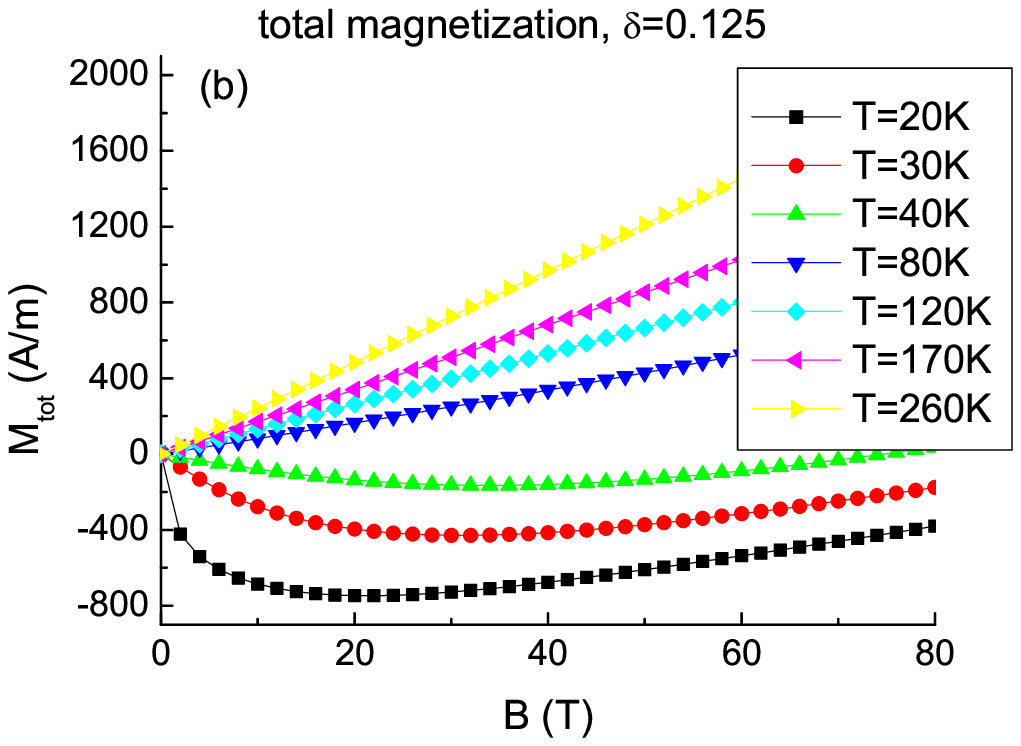} %
\includegraphics[width=2in] {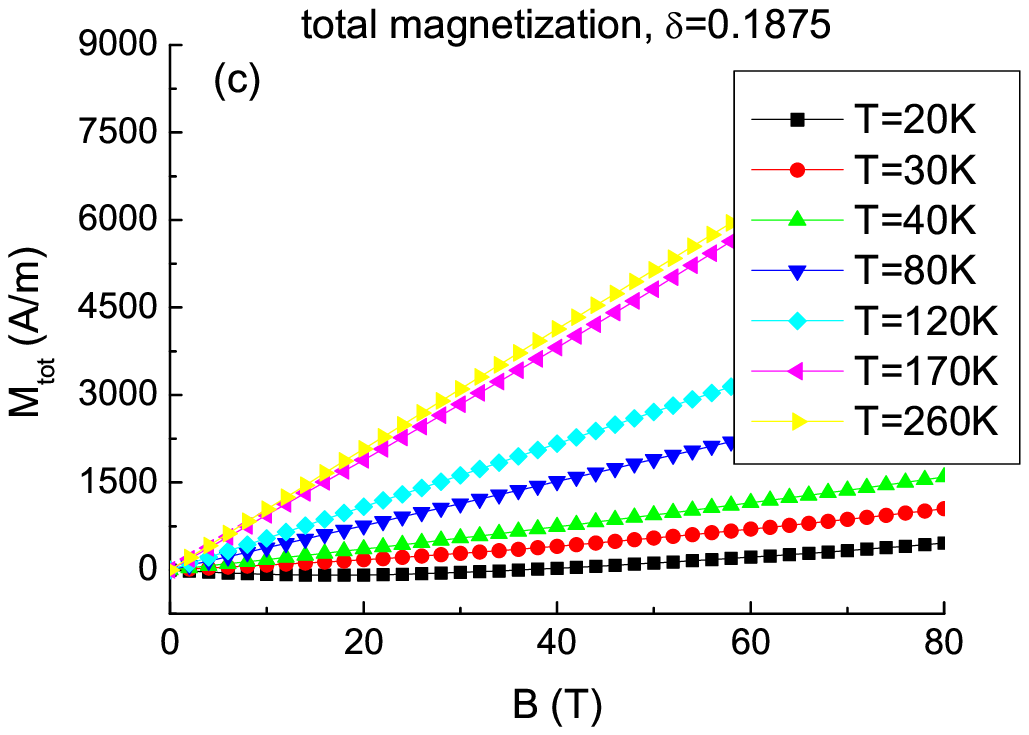}%
\includegraphics[width=2in]{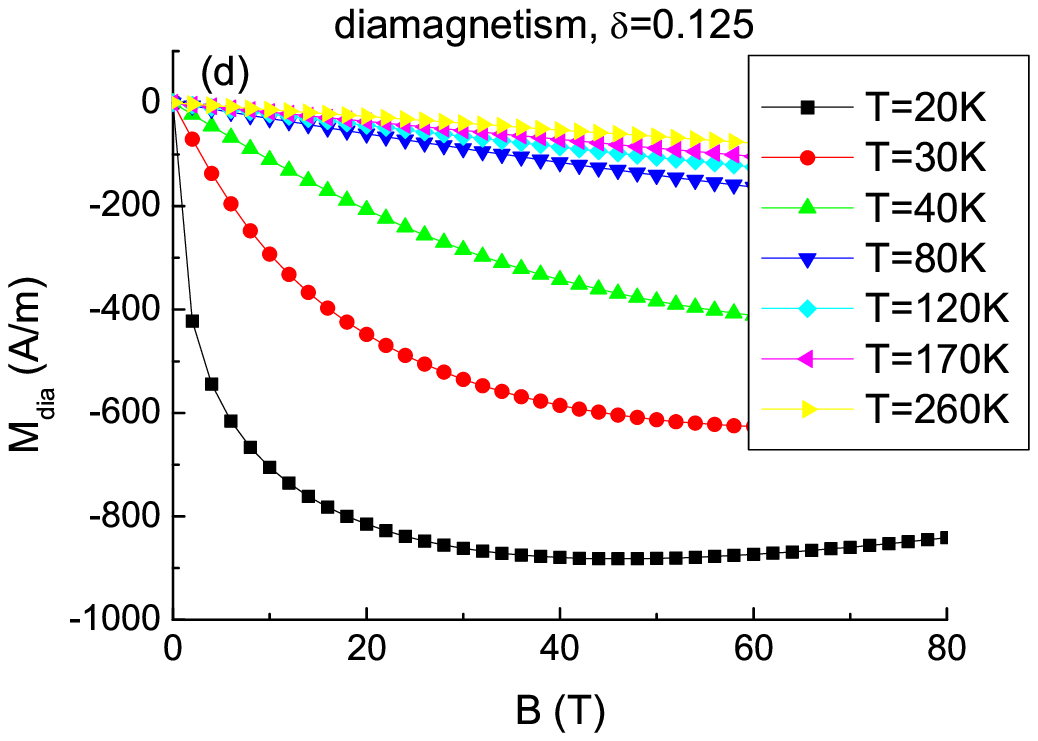}
\end{center}
\caption{The $B$ dependence of the total magnetization $M=M_{\mathrm{dia}%
}+M_{\mathrm{para}}$ at various doping concentrations: (a) $\protect\delta %
=0.078$, (b) $\protect\delta =0.125$, (c) $\protect\delta =0.188;$ (d) the
diamagnetism $M_{\mathrm{dia}}$ at $\protect\delta =0.125$. [From Ref.
\protect\cite{QW07}]}
\label{magnetization}
\end{figure*}

The magnetic field and temperature dependence of the total magnetization at
different doping concentrations as well as the diamagnetism part $M_{\mathrm{%
dia}}$ at $\delta =0.125$ are shown in Fig. \ref{magnetization} based on a
mean-field approximation in the mutual Chern-Simons description.\cite{QW07}
Note that $M_{\mathrm{para}}=\chi _{s}B$ with $\chi _{s}$ as the uniform
spin susceptibility to be discussed later in the weak field limit.

\subsubsection{Magneto-resistance}

Due to the deconfinement of spinons, the quasiparticles are no longer the
stable low-lying excitations in the LPP. The origin of the dissipation comes
mainly from the flow of the spinon vortices. In fact, the resistivity (\ref%
{rhov}) is similar to the flux-flow resistivity in a Type II superconductor
except that $n_{v}$ in general is not simply proportional to the external
magnetic field $B$. Namely, the spinon vortices can be spontaneously
(thermally) generated with $n_{v}\neq 0,$ such that $\rho \neq 0$ even at $%
B=0$. The resistivity $\rho (B)$ can be then expanded as\cite{QW07}
\begin{figure}[tbp]
\begin{center}
\includegraphics[width=3in] {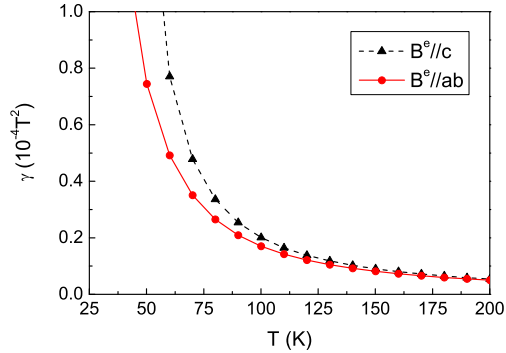}
\end{center}
\caption{The magneto-resistance coefficient $\protect\gamma $ vs temperature
for the magnetic field $B$ which is either perpendicular or transverse to
the $\mathrm{ab}$ plane. [From Ref. \protect\cite{QW07}]}
\label{drho}
\end{figure}
\begin{equation}
\rho (B)=\rho (0)\left[ 1+\gamma B^{2}+O(B^{4})\right]
\end{equation}%
where the odd power terms of $B$ vanish due to the symmetry $\rho (B)=\rho
(-B)$. Suppose that the dependence of the viscosity $\eta _{s}$ on $B$ is
negligible, then the quadratic coefficient $\gamma $ can be expressed as
\begin{equation}
\gamma =\frac{\rho (B)-\rho (0)}{\rho (0)B^{2}}\simeq \frac{n_{v}(B)-n_{v}(0)%
}{n_{v}(0)B^{2}}.  \label{nu}
\end{equation}%
The coefficient $\gamma _{\perp }$ and $\gamma _{\parallel }$, with the
external magnetic field $B$ perpendicular and parallel to the 2D plane,
respectively, can be calculated numerically as shown in Fig. \ref{drho}.\cite%
{QW07} An important prediction of the present theory, as shown by Fig. \ref%
{drho}, is that $\gamma _{\Vert }$ is \emph{comparable} to $\gamma _{\perp }$%
. This is a rather unusual case for a vortex-flow-induced resistivity, since
normally the in-plane vortices are always created by the \emph{perpendicular}
magnetic field in a Type II superconductor, where the vortex-flow-induced
resistivity only exhibits field-dependent magneto-resistivity for the
component of $B$ which is perpendicular to the plane. But in the present
theory, vortices are tied to the free spinons. Since the latter can be
created by the Zeeman term with the external magnetic field pointing at
\emph{any} directions, the former can thus be created by the in-plane field
as well. The present mean-field-type treatment of $\rho $ may not be
expected to be quantitatively accurate in view of possible corrections from
the fluctuations, but the existence of an \emph{anomalous} transverse
magneto-resistivity with $\gamma _{\Vert }$ comparable to $\gamma _{\perp }$
remains a very peculiar prediction.\cite{QW07}

\subsection{Upper Pseudogap Phase (UPP)}

At $T>T_{v}$, the bosonic holons will be in a non-degenerate regime where
the quantum coherence is destroyed by the excited spinons via $\mathbf{A}%
^{s} $. By contrast, the bosonic spinons can still maintain their quantum
coherence up to $T_{0}$ where the spin singlet pairing order parameter $%
\Delta ^{s}$ eventually vanishes. The UPP as characterized by $\Delta
^{s}\neq 0$ is defined at $T_{v}<T<T_{0}$, whose key features will be
described in the following.

\subsubsection{Phase diagram for the UPP}

\begin{figure}[tbp]
\begin{center}
\includegraphics[width=2.7in]
{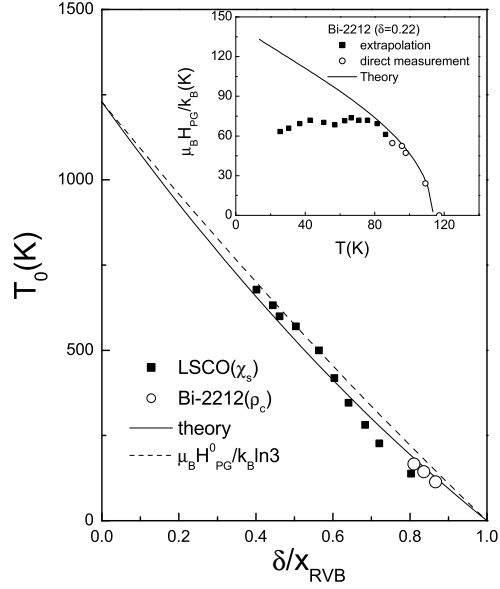}
\end{center}
\caption{The characteristic temperature $T_{0}$ of the UPP versus $\protect%
\delta /x_{\mathrm{RVB}}$. Solid line: the present theory; Full squares:
determined from the uniform spin susceptibility $\protect\chi _{s}$ in LSCO
compound;\protect\cite{LSCO} Open circles: determined from the c-axis
magneto-resistivity ($\protect\rho _{c})$ measurement in Bi-2212 compound;%
\protect\cite{pszeeman} The dashed line shows the scaling relation of the
zero-temperature critical field $H_{\mathrm{PG}}^{0}$ with $T_{0}$ as
predicted by the theory. Inset: the critical field $H_{\mathrm{PG}}$ as a
function of temperature at $\protect\delta =0.22$. The experiment data from
the c-axis transport in Bi-2212 (Ref. \protect\cite{pszeeman}) are also
shown by the open and full squares. [From Ref. \protect\cite{GW05}]}
\label{fT0}
\end{figure}

Based on the self-consistent solution of $H_{s}$, the characteristic
temperature $T_{0}$ at which $\Delta ^{s}\rightarrow 0$ is given by\cite%
{GW05}%
\begin{equation}
k_{B}T_{0}=\left( \frac{1-\frac{\delta }{2}}{\ln \frac{3-\delta }{1-\delta }}%
\right) J_{\mathrm{eff}}\text{ \ .}  \label{T0}
\end{equation}%
Figure \ref{fT0} shows $T_{0}$ (solid curve) as a function of doping with $%
J=1350$ \textrm{K.} The experimental data determined by the uniform spin
susceptibility measurement in $\mathrm{LSCO}$ \cite{scaling,LSCO} (see the
discussion in the next section) are shown by the full squares. Furthermore,
the open circles are independently determined from the c-axis transport \cite%
{pszeeman} in the overdoped regime.
Note that here $J_{\mathrm{eff}}=(1-2g\delta )$ vanishes at $\delta =$ $x_{%
\mathrm{RVB}}\equiv 1/\left( 2g\right) $ and the curve $T_{0}$ versus $%
\delta /x_{\mathrm{RVB}}$ in Fig. \ref{fT0} is not sensitive to the choice
of $g.$ Based on the above experimental data\cite{scaling,LSCO,pszeeman} $x_{%
\mathrm{RVB}}$ is fixed at $0.25$ such that $g=2$.

Due to the bosonic RVB origin of the UPP, the Zeeman effect of an external
magnetic field can effectively destroy the singlet pairing of spins in the
strong field limit, which is the only direct field effect on the RVB
background.

With incorporating the Zeeman term (\ref{zeeman}), one can obtain the
zero-temperature \textquotedblleft critical\textquotedblright\ field $%
H_{PG}^{0}$ at which $\Delta ^{s}$ vanishes:%
\begin{equation}
\mu _{B}H_{\mathrm{PG}}^{0}=\ln \left( \frac{3-\delta }{1-\delta }\right)
k_{B}T_{0}\text{ \ }  \label{h0}
\end{equation}%
In Fig. \ref{fT0}, $\ln 3\mu _{B}H_{PG}^{0}/k_{B}$ is plotted as the dashed
curve which scales with the zero-field $T_{0}$ fairly well, which means $\mu
_{B}H_{\mathrm{PG}}^{0}\simeq 1.1k_{B}T_{0}$. The temperature dependence of
the \textquotedblleft critical\textquotedblright\ field $H_{\mathrm{PG}}(T)$
is shown in the inset of Fig. \ref{fT0} at $\delta =0.22,$ together with the
experimental data obtained from the c-axis magneto transport measurements.%
\cite{pszeeman} We see that the \emph{high-temperature} experimental data
(open circles) fit the theoretical curve very well without any additional
adjustable parameter. Furthermore the zero-field $T_{0}$ determined by the
\emph{same }experiments is also in good agreement with the theory as shown
(open circles) in the main panel of Fig. \ref{fT0}. But one may notice that
the experimental $H_{\mathrm{PG}}(T)$ starts to deviate from the theoretical
curve in the inset (full squares) as the temperature is further lowered and
saturated to approximately the half of the predicted number (which implies $%
\mu _{B}H_{\mathrm{PG}}^{0}\simeq k_{B}T_{0}/2)$. However, such a deviation
occurs only for those data (full squares) which have been obtained by \emph{%
extrapolation} in the experimental measurement \cite{pszeeman} and therefore
may not be as reliable as the higher temperature ones (open squares) in the
inset of Fig. \ref{fT0}.

Finally, the three-dimensional phase diagram of the UPP, together with the
LPP and SC phase, in the parameter of magnetic field, doping concentration,
and the temperature is summarized in the right panel of Fig. \ref{phase3D}. $%
H_{\mathrm{PG}}$ vs. $T$ at $\delta =0.125$ is also plotted in Fig. \ref%
{HTPhD}$.$

\subsubsection{Uniform spin susceptibility}

\begin{figure}[tbp]
\begin{center}
\includegraphics[width=3in]{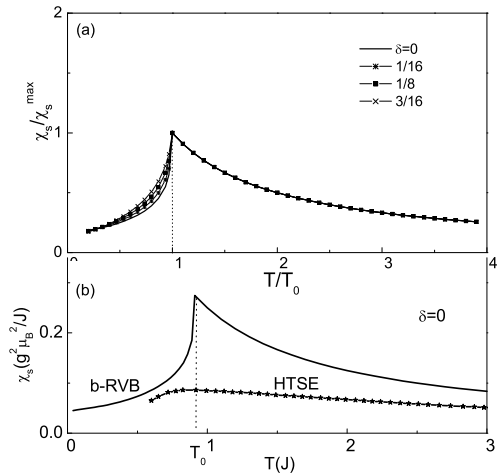}
\end{center}
\caption{(a) The calculated uniform spin susceptibility $\protect\chi _{s}$
scaled with the maximum $\protect\chi _{s}^{\mathrm{\max }}$ at $T_{0}$
versus $T/T_{0}$, which follows an approximately doping-dependent curve. (b)
The theoretical $\protect\chi _{s}$ at half-filling (solid) and the one
obtained by the high temperature series expansion (HTSE). The latter fits
the experimental scaling curves\protect\cite{scaling,LSCO} very well. [From
Ref. \protect\cite{GW05}]}
\label{fsus}
\end{figure}

The spin singlet pairing (RVB) nature of the UPP is clearly manifested in
the uniform spin susceptibility $\chi _{s}$ given in the main panel of Fig. %
\ref{fsus}(a) at different doping concentrations.\cite{QW05} Note that $\chi
_{s}$ reaches a maximum value $\chi _{s}^{max}$ at temperature $T_{0}$ where
the RVB order parameter $\Delta ^{s}$ vanishes. At $T>T_{0}$, $\chi _{s}$
follows a Curie-$1/T$ behavior as spins become free moments at the
mean-field level. The curves in Fig. \ref{fsus}(a) are presented as $\chi
_{s}/\chi _{s}^{\max }$ versus $T/T_{0}$, which approximately collapse onto
a single curve independent of doping. The comparison with experiment has
been discussed in Ref. \cite{GW05}.

In Fig. \ref{fsus}(b), the calculated $\chi _{s}$ versus $T$ at $\delta =0$
is shown together with the high temperature series expansion (HTSE) result%
\cite{HTSE}. It is noted that the experimental scaling curve actually
coincides with the half-filling HTSE very well.\cite{scaling,LSCO} Thus one
can clearly see the overall qualitative agreement between the bosonic RVB
theory and the experiment from Figs. \ref{fsus}(a) and (b). Note that the
mean-field $\chi _{s}$ deviates from the HTSE result prominently around $%
T_{0}$ where the latter is a much smoother function of $T.$ It reflects the
fact that $T_{0}$ is only a crossover temperature and the vanishing $\Delta
^{s}$ does not represent a true phase transition. Obviously, the amplitude
fluctuations beyond the mean-field $\Delta ^{s}$ have to be considered in
order to better describe $\chi _{s}$ in this regime. $T_{0}$ determined in
the mean-field theory is quite close to the HTSE result, indicating the
crossover temperature itself can still be reasonably decided by the
mean-field bosonic RVB description given above.

\subsubsection{Spin-lattice relaxation and spin-echo decay rates}

The NMR spin-lattice relaxation rate of nuclear spins is determined by\cite%
{GW05}
\begin{equation}
\frac{1}{T_{1}}=\frac{2k_{B}T}{g^{2}\mu _{B}^{2}N}\sum_{\mathbf{q}}F(\mathbf{%
q})^{2}\left. \frac{\chi _{zz}^{\prime \prime }(\mathbf{q},\omega )}{\omega }%
\right\vert _{\omega \rightarrow 0^{+}}  \label{NMRG1}
\end{equation}%
where the form factor $F(\mathbf{q})^{2}$ comes from the hyperfine coupling
between nuclear spin and spin fluctuations. Due to the fact that the $F(%
\mathbf{q})^{2}$ for planar $^{17}\mathrm{O(2)}$ nuclear spins vanishes at
the AF wave vector $\mathbf{Q}_{\mathrm{AF}}$, while for planar $^{63}%
\mathrm{Cu(2)}$ nuclear spins is peaked at $\mathbf{Q}_{\mathrm{AF}}$, a
combined measurement of $1/^{63}T_{1}$ and $1/^{17}T_{1}$ will thus provide
important information about the AF correlations at low frequency $\omega
\rightarrow 0$.

\begin{figure}[tbp]
\begin{center}
\includegraphics[width=2.5in]
{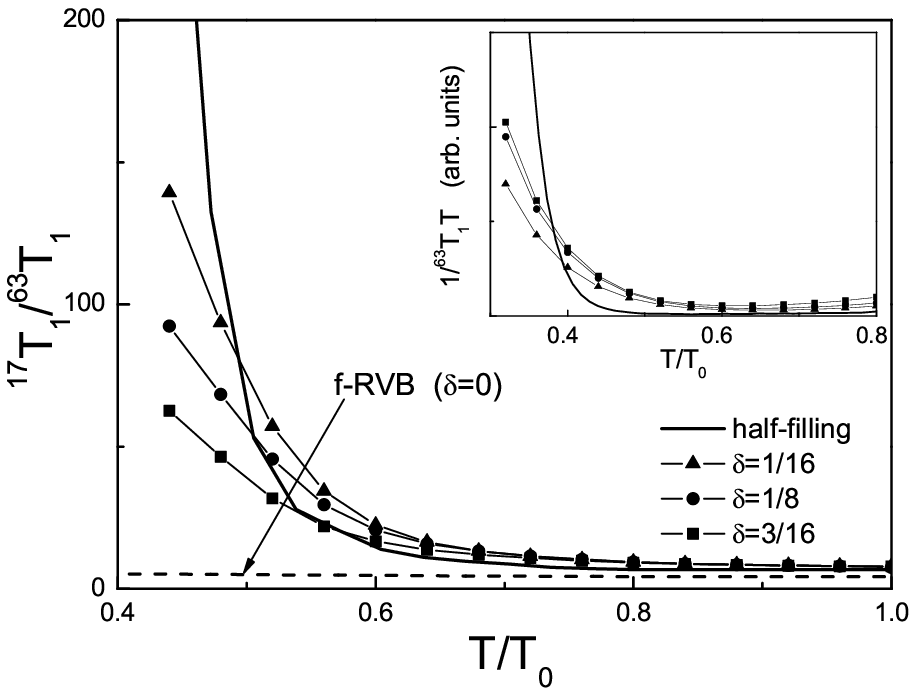}\includegraphics[width=2.5in] {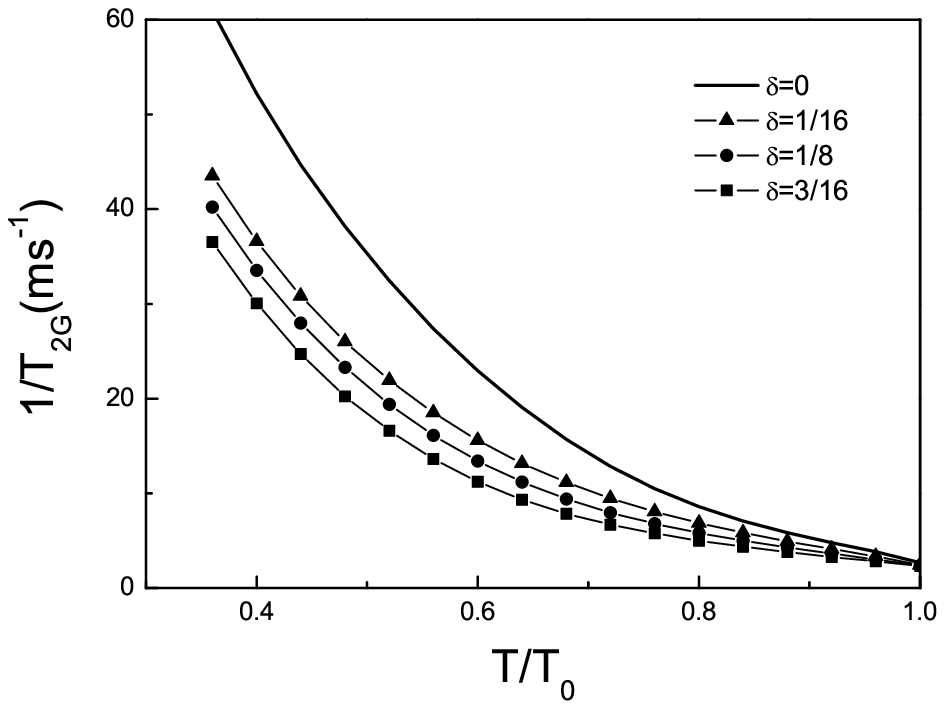}
\end{center}
\caption{Left panel: $^{17}T_{1}/^{63}T_{1}$ vs. temperature at different
doping concentrations in the upper pseudogap phase of the b-RVB state. The
dashed line shows the result of an f-RVB state ($\protect\pi $ flux phase)
at half-filling. The inset shows the non-Korringa behavior of $1/^{63}T_{1}T$
in the b-RVB state at various dopings. Right panel: $1/T_{2G}$ vs.
temperature in the upper pseudogap phase below $T_{0}$ at different doping
concentrations. [From Ref. \protect\cite{GW05}]}
\label{fnmr}
\end{figure}

Based on the mean-field equation,\cite{GW05} the calculated spin-lattice
relaxation rates, $1/^{63}T_{1}$ and $1/^{17}T_{1},$ for the planar copper
and oxygen nuclear spins are presented in the left panel of Fig. \ref{fnmr}.
It shows that the ratio $^{17}T_{1}/^{63}T_{1}$, which is a constant above $%
T_{0}$, starts to increase with reducing temperature below $T_{0}$. At lower
temperature, $T/T_{0}<0.5$, such a ratio arises sharply. For example, $%
^{17}T_{1}/^{63}T_{1}$ diverges at $\delta =0$ as a true AFLRO exists at $%
T=0;$ And it can still reach about $100$ in the low temperature limit at $%
\delta =0.125$, all qualitatively consistent with the experimental
observations. As pointed out above, such behavior clearly demonstrates that
strong low-lying AF correlations around $\mathbf{Q}_{\mathrm{AF}}$ develop
in the UPP, leading to the simultaneous enhancement of $1/^{63}T_{1}$ and
the cancellation in $1/^{17}T_{1}$. In the inset of the left panel in Fig. %
\ref{fnmr}, $1/^{63}T_{1}T$ has been plotted, which is also qualitatively
consistent with the experiment, but deviates from the conventional Korringa
behavior $1/^{63}T_{1}T\sim \mathrm{const}$ for a Fermi liquid system. By
contrast, the ratio $^{17}T_{1}/^{63}T_{1}$ in an f-RVB mean-field state
(the $\pi $ flux phase) at half-filling remains flat over the whole
temperature region as shown by the dashed line in Fig. \ref{fnmr},
indicating the absence of any significant AF correlations around $\mathbf{Q}%
_{\mathrm{AF}}$ in the pseudogap regime of the fermionic RVB state.

\begin{figure}[tbp]
\begin{center}
\includegraphics[width=3in]
{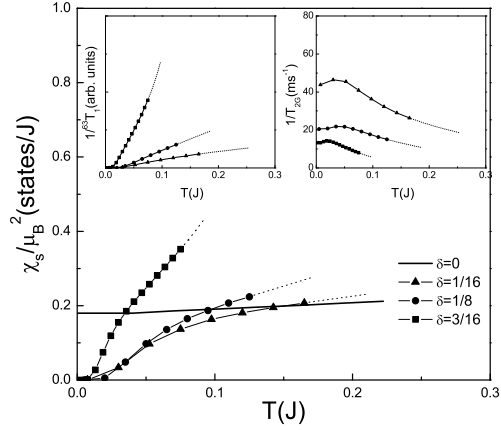}
\end{center}
\caption{Uniform spin susceptibility in the \emph{lower} pseudogap phase at
different dopings including half-filling. The left insert shows $%
1/^{63}T_{1} $ and the right insert $1/T_{2G}$ with the same symbols as in
the main panel. [From Ref. \protect\cite{GW05}]}
\label{6}
\end{figure}

The spin-echo decay rate $1/T_{2G}$, which is related to the static AF
correlations via the real part of spin susceptibility function, is also
examined in the right panel of Fig. \ref{fnmr}. It shows that $1/T_{2G}$
begins to increase with reducing temperature below $T_{0}$. Such behavior
has been also observed in the experiment, which once again clearly supports
the picture that the strong AF correlations start to develop in the UPP.

By comparison, the corresponding magnetic properties in the LPP are
presented Fig. \ref{6}. In the main panel, the uniform spin susceptibility
shows a true \textquotedblleft spin gap\textquotedblright\ behavior, in
contrast to the \textquotedblleft scaling\textquotedblright\ curve shown in
the UPP in Fig. \ref{fsus} where $\chi _{s}$ in the doped regime roughly
behaves like that at half-filling---in the latter case $\chi _{s}$ saturates
to a constant at $T=0$. In the LPP, these $\chi _{s}$'s can drop below that
at $\delta =0$ and vanish at\ $T=0$. Furthermore, $1/^{63}T_{1}$ also
decreases with temperature (see the left inset of Fig. \ref{6}), as opposed
to the behavior in the UPP, indicating the appearance of the spin gap over
whole momenta. On the other hand, although the low-energy spin fluctuations
are gapped, the static AF spin-spin correlations as described by the real
part of spin susceptibility function still remain, as reflected by $1/T_{2G}$
shown in the right inset of Fig. \ref{6}, where the monotonic increase of $%
1/T_{2G}$ in the UPP (Fig. \ref{fnmr}) is replaced by the saturation at the
LPP.

\subsection{Incoherent \textquotedblleft Normal State\textquotedblright :
Classical Regime}

As shown above, at $T>T_{0}$, the mean-field RVB order parameter $\Delta
^{s} $ vanishes and to the leading order of approximation the spins are
localized at the lattice sites with very weak correlations between them. The
residual AF superexchange coupling will come from $\left\vert \hat{\Delta}%
_{ij}^{s}\right\vert ^{2}$ which has been neglected in the minimal phase
string model.

In this regime, the charge dynamics will be highly nontrivial as governed by
$H_{h}$. According to Eq. (\ref{cond1}), an isolated spinon excitation will
behave like a $\pi $-flux tube as perceived by the holons and thus provides
a strong, unconventional charge scattering source. So at high temperature
where a lot of spinons ($\sim 1-\delta $) are thermally excited, one expects
a severe intrinsic frustration effect exerted from $A_{ij}^{s}$ on the
holons.

\subsubsection{Novel scattering mechanism for the charge carriers}

To see how the spin dynamics influences the charge degree of freedom via $%
A^{s}$, one may write down the propagator for $\mathbf{A}^{s}$ [Eq. (\ref%
{asc})]\cite{GW07}
\begin{eqnarray}
D_{\alpha \beta }^{A^{s}}(\mathbf{q},i\omega _{n}) &\equiv &\int_{0}^{\beta
}d\tau e^{i\omega _{n}\tau }\langle T_{\tau }A_{\alpha }^{s}(\mathbf{q,}\tau
)A_{\beta }^{s}(-\mathbf{q,0})\rangle  \nonumber \\
&=&-\left( \delta _{\alpha \beta }-\frac{q_{\alpha }q_{\beta }}{q^{2}}%
\right) \frac{4\pi ^{2}}{q^{2}a^{4}}\chi ^{zz}(\mathbf{q},i\omega _{n})
\label{pp}
\end{eqnarray}%
Define the local flux per plaquette (surrounding a lattice site) $\Phi _{%
{\small \square }}^{s}=a^{2}\mathbf{\hat{z}\cdot }\left( \nabla \times
\mathbf{A}^{s}\right) .$ Its total strength is generally determined by \
\begin{equation}
\left\langle \left( \Phi _{{\small \square }}^{s}\right) ^{2}\right\rangle
=\int d\omega \frac{1}{N}\sum_{\mathbf{q}}4\pi ^{2}S^{zz}(\mathbf{q},\omega )
\label{phi}
\end{equation}%
where the spin structure factor $S^{zz}(\mathbf{q},\omega )=\pi ^{-1}\left[
1+n\left( \omega \right) \right] \mathrm{Im}\chi ^{zz}(\mathbf{q},\omega )$.

In particular, at $T\gtrsim T_{0}$, there is no more significant AF
correlations among spins as $\Delta ^{s}=0$ and one finds $S^{zz}(\mathbf{q}%
,\omega )=\frac{1}{4\pi }\sqrt{\frac{(3-\delta )(1-\delta )}{3}}\delta
(\omega )$. Thus the corresponding gauge flux fluctuation becomes truly
\emph{static }with the weight $\sim \pi \sqrt{\frac{(3-\delta )(1-\delta )}{3%
}}$ concentrating at $\omega =0$. At low doping, $\sqrt{\left\langle \left(
\Phi _{_{\square }}^{s}\right) ^{2}\right\rangle }$ is comparable to the
simple picture that each excited spinon contribute to a flux of order of $%
\pi $, which represents the maximal frustration effect that the holons can
experience in the phase string model.\cite{GW07}
\begin{figure}[tbp]
\begin{center}
\includegraphics[width=3in]
{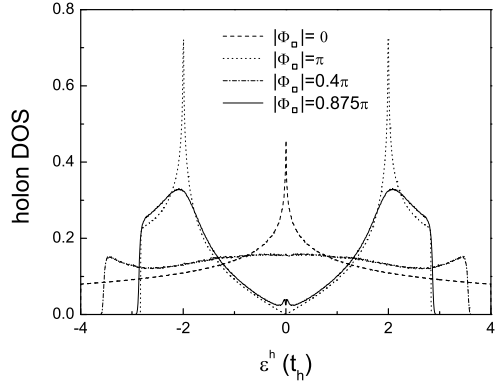}
\end{center}
\caption{Solid curve: the holon density of states (DOS) determined by $%
H_{h}\,$with the flux strength $\left\vert \Phi _{_{\square }}\right\vert
=(1-\protect\delta )\protect\pi $ at $\protect\delta =0.125$ and the
dash-dotted curve: with the reduced flux strength $\left\vert \Phi
_{_{\square }}\right\vert =0.4\protect\pi $. By comparison, the dashed curve
represents the flux-free limit, while the dotted curve corresponds to the
case in the presence of uniform $\protect\pi $ flux per plaquette. [From
Ref. \protect\cite{GW07}]}
\label{DOS}
\end{figure}

Figure \ref{DOS} illustrates\cite{GW07} how the holon density of states
(DOS) will get reshaped by $A^{s}$ or $\Phi _{{\small \square }}^{s}$. The
\emph{quenched} method is used to average over the static random flux
configurations for $\Phi _{{\small \square }}^{s}=\pm \left\vert \Phi _{%
{\small \square }}\right\vert $ on the lattice plaquette$.$ It shows how the
DOS drastically reshaped by the gauge field, i.e., the \emph{suppression} in
the high-energy (mid-band) DOS, as compared to the flux free case (dashed
curve). Note that the dotted curve in Fig. \ref{DOS} represents the DOS for
the case of a uniform $\pi $ flux per plaquette, which looks similar to the
random flux case $\left\vert \Phi _{{\small \square }}\right\vert =0.875$
except that the momenta remains well defined in a reduced Brillouin zone in
contrast to a strong mixing of momenta over a wide range by the scattering
effect in the latter.

\subsubsection{Optical conductivity}

\begin{figure}[tbp]
\begin{center}
\includegraphics[width=2.5in]
{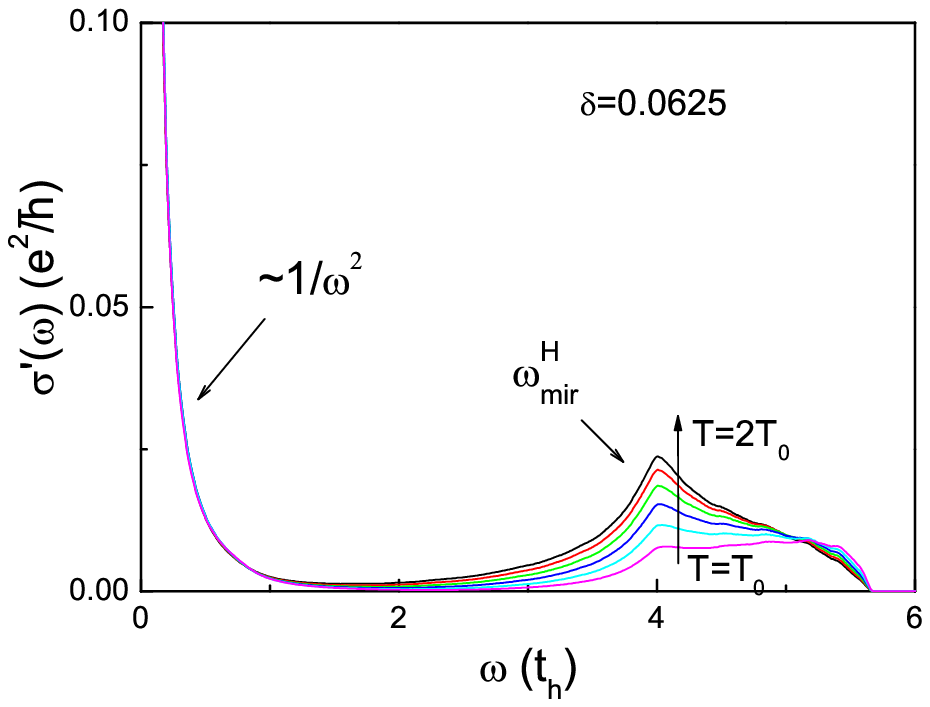}%
\includegraphics[width=2.5in]
{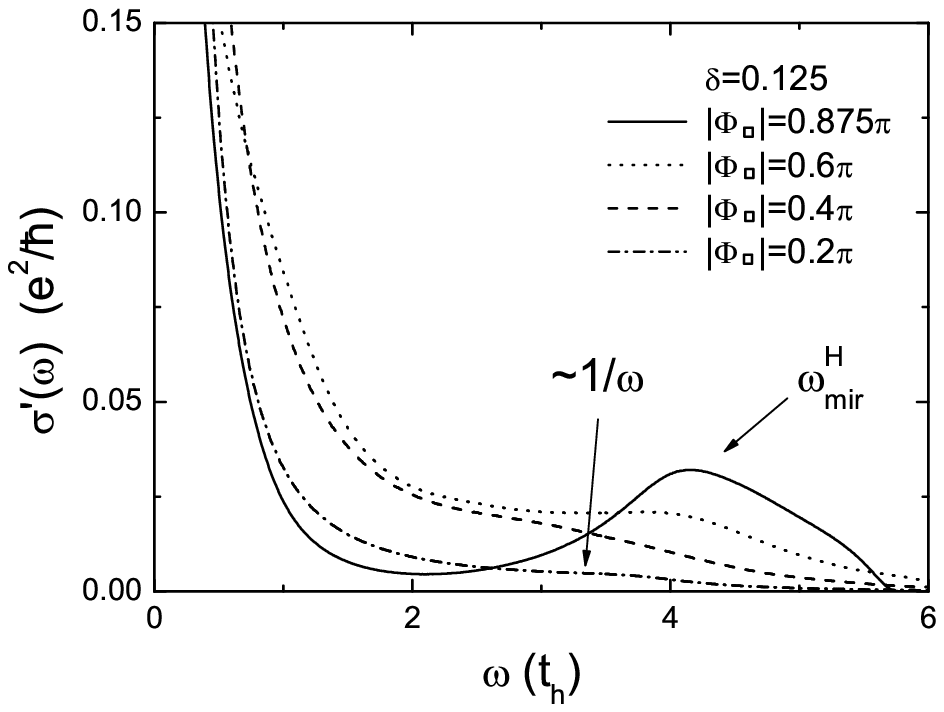}
\end{center}
\caption{Left: The real part of the optical conductivity at $\protect\delta %
=0.0625$ with $|\Phi _{{\protect\small \square }}|=(1-0.625)\protect\pi $.
Right: The optical conductivity at $\protect\delta =0.125$ with $|\Phi _{%
{\protect\small \square }}|$ chosen from $|\Phi _{{\protect\small \square }%
}|_{\max }=0.875\protect\pi $ to $0.2\protect\pi $ at a fixed $T=0.5t_{h}$.
[From Ref. \protect\cite{GW07}]}
\label{optical1}
\end{figure}

Figure \ref{optical1} shows the real part of the calculated optical
conductivity.\cite{GW07} The main feature of the spectral curves at various
temperatures $T\geq T_{0}$ is that there is generally a two-component
structure with a usual low-energy Drude component ($\sim 1/\omega ^{2}$) and
a mid-infrared resonancelike peak around the energy scale $\omega _{\mathrm{%
mir}}^{H}\sim 4t_{h}$. Furthermore, such a mid-infrared peak will actually
smoothly evolve into the $1/\omega $ behavior with reducing $\left\vert \Phi
_{_{\square }}\right\vert $, as clearly illustrated in the right panel of
Fig. \ref{optical1} at a fixed holon concentration $\delta =0.125$ where the
mid-infrared resonancelike peak at smaller $\left\vert \Phi _{_{\square
}}\right\vert $'s becomes softened and finally behaves like a $1/\omega $
tail in the regime $\sim 2t_{h}-4t_{h}$ with the weight shifting towards the
lower energy.

The origin of the mid-infrared resonance has been one of the most intriguing
optical properties in the underdoped cuprates. Normally a photon with the
momentum $\mathbf{q}\sim 0$ cannot excite a high-energy particle-hole pair
involving a large momentum transfer due to the momentum conservation law,
unless there is a scattering mechanism to strongly and widely smear the
momentum. This is difficult to realize in a conventional
electron-collective-mode coupling mechanism. The phase string model provides
an alternative scattering mechanism due to the strong correlation effect.

We have already seen that the effect of $A^{s}$ results in a double-peak
structure in the holon DOS (Fig. \ref{DOS}). In contrast to the uniform $\pi
$ flux case shown in the same figure, which also has a double-peak
structure, the high-energy inter-peak transition at $\mathbf{q}\rightarrow 0$
becomes possible in the random flux case due to the mixing between the small
and large momenta by the strong scattering via $A^{s}.$ This is the origin
for the mid-infrared peak found in Fig. \ref{optical1}.
\begin{figure}[tbp]
\begin{center}
\includegraphics[width=2.7in]
{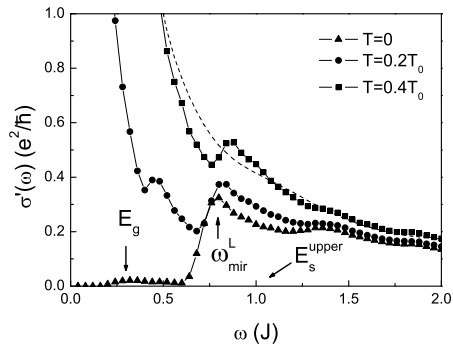}
\end{center}
\caption{Optical conductivity at different temperatures below $T_{0}$. A new
lower mid-infrared peak emerging at $\protect\omega _{\mathrm{mir}}^{L}$
which is weighted between the magnetic energy scales $E_{g}$ and $E_{s}^{%
\mathrm{upper}}$. [From Ref. \protect\cite{GW07}]}
\label{optical}
\end{figure}

Finally, as a comparison, the low-energy optical conductivity at $T<T_{0}$
can be obtained by using the perturbative method.\cite{GW07} Such an
approach is meaningful in the regime where the spin fluctuations are
substantially suppressed, which in turn results in the weak fluctuations of $%
\mathbf{A}^{s}$ according to Eq.\ (\ref{pp}). As shown in Fig. \ref{optical}%
, a prominent suppression of $\sigma ^{\prime }(\omega )$ at low-$\omega $
is present at $T=0$ with a second \textquotedblleft
mid-infrared\textquotedblright\ peak emerging around $\omega _{\mathrm{mir}%
}^{L}\sim 0.75J$ which sits somewhat between the two characteristic magnetic
energy scales$,$ $E_{g}$ and $E_{s}^{\mathrm{upper}}$, as marked in the
figure. Note that such a new energy scale in the low-$\omega $ optical
conductivity merely reflects some weighted energy scale based on the
magnetic $\mathrm{Im}\chi ^{zz}$. With the increase of temperature, the
\textquotedblleft gap\textquotedblright\ at low energy in $\sigma ^{\prime
}(\omega )$ is quickly filled up by the thermal excitations as shown in Fig. %
\ref{optical}. The lower \textquotedblleft mid-infrared\textquotedblright\
peak feature remains around $\omega _{\mathrm{mir}}^{L}$ at low temperature
throughout the LPP below $T_{v}.$ Note that $T_{v}$ is between $T_{c}$ and $%
T_{0},$ and the dashed curve at $T=0.4T_{0\text{ }}$is obtained by supposing
that $T>T_{v}$ where $\mathrm{Im}\chi ^{zz}$ behaves differently.\cite{GW05}
As compared to the solid curve at the same $T=0.4T_{0}$, which corresponds
to the case \emph{inside }the LPP, the overall difference is small except
for the vanishing the lower \textquotedblleft
mid-infrared\textquotedblright\ peak [Fig. \ref{optical}].

\subsubsection{Density-density correlation function}

The mid-infrared resonance peak of the $\mathbf{q}=0$ optical conductivity
has been attributed to a large-$\omega $ transition between the double peaks
of the holon DOS [Fig. \ref{DOS}]. In the following we discuss an
independent probe of such a peculiar DOS structure by studying the
density-density function at finite momentum $\mathbf{q}$ and energy $\omega
, $ and compare the results with the exact numerical calculations.

\begin{figure}[tbp]
\begin{center}
\includegraphics[width=3.6in]
{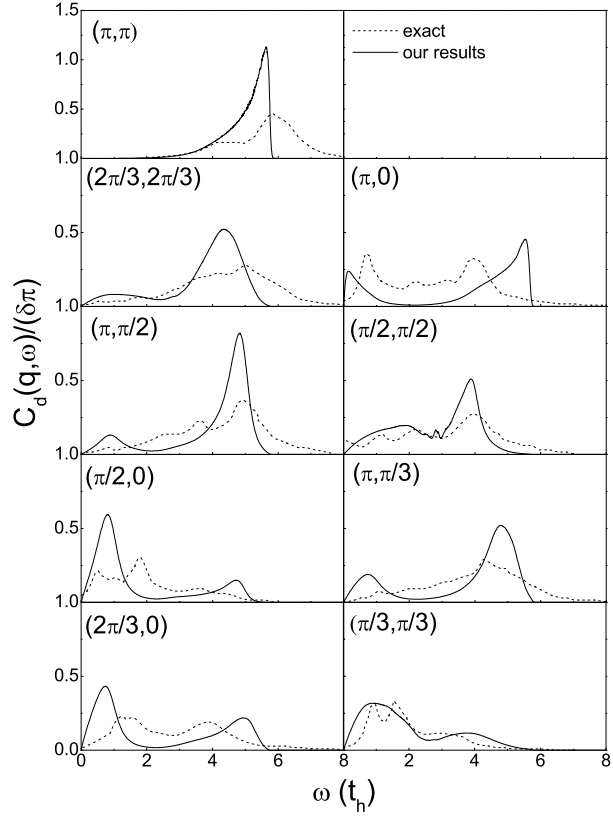}
\end{center}
\caption{The calculated density-density correlation function $C_{d}(\mathbf{q%
},\protect\omega )/\protect\delta \protect\pi $ at different momenta (solid
curves) with $T=0.5t_{h}$. The exact diagonalization results\protect\cite%
{exact2} are shown as dashed curves. [From Ref. \protect\cite{GW07}]}
\label{density}
\end{figure}

The imaginary part of the (retarded) density-density correlation function $%
C_{d}(\mathbf{q},\omega )$ is presented in Fig. \ref{density} (solid
curves), which evolves distinctively with different momenta. Note that $%
C_{d}(\mathbf{q},\omega )/\delta \pi $ is shown in the figure because it is
quantity roughly doping independent. For comparison, the exact
diagonalization results\cite{exact2} are presented as dotted curves. It is
interesting to see that the overall $\omega $-peak feature of the calculated
density-density correlation function is in qualitative and systematic
agreement with the numerical one at different $\mathbf{q}$'s without fitting
parameters (here $t$ is simply set at $t_{h}$ as the mid-infrared feature
peaks around $\sim 4t$ in the numerical calculation). Such a consistency
between the present effective theory and the exact diagonalization provides
another strong evidence that the gauge-coupling boson model (\ref{hh})
correctly captures the high-energy charge excitations in the $t$-$J$ model
and large-$U$ Hubbard model.

\subsubsection{Scattering rate}

Experimentally the scattering rate is normally defined by
\begin{equation}
\frac{1}{\tau (\omega )}=\left( \frac{\omega _{p}^{2}}{4\pi }\right) \mathrm{%
Re}\left[ \frac{1}{\sigma (\omega )}\right]  \label{1tao}
\end{equation}%
which is determined by the measured optical conductivity. Here $\omega _{p}$
denotes the plasma frequency, which in the present case is given by $\omega
_{p}=\sqrt{8\pi e^{2}\delta t_{h}}$.
\begin{figure}[tbp]
\begin{center}
\includegraphics[width=2.3in]{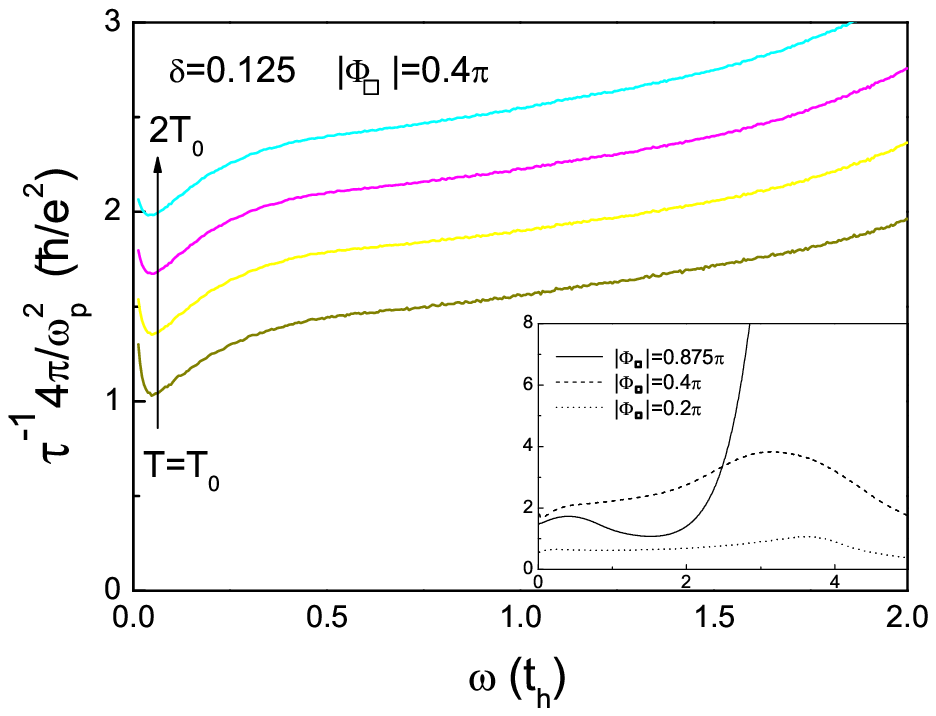} %
\includegraphics[width=2.4in]{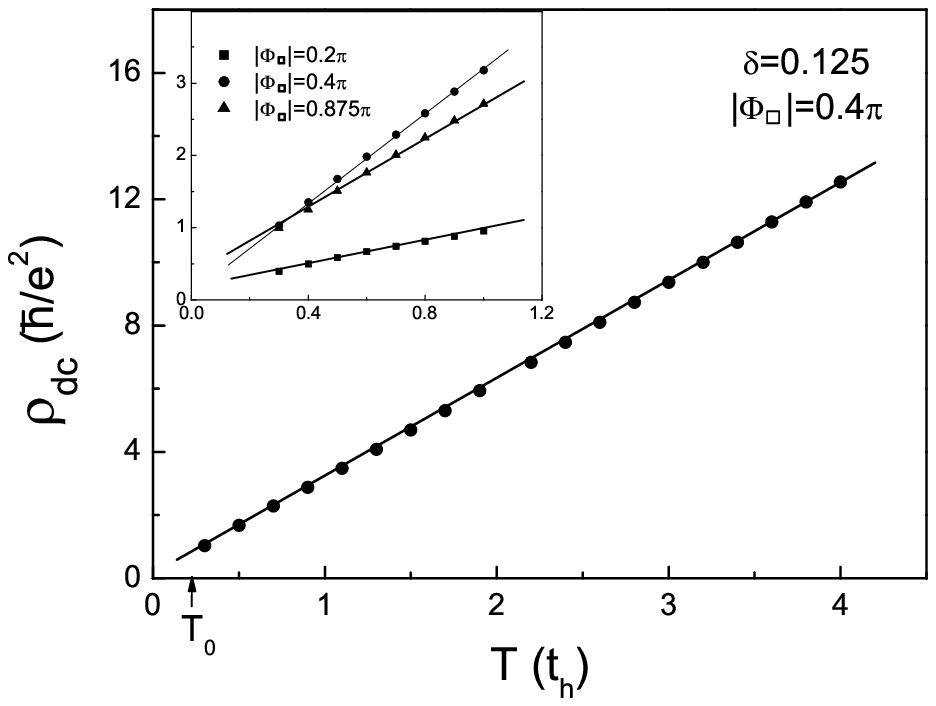}
\end{center}
\caption{Left: The scattering rate $1/\protect\tau (\protect\omega )$
defined by Eq. (\protect\ref{1tao}) at various temperatures between $%
T_{0}\simeq 0.25t_{h}$ and $2T_{0}$ which show a rough linear-$\protect%
\omega $ dependence over a wide range at $\protect\omega >k_{\mathrm{B}%
}T_{0} $. Inset: $1/\protect\tau (\protect\omega )$ vs. $\protect\omega $ at
different $\left\vert \Phi _{{\protect\small \square }}\right\vert $'s
corresponding to Fig. \protect\ref{optical1}. Right: The dc resistivity $%
\protect\rho _{\mathrm{dc}}=1/\protect\sigma ^{\prime }(\protect\omega )|_{%
\protect\omega \sim 0}$ as a function of temperature which is fit by a
straight line showing the linear-$T$ dependence. Inset: $\protect\rho _{%
\mathrm{dc}}$ at different $\left\vert \Phi _{{\protect\small \square }%
}\right\vert $'s which all show good linear-$T$ behavior with slightly
different slopes. [From Ref. \protect\cite{GW07}]}
\label{tao}
\end{figure}

In the left panel of Fig. \ref{tao}, $1/\tau (\omega )$ based on the
calculated $\sigma (\omega )$ is plotted as a function of $\omega $ in
different temperatures at $\delta =0.125$, with $\left\vert \Phi _{_{\square
}}\right\vert =0.4\pi $ which corresponds to the case where the high-$\omega
$ optical conductivity looks more like a $1/\omega $ behavior (Fig. \ref%
{optical1}). Here one finds that $1/\tau (\omega )$ increases monotonically
with $\omega $ and is roughly linear-$\omega $ dependent over a wide $\omega
$ region at $\omega >k_{\mathrm{B}}T_{0}$. Note that generally the $\omega $%
-dependence of $1/\tau (\omega )$ at higher energies is closely correlated
with the evolution of the aforementioned mid-infrared feature, as shown in
the inset of the left panel of Fig. \ref{tao}.

In particular, one sees a parallel shift of $1/\tau (\omega )$ with
increasing temperature at low-$\omega $, which implies a linear-temperature
dependence of the dc scattering rate. The dc scattering rate $1/\tau _{%
\mathrm{dc}}$ can be determined by extrapolating $1/\tau (\omega )$ to $%
\omega =0$. The obtained dc resistivity based on the Drude formula $\rho _{%
\mathrm{dc}}=\left( \omega _{p}^{2}/4\pi \right) \tau _{\mathrm{dc}}^{-1}$ $%
=1/\sigma ^{\prime }(0)$ is shown in the right panel of Fig. \ref{tao} which
is indeed quite linear over a very wide range of temperature at $T\geq T_{0}$%
.

It is important to note that $\sigma ^{\prime }(0)\propto $ $\beta $ over a
very wide range of the temperature at $T>T_{0}$ where the Bose distribution
factor $n(\xi _{m})\ll 1,$ \emph{i.e.}, in the classical regime of the
bosons ($\xi _{m}$ is the holon energy spectrum). The corresponding
scattering rate $\hslash /\tau _{\mathrm{dc}}\sim 0.7k_{B}T$ for the case
shown in the main right panel of Fig. \ref{tao}, whose slope is slightly $%
\left\vert \Phi _{_{\square }}\right\vert $ dependent as indicated in the
inset. Indeed, as discussed above, the bosonic degenerate regime for the
holons already ends up at $T_{v}$, \emph{i.e.,} at the boundary of the
LPP/SVP (Fig. \ref{phase3D}). At $T\geq T_{0}$, totally $1-\delta $ randomly
distributed $\pi $-flux tubes are perceived by the $\delta $ holons and the
latter behave like classical particles. One expects this anomalous transport
be smoothly connected to the Brinkman-Rice retracing path regime\cite{br} in
the large $T$ limit.

The dc scattering rate $\hslash /\tau _{\mathrm{dc}}\sim 2k_{B}T$ has been
previously obtained\cite{pali3} by the quantum Monte Carlo numerical method,
where the starting model is a system of interacting bosons coupled with
strong Gaussian fluctuations of the static gauge field of the strength $%
\left\langle \left( \Phi _{{\small \square }}^{s}\right) ^{2}\right\rangle .$
Note that $\left\langle \left( \Phi _{{\small \square }}^{s}\right)
^{2}\right\rangle $ used in the Monte Carlo simulation\cite{pali3} is about
the same order of magnitude as in the above case and, in particular, it is
\emph{temperature independent} in contrast to a linear-$T$ dependent $%
\left\langle \left( \Phi _{{\small \square }}^{s}\right) ^{2}\right\rangle $
predicted in the slave-boson \textrm{U(1) }gauge theory\cite{lee3} which was
the original motivation for such a Monte Carlo study.

\subsection{Low-Doping AF State: Beyond the Minimal Model}

At half-filling, the antiferromagnetism can be well described by $H_{s}$.
However, even in the presence of a very dilute hole concentration, an AFLRO
would be immediately destroyed due to the opening up of a spin gap $%
E_{g}\propto \delta J$ as predicted by the minimal phase string model, as
illustrated by the phase diagram in Fig. \ref{phase3D}. In the following we
discuss a modified phase diagram by taking into account of a new topological
excitation in this regime.

The motion of holes will generally induce the irreparable phase string
effect (Sec. 2.2.). In the dilute limit of the hole concentration, the phase
string effect should mainly influence the hole dynamics, without drastically
affecting the spin part which is AFLRO ordered in the ground state. It turns
out that the holes can be self-localized by the phase string effect here.%
\cite{WMST01,KW03,KW05} Without the condensation of the holons, then a spin
gap $E_{g}\propto \delta J$ will no longer exist in this dilute doping
regime.

\begin{figure}[tbp]
\begin{center}
\includegraphics[width=2.6in]{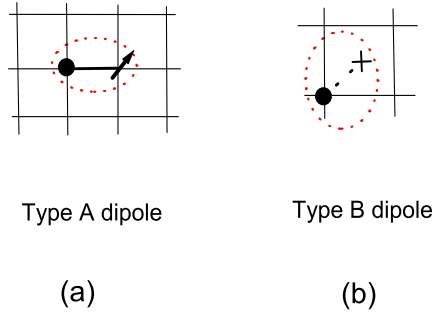}
\end{center}
\caption{(a) Type A dipole with a holon and a spin sitting at two poles at
nearest-neighboring lattice sites; (b) Type B dipole with one pole at the
center of a plaquette. [From Ref. \protect\cite{KW-1-03}]}
\label{dipole}
\end{figure}

Mathematically, a Z$_{2}$ topological excitation\cite{SF} (meron) is allowed
by the phase string model\cite{KW03,KW-1-03}
\begin{equation}
b_{i\sigma }\rightarrow b_{i\sigma }e_{i}^{i\frac{\sigma }{2}\vartheta
_{i}^{k}}  \label{bmeron}
\end{equation}%
\begin{equation}
h_{i}^{\dagger }\rightarrow h_{i}^{\dagger }e^{i\frac{1}{2}\vartheta
_{i}^{k}}  \label{hmeron1}
\end{equation}%
where $\vartheta _{i}^{k}=\pm \mathop{\rm Im}\ln (z_{i}-z_{k}^{0})$ with the
core position $z_{k}^{0}$ either inside a plaquette or on a lattice site.
Such a meron can be \textquotedblleft nucleated\textquotedblright\ from the
AF state where the spinons are Bose condensed with $\left\langle b_{i\sigma
}\right\rangle \neq 0$ and the holon becomes a topological vortex with a
logarithmically divergent energy.\cite{KW03,KW-1-03} A holon must be then
forced to be \emph{\textquotedblleft confined\textquotedblright } to a meron
to form a truly stable hole object, known as a \emph{hole dipole}$.$\cite%
{KW03,KW-1-03} Two typical hole-dipoles of minimal size are sketched in Fig. %
\ref{dipole}$:$
It can be shown that the effective mass of the induced meron is infinity
such that the hole-dipole object is self-trapped in space.\cite{KW-1-03,KW05}
\begin{figure}[tbp]
\begin{center}
\includegraphics[width=3in]{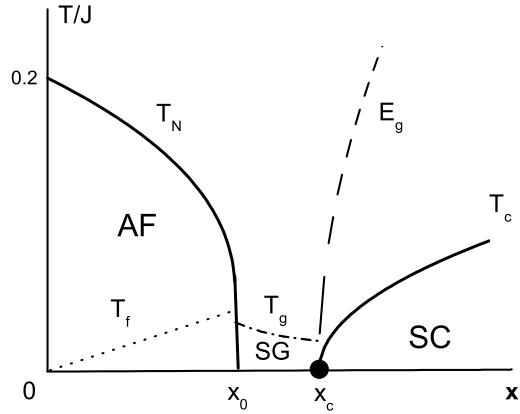}
\end{center}
\caption{Phase diagram at low doping $x$: a dual confinement-deconfinement
occurs at a quantum critical point $x_{c}\simeq 0.043$. The N\'{e}el
temperature $T_{N}$ vanishes at $x_{0}\simeq 0.03$. $T_{f}$ and $T_{g}$
denote characteristic spin freezing temperatures, and $T_{c}$ is the
superconducting transition temperature. [From Ref. \protect\cite{KW03}]}
\label{phase-ld}
\end{figure}

In such a self-localization regime, the kinetic energy of the holes is
suppressed. Without the balance from the kinetic energy, the low-energy
physics in this regime will be determined by potential energies. The latter
will then decide various competing orders in this low-doping insulating
phase. The AFLRO should persist if a weak\emph{\ interlayer} coupling is
considered.\cite{KW03} But the freedom in the directions of the hole-dipole
moment will lead to the reduction of the N\'{e}el temperature $T_{N}$ as
shown in Fig. \ref{phase-ld}. Based on the hole-dipole picture and the
renormalization group (RG) calculation,\cite{KW03} the critical doping $%
x_{0}\simeq 0.03$. Beyond $x_{0}$ or $T_{N},$ the system is in a cluster
spin glass phase with the dipole moments being quenched randomly in space.
With the further increase of doping, the sizes of hole dipoles will get
larger and larger, and eventually a deconfinement can occur at a critical
doping $\delta =x_{c}\simeq 0.043,$ beyond which single holons will be
unbound from their anti-vortex partners and experience a Bose condensation
leading to a finite $E_{g}\propto \sqrt{\delta -x_{c}}J$.\cite{KW03}

Some interesting properties including the thermopower,
variable-range-hopping resistivity, and dielectric constant, etc., in this
self-localization regime have been also discussed.\cite{KW05} Furthermore, a
possible stripe instability has been explored. Note that in the above it is
assumed that the hole dipoles are self-trapped \emph{uniformly }in space.
However, if there is no impurities or disorders, the uniform distribution of
the self-trapped hole-dipoles may not be stable against the formation of the
stripes due to the long-range dipole-dipole interaction.\cite{KW-1-03,YK05}

So in the low-doping regime, the minimal phase string model in Eqs. (\ref{hh}%
) and (\ref{hs}) can be generalized to

\begin{eqnarray}
H_{h} &=&-t_{h}\sum_{\langle ij\rangle }\left( e^{iA_{ij}^{s}+\Omega
_{ij}}\right) h_{i}^{\dagger }h_{j}+H.c.  \label{mhh} \\
H_{s} &=&-J_{s}\sum_{\langle ij\rangle \sigma }\left( e^{i\sigma \left[
A_{ij}^{h}-\Omega _{ij}\right] }\right) b_{i\sigma }^{\dagger }b_{j-\sigma
}^{\dagger }+H.c.  \label{mhs}
\end{eqnarray}%
to include a Z$_{2}$ gauge field $\Omega _{ij}$:
\begin{equation}
\sum_{\square }\Omega _{ij}=\QATOPD\{ . {{\pm \pi ,}}{{0,}}  \label{z2}
\end{equation}%
which is allowed by the general construction of the phase string model and
compatible with the bosonic RVB pairing $\Delta ^{s}$. Normally the core
energy of a Z$_{2}$ vortex is too big in the superconducting phase, but a Z$%
_{2}$ vortex excitation can become important at $\delta \leq x_{c}$ in the
AF state. Furthermore, a quasipaticle excitation discussed in Sec. 4.1.7.
may be equivalently considered as a bound state of a spinon, a holon, and a Z%
$_{2}$ vortex. The details will be presented elsewhere.

\section{Synthesis and Perspectives}

In this brief review, I have surveyed a systematic effort in the study of a
doped antiferromagnet, which may be relevant to the mechanism of the high-$%
T_{c}$ cuprates. The core of this approach lies in the so-called phase
string effect, which has been mathematically identified based on the $t$-$J$
model. It is by nature a \emph{frustration effect} induced by the motion of
doped holes in an antiferromagnetic spin background.

Such a frustration effect on the spin degrees of freedom differs
fundamentally from an ordinary frustrated antiferromagnet in the presence of
geometrically \textquotedblleft mismatched\textquotedblright\ spin
interactions, \emph{e.g., }the next nearest neighbor superexchange coupling.
The key distinction is that the frustration in the former is dynamically
driven and mutual between the spin and charge degrees of freedom. Namely the
extent that the spins get frustrated crucially depends on the charge
behavior and \emph{vise versa}. In different doping, temperature, magnetic
field, or other parameter regimes, the spin and charge parts will then
behave differently in order to minimize the \emph{total} free energy. For
example, in the dilute hole limit, the superexchange energy of spins will
dominate and with maintaining longer range antiferromagnetic correlations
the kinetic energy of doped holes can get severely frustrated by the phase
string effect, resulting in their self-localization at low temperature; At
higher doping, to gain the kinetic energy of the doped holes, however, the
spin correlations can be \textquotedblleft forced\textquotedblright\ to
become short-ranged via the phase string effect, and the spin background
becomes a spin liquid state. The superconducting phase coherence and nodal
quasiparticle excitation are protected by the spin gap of such a spin liquid
state.

The mathematical description of the phase string effect is rather simple,
which is basically represented by a sequence of signs [Eq. (\ref{pstring})]
picked up by the nearest neighboring hoppings of the holes in a Heisenberg
spin background. It depends on the path of the hole hopping as well as the
polarizations of those spins exchanged with the hole during its motion. It
is thus both geometric and dynamic, which weights each motion path of the
holes. We have seen that such a phase string is \emph{irreparable} in the
sense that the system cannot generate other signs to compensate it at each
path. In fact, the Heisenberg superexchange interaction respects the
Marshall signs, so a phase string as the disordered Marshall signs caused by
hopping cannot be \textquotedblleft self-healed\textquotedblright\ through
the superexchange process.

Such an irreparable phase string effect identified as the most important
frustration effect in the doped antiferromagnet is singularly sensitive to
any perturbative treatment, since Eq.\ (\ref{pstring}) will change sign for
a fluctuation with merely one additional or less $\downarrow $ spin
exchanged with the hole on a given path, no matter how long the path is.
Fortunately a unitary transformation exists in the $t$-$J$ model which can
precisely keep track of such a phase string because it essentially involves
the counting of the exchanges occurring between the holes and spins during
the their travelling. Then in the new representation after the unitary
transformation, known as the phase string formalism, the $t$-$J$ model
presumably becomes less singular and perturbatively treatable.

The $t$-$J$ model in the \emph{exact }phase string formalism is a
topological gauge model, in which the phase string effect is precisely
described by a pair of mutual Chern-Simons gauge fields in two dimensions.
Without these gauge fields, the model reduces to a full bosonic one free
from any \textquotedblleft sign problem\textquotedblright . In other words,
the nontrivial spin and charge dynamics will be governed by the topological
gauge fields which precisely reflect the phase string effect. Thus, the exact%
\emph{\ }phase string formalism of the $t$-$J$ model provides a unique
starting point to study the doped antiferromagnet.

Based on the precise topological gauge structure and the good understanding
of the half-filling phase, an effective minimal phase string model working
for small doping can be then constructed as given by Eqs. (\ref{hh}) and (%
\ref{hs}). Despite its novel looking, this is a rather simple model where
two \emph{bosonic} matter fields, spinless holons and neutral spinons,
interact with each other by perceiving the opposite species as a $\pi $ flux
fluxoid, and the spinons form the RVB pairing whose amplitude is
self-consistently determined.
\begin{figure}[tbp]
\begin{center}
\includegraphics[width=3in]{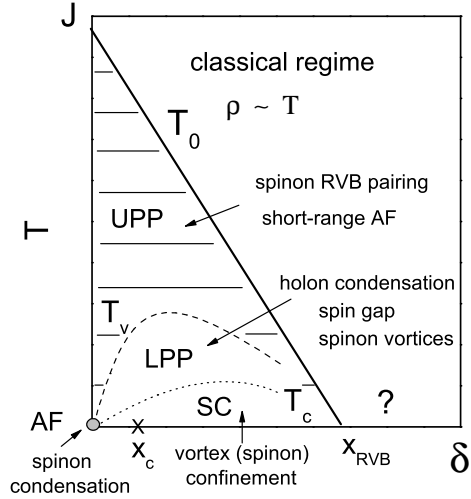}
\end{center}
\caption{The global phase diagram of the minimal phase string model. }
\label{phase-sum}
\end{figure}

Figure \ref{phase-sum} summaries the global phase diagram of this minimal
model. The \textquotedblleft normal state\textquotedblright\ at $T>T_{0}$ is
an incoherent \textquotedblleft classical\textquotedblright\ regime for both
spinons and holons, where the spinons are weakly correlated, while the
diffusive holons are maximally scattered by the $\pm \pi $ flux fluxoids
bound to the randomly distributed spinons, leading to a linear-temperature
scattering rate. At $T\leq T_{0}$, the spinons start to form the singlet RVB
pairing, and the short-range antiferromagnetic correlations become enhanced
with the reducing temperature, as clearly indicated by the NMR spin
relaxation rate and spin-echo decay rate. This regime is known as the upper
pseudogap phase, which is superexchange energy driven and continuously
evolves into the AFLRO state at half-filling and $T=0$. The holons remain
diffusive in this phase, although the scattering from the gauge field gets
reduced with decreasing temperature as more and more spinons are RVB paired
with the cancellation of their $\pm \pi $ fluxoids. Eventually at $T\leq
T_{v}$, the gauge flux is reduced so much that the bosonic coherence between
the holons can be finally established, which defines the lower pseudogap
phase that is obviously kinetic energy driven. In this phase, with the holon
condensation, a spin gap is opened up for the low-lying antiferromagnetic
fluctuations, with the weight being pushed up and concentrated at a
resonancelike peak at $E_{g}$ around $\mathbf{Q}_{\mathrm{AF}}=(\pi ,\pi )$,
and the spin correlation length gets \textquotedblleft
truncated\textquotedblright\ at a finite scale comparable to the average
hole-hole distance. A very unique feature in this regime is the presence of
a lot of spontaneous spinon vortices, composites of isolated spinons locking
with current vortices, which are responsible for the Nernst effect, residual
diamagnetism, and spin Hall effect, etc. In this peculiar phase, the Cooper
pair amplitude is finite because of the RVB pairing of the spinons and Bose
condensation of the holons. But the phase is disordered as the proliferation
of the spinon vortices. Eventually, at a lower temperature $T_{c}$, the
binding of vortices-antivortices or the confinement of the spinons will
finally lead to the superconducting phase coherence. It turns out that both
the kinetic energy of the holons and superexchange energy of the spinons are
generally benefited from this phase transition. The nodal quasiparticles
also become coherent in the superconducting phase as the result of the phase
coherence and spinon confinement. The non-BCS neutral $S=1$ spin excitation
is similar to the lower pseudogap phase with a reasonancelike structure at $%
E_{g}$ and the suppression of antiferromagnetic correlations below $E_{g}$,
as caused by the holon condensation.

Such a top-down phase diagram demonstrates an amazing richness of the
minimal phase string model, which covers almost all the interesting regimes
of the high-$T_{c}$ cuprate superconductors, except for the very underdoped
regime where the superconducting phase ends as well as the overdoped regime
where the upper pseudogap terminates, which is question-marked in Fig. \ref%
{phase-sum}.

At half filling, with the vanishing gauge fields, the minimal model does
produce an AFLRO state with a very precise variational ground-state energy.
But once away from half-filling, the holon condensation will force the
opening up of a spin gap $E_{g}\propto \delta J$ and thus the disappearance
of the AFLRO. But this is unphysical at sufficiently low doping where the
long-range antiferromagnetic correlations should remain dominant. Indeed, in
this regime a topological Z$_{2}$ vortex can become a low-lying excitation
once being bound to a holon to form a localized composite object, known as a
hole dipole. In this regime the correct low-energy phase string model is
given by Eqs. (\ref{mhh}) and (\ref{mhs}), and a hole dipole in Fig. \ref%
{dipole} can be regarded as the realization of the dual holon confinement in
the antiferromagnetic phase in contrast to the spinon confinement in the
superconducting phase.

In the overdoped regime where the spinon RVB pairing disappears at $T=0$,
the minimal phase string model should be also modified. A possibility is for
the Z$_{2}$ vortex to be bound with a bosonic spinon such that two gauge
fields are effectively cancelled in Eq. (\ref{mhh}), in favor of the holon
condensation as well as the kinetic energy. In this way, the bosonic spinons
will be effectively turned into fermions and a Fermi liquid state may be
recovered if the bosonic holons remain condensed in the high-doping regime.
So the phase string model may simply reduce to the slave-boson mean-field
description at $\delta \gtrsim x_{\mathrm{RVB}}$.

Finally we remark on that throughout this paper, only the nearest
neighboring hopping of the $t$-$J$ model is considered, which is related to
the origin of the singular phase string effect. However, the phase string
effect will get qualitatively modified in the presence of the next nearest
neighbor hopping process. This is an important issue which has not been
touched upon so far. Just like the phase string effect is very singular in
the original $t$-$J$ model, the next nearest neighbor hopping term will also
become singular in the phase string formalism. It is thus expected to be
important to interpret the detailed experimental measurements in the
cuprates, \emph{e.g., }the asymmetry in the hole- and electron-doped
cuprates.

\section{Acknowledgements}

This work is partially supported by the NSFC grants.

\end{document}